\documentclass[11pt,a4paper]{book}
\usepackage[latin1]{inputenc}
\usepackage[T1]{fontenc}
\usepackage{hyperref}
\usepackage[nottoc]{tocbibind} 
\usepackage{fancyhdr}
\usepackage{minitoc}
\usepackage{graphicx}
\usepackage{color}
\usepackage{bm} 
\usepackage{amssymb} 
\usepackage{mathrsfs} 

\hypersetup{pdftitle={An introduction to the theory of rotating relativistic stars},
  pdfsubject={CompStar 2010 School},
  pdfauthor={Eric Gourgoulhon <eric.gourgoulhon@obspm.fr>},
  pdfkeywords={LaTeX},
  colorlinks=true
}
\newcommand{\encadre}[1]{\fbox{$\displaystyle #1$}}
\newcommand{\der}[2]{\frac{\partial #1}{\partial #2}}
\newcommand{\dder}[3]{\frac{\partial^2 #1}{\partial #2\partial #3}}
\newcommand{\dderr}[2]{\frac{\partial^2 #1}{\partial {#2}^2}}
\newcommand{\dert}[2]{{\partial #1}/{\partial #2}}

\newcommand{\w}[1]{\bm{#1}}
\newcommand{\vv}[1]{\vec{\w{#1}}}

\newcommand{\vp}[1]{\overrightarrow{#1}}

\newcommand{\uu}[1]{\underline{\w{#1}}}

\newcommand{\wnab}{\w{\nabla}}
\newcommand{\dd}{\bm{\mathrm{d}}}

\newcommand{\be}{\begin{equation}}
\newcommand{\ee}{\end{equation}}
\newcommand{\bea}{\begin{eqnarray}}
\newcommand{\eea}{\end{eqnarray}}

\newcommand{\M}{\mathscr{M}}
\newcommand{\T}{\mathscr{T}}
\newcommand{\Sp}{\mathscr{S}}
\newcommand{\vxi}{\vv{\xi}}
\newcommand{\vchi}{\vv{\chi}}

\newcommand{\Lie}[1]{\mathcal{L}_{#1}\,}
\newcommand{\Liec}[1]{{\mathcal{L}}_{\vv{#1}}\,}
\newcommand{\vpar}{\vv{\partial}}
\newcommand{\ph}{\varphi}
\newcommand{\Df}{\mathcal{D}}
\newcommand{\Li}{\mathscr{L}}
\newcommand{\Pp}{\mathscr{P}}

\newcommand{\R}{\mathbb{R}}
\newcommand{\Obs}{\mathcal{O}}
\newcommand{\vep}{\varepsilon}

\newcommand{\rr}{{\bar r}}
\newcommand{\defin}[1]{\textbf{\itshape #1}}

\newenvironment{remark}[1][]%
{\begin{description} \item[{\small Remark #1:}]\small}%
{\end{description}}

{\begin{description} \item[{\small Example #1:}]\small}%
{\end{description}}

{\begin{description} \item[{\small Examples:}]\small}%
{\end{description}}

{\par\smallskip\noindent\textbf{Historical note: }\em }{\par\smallskip}

\begin{document}



\begin{titlepage}
\ \\[2cm]
\begin{center}
\textcolor{blue}{\textbf{\Huge An introduction}} \\[1cm]
\textcolor{blue}{\textbf{\Huge to the theory of}} \\[1cm]
\textcolor{blue}{\textbf{\Huge rotating relativistic stars}} \\[2cm]
{\Large \slshape Lectures at the COMPSTAR 2010 School \\[5mm]
 GANIL, Caen, France, 8-16 February 2010} \\[2cm]
{\Large \'Eric Gourgoulhon} \\[1cm]
{Laboratoire Univers et Th\'eories, \\
UMR 8102 du CNRS, Observatoire de Paris, \\
Universit\'e Paris Diderot - Paris 7 \\
92190 Meudon, France} \\
\href{mailto:eric.gourgoulhon@obspm.fr}{\texttt{eric.gourgoulhon@obspm.fr}}\\[3cm]
20 September 2011
\end{center}
\end{titlepage}

\dominitoc

\tableofcontents


%
%

\chapter*{Preface} \label{s:ava}
\addstarredchapter{Preface}

These notes are the written version of lectures given at the Compstar 2010 
School\footnote{\url{http://ipnweb.in2p3.fr/compstar2010/}} held in Caen on 
8-16 February 2010. They are intended to introduce the theory of rotating stars in general relativity. Whereas the application is clearly towards neutron stars (and associated 
objects like strange quark stars), these last ones are not discussed here, except for illustrative purposes. 
Instead the focus is put on the theoretical foundations, with a detailed discussion 
of the spacetime symmetries, the choice of coordinates and the derivation of the 
equations of structure from the Einstein equation. The global properties of rotating stars (mass, angular momentum, redshifts, orbits, etc.) are also introduced. 
These notes can be fruitfully complemented by N.~Stergioulas' review \cite{Sterg03} and by J.L.~Friedman \& N.~Stergioulas' textbook \cite{FriedS10}. 
In addition, the textbook by P.~Haensel, A.~Y.~Potekhin \& D.~G.~Yakovlev \cite{HaensPY07} provides a good introduction to rotating neutron star models computed by the method exposed here. 

The present notes are limited to perfect fluid stars. For more 
complicated physics (superfluidity, dissipation, etc.), see N.~Andersson \& G.~L.~Comer's review \cite{AnderG07}. 
Magnetized stars are treated in J.~A.~Pons' lectures \cite{Pons10}. 
Moreover, the stability of rotating stars is not investigated here. 
For a review of this rich topic, see N.~Stergioulas' article \cite{Sterg03} or L.~Villain's one \cite{Villa06}. 

To make the exposure rather self-contained, a brief overview of general relativity and the 3+1 formalism is given in Chap.~\ref{s:ein}. For the experienced reader, this can be considered as a mere presentation of the notations used in the text.
Chap.~\ref{s:sym} is entirely devoted to the spacetime symmetries relevant for 
rotating stars, namely stationarity and axisymmetry. The choice of coordinates is also discussed in this chapter. The system of partial differential equations governing the
structure of rotating stars is derived from the Einstein equation in 
Chap.~\ref{s:eer}, where its numerical resolution is also discussed.
Finally, Chap.~\ref{s:glo} presents the global properties of rotating stars, some of them being directly measurable by a distant observer, like the 
redshifts. The circular orbits around the star are also discussed in this chapter.

\section*{Acknowledgements}

I would like to warmly thank Jérôme Margueron for the perfect organization of the Compstar 2010 School, as well as Micaela Oertel and Jérôme Novak for their help in the preparation of the \texttt{nrotstar} code described in Appendix~\ref{s:lor}. I am also very grateful to John Friedman for his careful reading of the first version of these notes and his suggestions for improving both the text and the content. Some of the studies reported here are the fruit of a long and so pleasant collaboration with Silvano Bonazzola, Pawe\l{} Haensel, Julian Leszek Zdunik, Dorota Gondek-Rosi\'nska and, more recently, Micha\l{} Bejger.
To all of them, I express my deep gratitude. 

Corrections and suggestions for improvement are welcome at
\href{mailto:eric.gourgoulhon@obspm.fr}{\texttt{eric.gourgoulhon@obspm.fr}}.  

%
%

\chapter{General relativity in brief} \label{s:ein}


\minitoc
\vspace{1cm}


\section{Geometrical framework}

\subsection{The spacetime of general relativity} \label{s:ein:gr_space_time}

Relativity has performed the fusion of \emph{space} and \emph{time}, two notions
which were completely distinct in Newtonian mechanics. This gave rise to 
the concept of \emph{spacetime}, on which both the special and general theory 
of relativity are based. Although this is not particularly fruitful (except for
contrasting with the relativistic case), one may also speak of \emph{spacetime} 
in the Newtonian framework: the Newtonian spacetime $\M$ is nothing but the 
affine space $\R^4$, foliated by the hyperplanes $\Sigma_t$ 
of constant absolute time $t$:
these hyperplanes represent the ordinary 3-dimensional space at successive instants.
The foliation $(\Sigma_t)_{t\in\R}$
is a basic structure of the Newtonian spacetime and does not depend
upon any observer. The \defin{worldline} $\mathcal{L}$ of a particle is the curve 
in $\M$ generated by the successive positions of the particle.
At any point $A\in\mathcal{L}$, 
the time read on a clock moving along $\mathcal{L}$ is simply the
parameter $t$ of the hyperplane $\Sigma_t$ that intersects $\mathcal{L}$
at $A$.

The spacetime $\M$ of special relativity is the 
same mathematical space as the
Newtonian one, i.e. the affine space $\R^4$. The major difference with
the Newtonian case is that there does not exist any privileged foliation
$(\Sigma_t)_{t\in\R}$. Physically this means that the notion of
absolute time is absent in special relativity. 
However $\M$ is still endowed with some absolute structure: 
the \defin{metric tensor} $\w{g}$ and the
associated \defin{light cones}. The metric tensor is a symmetric bilinear form $\w{g}$
on $\M$, which defines the scalar product of vectors. The null (isotropic) 
directions of $\w{g}$ give the worldlines of photons (the light cones). Therefore
these worldlines depend only on the absolute structure $\w{g}$ and not, for instance,
on the observer who emits the photon.

The spacetime $\M$ of general relativity differs from both Newtonian and special relativistic spacetimes, in so far as it is no longer the affine space
$\R^4$ but a more general mathematical structure, namely a \emph{manifold}.
A \defin{manifold of dimension 4} is a topological space such that around each point 
there exists a neighbourhood which is 
homeomorphic to an open subset of $\R^4$.
This simply means that, locally, one can label the points of $\M$ in a
continuous way by 4 real numbers $(x^\alpha)_{\alpha\in\{0,1,2,3\}}$
(which are called \defin{coordinates}). To cover the full $\M$, several different
coordinate patches (\defin{charts} in mathematical jargon) can be required. 

\begin{figure}
\centerline{\includegraphics[width=0.7\textwidth]{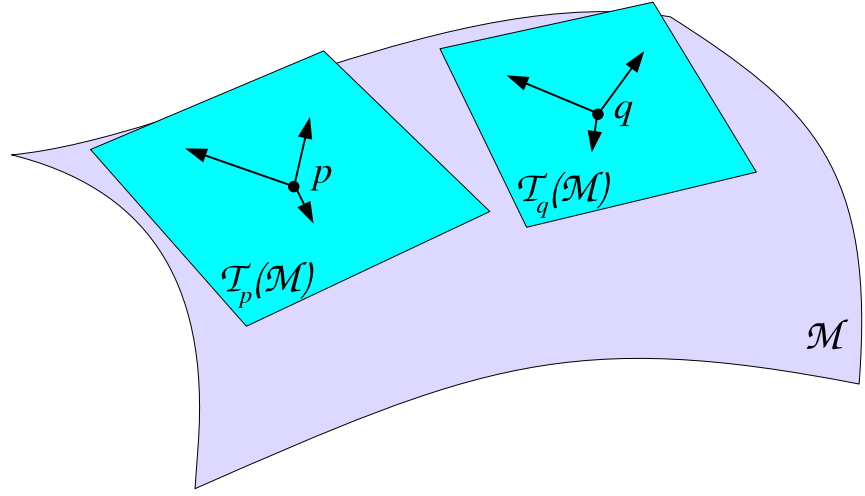}}
\caption{\label{f:esp_tangent} \small
The vectors at two points $p$ and $q$ on the
spacetime manifold $\M$ belong to two different vector spaces: 
the tangent spaces $\T_p(\M)$ and $\T_q(\M)$.}
\end{figure}

Within the manifold structure the definition of vectors is not as trivial
as within the affine structure of the Newtonian and special relativistic
spacetimes. Indeed, only infinitesimal vectors connecting 
two infinitely close points can be defined a priori on a manifold. At a given
point $p\in\M$, the set of such vectors generates a 4-dimensional vector
space, which is called the \defin{tangent space} at the point $p$ and is
denoted by $\T_p(\M)$. The situation is therefore different from the Newtonian
or special relativistic one, for which the very definition of an affine space
provides a unique global vector space. On a manifold
there are as many vector spaces as points $p$ (cf. Fig.~\ref{f:esp_tangent}).

Given a coordinate system $(x^\alpha)$ around some point $p\in\M$, one can define a
vector basis $(\vpar_\alpha)_{\alpha\in\{0,1,2,3\}}$ of the tangent space $\T_p(\M)$ by asking that the components of the infinitesimal vector $\vp{p p}'$ that connect neighbouring points 
$p$ (coordinates $(x^\alpha)$) and $p'$ (coordinates $(x^\alpha+dx^\alpha)$) are 
the infinitesimal increments $dx^\alpha$ in the coordinates when moving from $p$ to
$p'$:
\be \label{e:ein:ppp}
    \vp{p p}' = \sum_{\alpha = 0}^3 dx^\alpha \, \vpar_\alpha 
      = dx^\alpha \, \vpar_\alpha .
\ee 
For the second equality in the above equation, we have used \defin{Einstein summation convention} on repeated indices, which amounts to omitting the $\Sigma$ sign. We shall make a 
systematic use of this convention in these notes. 
The vector basis $(\vpar_\alpha)$ is called the \defin{natural basis associated with the coordinates $(x^\alpha)$}. The notation $\vpar_\alpha$ is inspired by 
$\partial/\partial x^\alpha$ and is a reminiscence of the intrinsic definition of vectors on a manifold as differential operators acting on scalar fields (see e.g. \cite{Wald84}).

For any vector $\vv{v}\in\T_p(\M)$ (not necessarily an infinitesimal one), one defines the components $(v^\alpha)$ of $\vv{v}$ in the basis $(\vpar_\alpha)$ according to the expansion
\be
  \vv{v} = v^\alpha \, \vpar_\alpha . 
\ee

\subsection{Linear forms}

Besides vectors, important objects on the manifold $\M$ are linear forms. 
At each point $p\in\M$, a \defin{linear form} is a an application\footnote{We are using the same bra-ket notation as in quantum mechanics to denote the action of a linear form on a vector.} 
\be 
	\begin{array}{rccl}
	\w{\omega}: & \T_p(\M) & \longrightarrow & \R \\
		& \vv{v} & \longmapsto & \langle \w{\omega}, \vv{v} \rangle 
	\end{array}
\ee
that is linear: 
$\langle\w{\omega},\lambda \vv{v} + \vv{u}\rangle =  \lambda \langle\w{\omega},\vv{v}\rangle +  \langle\w{\omega},\vv{u}\rangle$. The set of all linear forms at $p$ constitutes a 4-dimensional vector
space, which is called the \defin{dual space of $\T_p(\M)$} and denoted by
$\T_p(\M)^*$. A field of linear forms on $\M$ is called a \defin{1-form}. 
Given the natural basis $(\vpar_\alpha)$ of $\T_p(\M)$ associated with the coordinates 
$(x^\alpha)$, there is a unique basis of $\T_p(\M)^*$, denoted by $(\dd x^\alpha)$, such that
\be \label{e:ein:dual_basis}
  \encadre{ \langle \dd x^\alpha ,\vpar_\beta\rangle = \delta^\alpha_{\ \, \beta} } ,  
\ee
where $\delta^\alpha_{\ \, \beta}$ is the \defin{Kronecker symbol} : 
$\delta^\alpha_{\ \, \beta} = 1$ if $\alpha=\beta$ and $0$ otherwise. 
The basis $(\dd x^\alpha)$ is called the \defin{dual basis} of the basis 
$(\vpar_\alpha)$. The notation $(\dd x^\alpha)$ stems from the fact that if we apply 
the linear form $\dd x^\alpha$ to the infinitesimal displacement vector 
(\ref{e:ein:ppp}), we get nothing but the number $dx^\alpha$: 
\be \label{e:ein:dxa_dxa}
    \langle\dd x^\alpha,\vp{p p}'\rangle = \langle\dd x^\alpha , dx^\beta \, \vpar_\beta
    \rangle
    = dx^\beta \langle \dd x^\alpha , \vpar_\beta \rangle = dx^\beta \, \delta^\alpha_{\ \, \beta}
    = dx^\alpha . 
\ee
The dual basis can be used to expand any linear form $\w{\omega}$, thereby defining its
components $\omega_\alpha$: 
\be \label{e:ein:def_comp_form}
  \w{\omega} = \omega_\alpha \, \dd x^\alpha . 
\ee
In terms of components, the action of a linear form on a vector takes then a very simple form:
\be
  \encadre{ \langle\w{\omega},\vv{v}\rangle  = \omega_\alpha v^\alpha }. 
\ee 

Given a smooth \defin{scalar field} on $\M$, namely a smooth application 
$f:\;  \M \rightarrow \R$, there exists a 1-form canonically associated with it, called
the \defin{gradient of $f$} and denoted $\dd f$. It is the unique linear form which, once
applied to the infinitesimal displacement vector $\vp{p p}'$, gives the change in 
$f$ between points $p$ and $p'$: 
\be \label{e:ein:df}
  df := f(p') - f(p) = \dd f (\vp{p p}') . 
\ee
Since $df = \dert{f}{x^\alpha} \, dx^\alpha$, formula (\ref{e:ein:dxa_dxa}) implies that the components of the gradient with respect to the dual basis are nothing but the partial derivatives of $f$ with respect to the coordinates $(x^\alpha)$ : 
\be
  \dd f = \der{f}{x^\alpha} \, \dd x^\alpha . 
\ee

\begin{remark}
In non-relativistic physics, the gradient is very often 
considered as a vector and not as a 1-form. 
This is so because one associates implicitly a vector
$\vec{\omega}$ to any 1-form $\omega$ via the Euclidean scalar product 
of $\R^3$: $\forall \vec{v}\in \R^3,\ \langle \omega,\vec{v}\rangle = \vec{\omega}\cdot\vec{v}$.
Accordingly, formula (\ref{e:ein:df}) is rewritten as
$df = \vec{\nabla} f \cdot \overrightarrow{p p}'$. But one shall keep in mind
that, fundamentally, the gradient is a linear form and not a vector.
\end{remark}

\subsection{Tensors}  \label{s:tensors}

In relativistic physics, an abundant use is made
of linear forms and their generalizations: the tensors.
A \defin{tensor of type} $(k,\ell)$, also called \defin{tensor $k$ times
contravariant and $\ell$ times covariant}, is an application 
\be \label{e:def_tensor}
	\begin{array}{rccl}
	\w{T}: & \underbrace{\T_p(\M)^*\times\cdots\times\T_p(\M)^*}_{k {\ \rm times}}
	\times \underbrace{\T_p(\M)\times\cdots\times\T_p(\M)}_{\ell {\ \rm times}}
	& \longrightarrow & \R  \\
	& (\w{\omega}_1,\ldots,\w{\omega}_k,\vv{v}_1,\ldots,\vv{v}_\ell) 
		& \longmapsto & 
	\w{T}(\w{\omega}_1,\ldots,\w{\omega}_k, \vv{v}_1,\ldots,\vv{v}_\ell)
	\end{array}
\ee
that is linear with respect to each of its arguments. The integer $k+\ell$ is
called the \defin{valence} of the tensor. Let us recall the canonical duality
$\T_p(\M)^{**}=\T_p(\M)$, which means that every vector $\vv{v}$ can be considered
as a linear form on the space $\T_p(\M)^*$, via
$\vv{v}:\; \T_p(\M)^*\rightarrow \R$, 
$\w{\omega}\mapsto \langle \w{\omega},\vv{v}\rangle$.
Accordingly a vector is a tensor of type $(1,0)$. A linear form is a
tensor of type $(0,1)$. A tensor of type $(0,2)$ is called a
\defin{bilinear form}. It maps couples of vectors to real numbers, in a linear way for each
vector. 

Given the natural basis $(\vpar_\alpha)$ associated with some coordinates $(x^\alpha)$
and the corresponding dual basis $(\dd x^\alpha)$, we
can expand any tensor $\w{T}$ of type $(k,\ell)$ as
\be \label{e:def_components}
	\encadre{\w{T} = T^{\alpha_1\ldots\alpha_k}_{\qquad\ \; \beta_1\ldots\beta_\ell}
		\; \vpar_{\alpha_1} \otimes \ldots \otimes \vpar_{\alpha_k} 
                \otimes
		\dd x^{\beta_1} \otimes \ldots \otimes \dd x^{\beta_\ell} } ,
\ee
where the \defin{tensor product} $ \vpar_{\alpha_1} \otimes \ldots \otimes \vpar_{\alpha_k} \otimes
\dd x^{\beta_1} \otimes \ldots \otimes \dd x^{\beta_\ell}$ is the tensor of
type $(k,\ell)$ for which the image of  $(\w{\omega}_1,\ldots,\w{\omega}_k,\vv{v}_1,\ldots,\vv{v}_\ell)$ as in 
(\ref{e:def_tensor}) is the real number
\[
	\prod_{i=1}^k \langle \w{\omega}_i,\vpar_{\alpha_i}\rangle \;\times\; 
	\prod_{j=1}^\ell \langle \dd x^{\beta_j},\vv{v}_j\rangle .
\]
Notice that all the products in the above formula are simply products in $\R$.
The $4^{k+\ell}$ scalar coefficients  $T^{\alpha_1\ldots\alpha_k}_{\qquad\ \; \beta_1\ldots\beta_\ell}$ in (\ref{e:def_components}) are called the \defin{components
of the tensor $\w{T}$ with respect the coordinates $(x^\alpha)$}. 
These components are unique and fully characterize
the tensor $\w{T}$. 

\begin{remark}
The notation $v^\alpha$ and $\omega_\alpha$ already introduced for the components 
of a vector $\vv{v}$ or a linear form $\w{\omega}$ are of course the 
particular cases $(k,\ell)=(1,0)$ and $(k,\ell)=(0,1)$ of the general definition given above.
\end{remark}

\subsection{Metric tensor}

As for special relativity, the fundamental structure given on the spacetime manifold $\M$ 
of general relativity is the \defin{metric tensor} $\w{g}$. 
It is now a field on $\M$: at each point $p\in\M$, $\w{g}(p)$ is a
bilinear form acting on vectors in the tangent space $\T_p(\M)$:
\be 
	\begin{array}{rccl}
	\w{g}(p): & \T_p(\M)\times\T_p(\M) & \longrightarrow & \R \\
		& (\vv{u},\vv{v}) & \longmapsto & \w{g}(\vv{u},\vv{v}) .
	\end{array}
\ee
It is demanded that 
\begin{itemize}
\item $\w{g}$ is \defin{symmetric}: $\w{g}(\vv{u},\vv{v}) = \w{g}(\vv{v},\vv{u})$; 
\item $\w{g}$ is \defin{non-degenerate}: 
any vector $\vv{u}$ that is orthogonal to all vectors 
($\forall \vv{v}\in\T_p(\M),\quad  \w{g}(\vv{u},\vv{v}) = 0$) is necessarily the null vector; 
\item $\w{g}$ has a \defin{signature} $(-,+,+,+)$: in any basis of $\T_p(\M)$ where the components $(g_{\alpha\beta}$) form a diagonal matrix, there is necessarily one negative component ($g_{00}$ say) and three positive ones ($g_{11}$, $g_{22}$ and $g_{33}$ say). 
\end{itemize}
\begin{remark}
An alternative convention for the signature of $\w{g}$ exists in the literature (mostly in 
particle physics): $(+,-,-,-)$. It is of course equivalent to the convention adopted
here (via a mere replacement of $\w{g}$ by $-\w{g}$). The reader is warned that the two signatures may lead to different signs in some formulas (not in the values of
measurable quantities !). 
\end{remark}

The properties of being symmetric and non-degenerate are typical of a \defin{scalar product}, thereby justifying the notation
\be
	\encadre{ \vv{u}\cdot\vv{v} := \w{g}(\vv{u},\vv{v}) = g_{\alpha\beta} \, u^\alpha v^\beta }
\ee
that we shall employ throughout. The last equality in the above equation involves
the components $g_{\alpha\beta}$ of $\w{g}$, which according to 
definition (\ref{e:def_components}), are given by 
\be
  \w{g}=g_{\alpha\beta} \, \dd x^\alpha\otimes \dd x^\beta  . 
\ee

The isotropic directions of $\w{g}$ give the local \defin{light cones}:
a vector $\vv{v}\in\T_p(\M)$ satisfying $\vv{v}\cdot\vv{v} = 0$
is tangent to a light cone and called a
\defin{null} or \defin{lightlike} vector.
If $\vv{v}\cdot\vv{v} < 0$, the vector $\vv{v}$ is said \defin{timelike}
and if $\vv{v}\cdot\vv{v} > 0$, $\vv{v}$ is said \defin{spacelike}.

\subsection{Covariant derivative} \label{s:ein:cov_deriv}

Given a vector field $\vv{v}$ on $\M$, it is natural to consider the variation of $\vv{v}$ between
two neighbouring points $p$ and $p'$. But one faces the following problem: from the manifold structure alone, the variation of $\vv{v}$ cannot be defined by a formula analogous to 
that used for a scalar field [Eq.~(\ref{e:ein:df})], namely
$d\vv{v} = \vv{v}(p') - \vv{v}(p)$, because $\vv{v}(p')$ and $\vv{v}(p)$ belong to different
vector spaces: $\T_{p'}(\M)$ and $\T_p(\M)$ respectively (cf. Fig.~\ref{f:esp_tangent}). 
Accordingly the subtraction $\vv{v}(p') - \vv{v}(p)$ is not well defined !
The resolution of this problem is provided by introducing an \defin{affine connection} on the manifold, i.e. of a operator $\wnab$ that, at any point $p\in\M$, 
associates to any infinitesimal displacement vector $\vp{p p}'\in\T_p(\M)$ a vector of
$\T_p(\M)$ denoted $\wnab_{\vp{p p}'} \vv{v}$  that one defines as the \defin{variation of $\vv{v}$ between $p$ and $p'$}:
\be
    d\vv{v} = \wnab_{\vp{p p}'} \vv{v} .
\ee
The complete definition of $\wnab$ allows for any vector $\vv{u}\in\T_p(\M)$ instead of
$\vp{p p}'$ and includes all the properties of a derivative operator (Leibniz rule,
etc...), which we shall not list here (see e.g. Wald's textbook \cite{Wald84}). 
One says that $\vv{v}$ is \defin{transported parallelly to itself} between $p$ and $p'$
iff $d\vv{v}=0$.

The name \emph{connection} stems from the fact that, by providing the definition of
the variation $d\vv{v}$, the operator $\wnab$ \emph{connects} the adjacent tangent spaces 
$\T_p(\M)$ and $\T_{p'}(\M)$.
From the manifold structure alone, there exists an infinite number of possible 
connections and none is preferred. Taking account the metric tensor $\w{g}$ changes the
situation: there exists a unique connection, called the 
\defin{Levi-Civita connection}, such that the tangent vectors
to the geodesics with respect to $\w{g}$ are transported parallelly to themselves
along the geodesics. In what follows, we will make use only of the Levi-Civita
connection.

Given a vector field $\vv{v}$ and a point $p\in\M$, we can consider the
type $(1,1)$ tensor at $p$ denoted by $\w{\nabla}\vv{v}(p)$ and defined by
\be \label{e:def_nabv_p}
	\begin{array}{cccc}
	\w{\nabla}\vv{v}(p) \ : & \T_p(\M)^*\times\T_p(\M) & \longrightarrow & \R \\
		& (\w{\omega},\vv{u}) & \longmapsto & 
	\w{\omega}(\w{\nabla}_{\vv{u}} \vv{v})
	\end{array} .
\ee
By varying $p$ we get a type
$(1,1)$ tensor field denoted $\w{\nabla}\vv{v}$ and called the \defin{covariant 
derivative} of $\vv{v}$.

The covariant derivative is extended to any tensor
field by (i) demanding that for a scalar field
$\w{\nabla} f := \dd f$ and (ii) using the Leibniz rule.
As a result, the covariant derivative of a tensor field $\w{T}$ of type $(k,\ell)$ is
a tensor field $\w{\nabla}\w{T}$ of type $(k,\ell+1)$.
Its components with respect a given coordinate system $(x^\alpha)$
are denoted 
\be
\nabla_\gamma T^{\alpha_1\ldots\alpha_k}_{\qquad\ \; \beta_1\ldots\beta_\ell}
	:= 
(\w{\nabla}\w{T})^{\alpha_1\ldots\alpha_k}_{\qquad\ \; \beta_1\ldots\beta_\ell\gamma}
\ee
(note the position of the index $\gamma$ !) and are given by
\bea
\nabla_\gamma T^{\alpha_1\ldots\alpha_k}_{\qquad\ \; \beta_1\ldots\beta_\ell}&=&
 \der{}{x^\gamma} T^{\alpha_1\ldots\alpha_k}_{\qquad\ \; \beta_1\ldots\beta_\ell} 
+ \sum_{i=1}^k \Gamma^{\alpha_i}_{\ \, \gamma\sigma}\; T^{\alpha_1\ldots
\!{{{\scriptstyle i\atop\downarrow}\atop \scriptstyle\sigma}\atop\ }\!\!
\ldots\alpha_k}_{\qquad\ \ \ \  \  \  \beta_1\ldots\beta_\ell} \nonumber \\
& & -  \sum_{i=1}^\ell \Gamma^\sigma_{\ \, \gamma\beta_i} \; 
T^{\alpha_1\ldots\alpha_k}_{\qquad\ \; \beta_1\ldots
\!{\ \atop {\scriptstyle\sigma \atop {\uparrow\atop \scriptstyle i}} }\!\!
\ldots\beta_\ell}  ,	\label{e:cov_derivT_comp}
\eea 
where the coefficients $\Gamma^\alpha_{\ \, \gamma\beta}$ are the
\defin{Christoffel symbols} of the metric $\w{g}$ with respect to the coordinates $(x^\alpha)$.
They are expressible in terms of the partial derivatives of the components of 
the metric tensor, via
\be \label{e:ein:Christoffel}
	 \Gamma^\alpha_{\ \, \gamma\beta} := \frac{1}{2} g^{\alpha\sigma}
	\left( \der{g_{\sigma\beta}}{x^\gamma} + \der{g_{\gamma\sigma}}{x^\beta}
	- \der{g_{\gamma\beta}}{x^\sigma} \right) , 
\ee
where $g^{\alpha\beta}$ stands for the components of the inverse matrix\footnote{Since $\w{g}$
is non-degenerate, its matrix $(g_{\alpha\beta})$ is always invertible, whatever the coordinate system.} of $(g_{\alpha\beta})$: 
\be \label{e:ein:inv_metric}
  g^{\alpha\mu} g_{\mu\beta} = \delta^\alpha_{\ \, \beta}.
\ee

A distinctive feature of the Levi-Civita connection is that
\be \label{e:nabla_g_zero}
	\encadre{ \w{\nabla} \w{g} = 0 }. 
\ee

Given a vector field $\vv{u}$ and a tensor field $\w{T}$ of type $(k,\ell)$, 
we define the \defin{covariant derivative of $\w{T}$ along $\vv{u}$} as 
\be
	\w{\nabla}_{\vv{u}} \, \w{T} := \w{\nabla}\w{T}
        (\underbrace{.,\ldots,.}_{k+\ell\ {\rm slots}},\vv{u}) .
\ee
Note that $\w{\nabla}_{\vv{u}} \, \w{T}$ is a tensor of the same type $(k,\ell)$
as $\w{T}$ and that its components are
\be
	\left(\w{\nabla}_{\vv{u}} \, \w{T}
	\right)^{\alpha_1\ldots\alpha_k}_{\qquad\ \; \beta_1\ldots\beta_\ell}
	= u^\mu \nabla_\mu T^{\alpha_1\ldots\alpha_k}_{\qquad\ \; \beta_1\ldots\beta_\ell} .
\ee

\section{Einstein equation} \label{s:ein:ee}

General relativity is ruled by the Einstein equation: 
\be \label{e:ein:ee}
	\encadre{ \w{R} - \frac{1}{2} R \, \w{g} = \frac{8\pi G}{c^4} \w{T} },  
\ee
where $\w{R}$ is the Ricci tensor associated with the Levi-Civita connection $\wnab$
(to be defined below), 
$R := \mathrm{tr}_{\w{g}} \w{R} = g^{\mu\nu} R_{\mu\nu}$ is the trace of $\w{R}$
with respect to $\w{g}$ and $\w{T}$ is the energy-momentum tensor of matter and electromagnetic field\footnote{Or any kind of field which are present in spacetime.} (to be defined below). 
The constants $G$ and $c$ are respectively Newton's gravitational constant and the speed of light. 
In the rest of this lecture, we choose units such that
\be
      G = 1 \qquad\mbox{and}\qquad c = 1 . 
\ee

The \defin{Ricci tensor} $\w{R}$ is a part of the \defin{Riemann tensor} $\w{\mathcal{R}}$, a 
type $(1,3)$ tensor 
which describes the \defin{curvature} of $\wnab$, i.e. the tendency of two initially parallel 
geodesics to deviate from each other. Equivalently the Riemann tensor measures the lack of 
commutativity of two successive covariant derivatives:
\be
     \nabla_\mu\nabla_\nu  v^\alpha
        - \nabla_\nu\nabla_\mu v^\alpha
        = \mathcal{R}^\alpha_{\ \, \beta\mu\nu} \, v^\beta  .
\ee
The Ricci tensor $\w{R}$ is defined as the following trace of the Riemann tensor:
\be
  R_{\alpha\beta} := \mathcal{R}^\mu_{\ \, \alpha\mu\beta} .
\ee
It is expressible in terms of the derivatives of the components of the metric tensor with respect
to some coordinate system $(x^\alpha)$ as
\be \label{e:ein:Ricci}
	R_{\alpha\beta}  = 
	\der{\Gamma^\mu_{\ \, \alpha\beta}}{x^\mu} 
	- \der{\Gamma^\mu_{\ \, \alpha\mu}}{x^\beta} 
	+ \Gamma^\mu_{\ \, \alpha\beta} \Gamma^\nu_{\ \, \mu\nu} 
	- \Gamma^\nu_{\ \, \alpha\mu} \Gamma^\mu_{\ \, \nu\beta} ,
\ee
where the $\Gamma^\mu_{\ \, \alpha\beta}$ have to be replaced by their expression 
(\ref{e:ein:Christoffel}). Although this not obvious on the above expression, the Ricci tensor is
symmetric : $R_{\beta\alpha} = R_{\alpha\beta}$. 

Having defined the left-hand side of Einstein equation (\ref{e:ein:ee}), let us focus on the 
right-hand side, namely the \defin{energy-momentum tensor} $\w{T}$. 
To define it, let us introduce a generic observer $\Obs$ in spacetime. It is described by its
worldline $\mathscr{L}$, which is a timelike curve in $\M$, and the future-directed unit vector
$\vv{u}$ tangent to $\mathscr{L}$ (the so-called observer \defin{4-velocity}): 
\be
  \vv{u}\cdot\vv{u} = -1 . 
\ee
An important operator is the projector $\w{\bot}$ onto the 3-dimensional vector space $E_{\w{u}}$ orthogonal
to $\vv{u}$, which can be considered as the local rest space of the observer $\Obs$.  
$\w{\bot}$ is a type $(1,1)$ tensor, the components of which are
\be \label{e:ein:comp_ortho_proj}
      \bot^\alpha_{\ \, \beta} = \delta^\alpha_{\ \, \beta} + u^\alpha u_\beta . 
\ee
In particular, $\bot^\alpha_{\ \, \beta} u^\beta = u^\alpha + u^\alpha u_\beta u^\beta 
  =  u^\alpha + u^\alpha (-1) = 0$. 
The energy-momentum tensor $\w{T}$ is the bilinear form defined by the following properties which
must be valid for any observer $\Obs$: 
\begin{itemize}
\item the \emph{energy density} measured by $\Obs$ is 
\be \label{e:ein:E_Tuu}
  E = \w{T}(\vv{u},\vv{u}) = T_{\mu\nu} \, u^\mu u^\nu \ ; 
\ee
\item the \emph{momentum density} measured by $\Obs$ is the 1-form $\w{p}$ the components of 
which are 
\be \label{e:ein:pa_T}
  p_\alpha = - T_{\mu\nu} \, u^\nu \bot^\mu_{\ \, \alpha} \ ; 
\ee
\item the \emph{energy flux ``vector''}\footnote{For an electromagnetic field, this is the
Poynting vector.} measured by $\Obs$ is the 1-form $\w{\varphi}$ the 
components of which are 
\be \label{e:ein:phia_T}
  \varphi_\alpha = - T_{\mu\nu} \, u^\mu \bot^\nu_{\ \, \alpha} \ ; 
\ee
\item the \emph{stress tensor} measured by $\Obs$ is the tensor $\w{S}$ the components of 
which are 
\be \label{e:ein:Sab_T}
  S_{\alpha\beta} = T_{\mu\nu}  \bot^\mu_{\ \, \alpha} \bot^\nu_{\ \, \beta} . 
\ee
\end{itemize}
A basic property of the energy-momentum tensor is to be symmetric. Anyway, it could not be
otherwise since the left-hand side of Einstein equation (\ref{e:ein:ee}) is symmetric. 
From Eqs.~(\ref{e:ein:pa_T}) and (\ref{e:ein:phia_T}), the symmetry of $\w{T}$
implies the equality of the momentum density and the energy flux 
form\footnote{Restoring units where $c\not=1$, this relation must read 
$\w{p} = c^{-2} \w{\varphi}$.}: 
$\w{p} = \w{\varphi}$ . This reflects the equivalence of mass and energy in relativity. 

Another property of $\w{T}$ follows from the following geometric property, known as 
\defin{contracted Bianchi identity}: 
\be
  \nabla^\mu \left( R_{\alpha\mu} - \frac{1}{2} R \, g_{\alpha\mu} \right) = 0 , 
\ee
where $\nabla^\mu := g^{\mu\nu} \nabla_\nu$. The Einstein equation (\ref{e:ein:ee}) implies
then the vanishing of the divergence of $\w{T}$: 
\be \label{e:ein:divT_zero}
   \encadre{ \nabla^\mu T_{\alpha\mu} = 0 } . 
\ee
This equation is a local expression of the conservation of energy and momentum. 

An astrophysically very important case of energy-momentum tensor is that of a
\defin{perfect fluid}:
\be \label{e:ein:T_perfect_fluid}
  \encadre{ \w{T} = (\vep + p) \uu{u}\otimes \uu{u} + p\, \w{g} } , 
\ee
where $\uu{u}$ is the 1-form associated by metric duality\footnote{The components of 
$\uu{u}$ are $u_\alpha = g_{\alpha\beta} u^\beta$.} to a unit timelike vector field $\vv{u}$
representing the fluid 4-velocity, $\vep$ and $p$ are two scalar fields, representing respectively 
the energy density and the pressure, both in the fluid frame. Indeed, plugging 
Eq.~(\ref{e:ein:T_perfect_fluid}) into Eqs.~(\ref{e:ein:E_Tuu})-(\ref{e:ein:Sab_T}) leads
respectively to the following measured quantities by an observer comoving with the fluid: 
\[
  E = \vep,\quad p_\alpha = 0, \quad, \varphi_\alpha = 0, \quad S_{\alpha\beta} = p \, \bot_{\alpha\beta} . 
\]

\section{3+1 formalism} \label{s:ein:3p1}

The 3+1 decomposition of Einstein equation is the standard formulation used in numerical relativity \cite{Alcub08,BaumgS10,Gourg07a}. 
It is also very useful in the study of 
rotating stars. Therefore we introduce it briefly here (see e.g. \cite{Gourg07a} for an extended presentation).

\begin{figure}
\centerline{\includegraphics[width=0.5\textwidth]{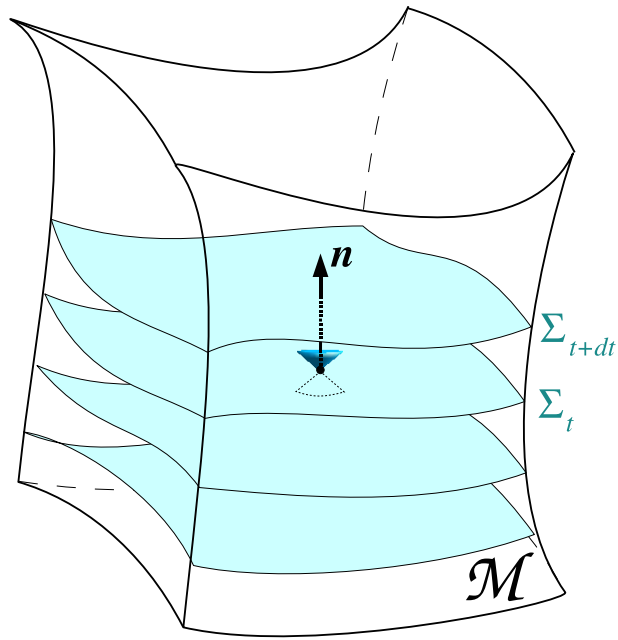}}
\caption[]{\label{f:ein:foliat} \small
Foliation of the spacetime $\M$ by a family of spacelike hypersurfaces
$(\Sigma_t)_{t\in\R}$.}
\end{figure}

\subsection{Foliation of spacetime}

The 3+1 formalism relies on a slicing of spacetime by a family of spacelike hypersurfaces
$(\Sigma_t)_{t\in\R}$. \defin{Hypersurface} means that $\Sigma_t$ is a 3-dimensional submanifold of $\M$, 
\defin{spacelike} means that every vector tangent to $\Sigma_t$ is spacelike and 
\defin{slicing} means (cf. Fig.~\ref{f:ein:foliat})
\be \label{e:ein:M_Sigmat}
    \M = \bigcup_{t\in\R} \Sigma_t . 
\ee
with $\Sigma_t \cap \Sigma_{t'} = \emptyset$ for $t\not=t'$. 
Not all spacetimes allow for a global foliation as in Eq.~(\ref{e:ein:M_Sigmat}), 
but only those
belonging to the class of the so-called \defin{globally hyperbolic spacetimes}\footnote{Cf.
\cite{ChoquY80} or \S~3.2.1 of 
\cite{Gourg07a} for the precise definition.}. However this class is large enough to encompass
spacetimes generated by rotating stars.  

At this stage, $t$ is a real parameter labelling the hypersurfaces $\Sigma_t$, to be identified later with 
some ``coordinate time''. 
We may consider $t$ as a scalar field on $\M$. It gradient $\wnab t = \dd f$ is then a
1-form that satisfies $\langle \wnab t,\vv{v}\rangle=0$ for any vector $\vv{v}$ tangent to $\Sigma_t$. 
This is direct consequence of $t$ being constant on $\Sigma_t$. Equivalently the vector
$\vv{\nabla} t$ associated to the 1-form $\wnab t$ by metric duality (i.e. the vector
whose components are $\nabla^\alpha t = g^{\alpha\mu} \nabla_\mu t$) 
is normal to $\Sigma_t$: 
$\vv{\nabla}t \cdot \vv{v}=0$ for any vector $\vv{v}$ tangent to $\Sigma_t$.
Since $\Sigma_t$ is spacelike, it possesses at each point a unique unit timelike normal vector 
$\vv{n}$, which is future-oriented (cf. Fig.~\ref{f:ein:foliat}): 
\be
  \vv{n}\cdot\vv{n}= -1 .
\ee
The two normal vectors $\vv{\nabla} t$ and $\vv{n}$ are necessarily colinear:
\be \label{e:ein:def_N}
  \encadre{ \vv{n} = - N \vv{\nabla} t }. 
\ee
The proportionality coefficient $N$ is called the \defin{lapse function}. 
The minus sign is chosen so that $N\geq 0$ 
if the scalar field $t$ is increasing towards
the future. 

Since each hypersurface $\Sigma_t$ is assumed to be spacelike, the metric $\w{\gamma}$ 
induced\footnote{$\w{\gamma}$ is nothing but the restriction of $\w{g}$ to $\Sigma_t$.} 
by $\w{g}$ onto $\Sigma_t$ is \defin{definite positive}, i.e. 
\be
  \forall \vv{v}\in\T_p(\Sigma_t),\quad \vv{v}\not=0 \Rightarrow 
      \w{\gamma}(\vv{v},\vv{v}) >  0 .
\ee 
Considered as a tensor field on $\M$, the components of $\w{\gamma}$ are given in terms of the
components of the normal via
\be
  \gamma_{\alpha\beta} = g_{\alpha\beta} + n_\alpha n_\beta . 
\ee
Note that if we raise the first index via $g^{\alpha\beta}$, we get the components of the 
orthogonal projector onto $\Sigma_t$ [compare with (\ref{e:ein:comp_ortho_proj})] :
\be 
      \gamma^\alpha_{\ \, \beta} = \delta^\alpha_{\ \, \beta} + n^\alpha n_\beta ,  
\ee
which we will denote by $\vv{\gamma}$. 

Let $\w{D}$ be the Levi-Civita connection in $\Sigma_t$ 
associated with the metric $\w{\gamma}$. It is expressible in terms of the spacetime connection
$\wnab$ and the orthogonal projector $\vv{\gamma}$ as 
\be
D_\rho T^{\alpha_1\ldots\alpha_p}_{\ \qquad\beta_1\ldots\beta_q}
		= \gamma_{\ \ \, \mu_1}^{\alpha_1} \, \cdots 
		 \gamma_{\ \ \, \mu_p}^{\alpha_p} \,
		  \gamma_{\ \ \, \beta_1}^{\nu_1} \, \cdots
		  \gamma_{\ \ \, \beta_q}^{\nu_q} \,
		  \gamma_{\ \ \, \rho}^{\sigma} \, \nabla_\sigma
		  T^{\mu_1\ldots\mu_p}_{\ \qquad\nu_1\ldots\nu_q}  .		  
\ee

\subsection{Eulerian observer or ZAMO}

Since $\vv{n}$ is a unit timelike vector, we may consider the family of observers whose 
4-velocity is $\vv{n}$. Their worldlines are then the field lines of $\vv{n}$ and are 
everywhere orthogonal to the hypersurfaces $\Sigma_t$. These observers are called the
\defin{Eulerian observers}. In the context of rotating stars or black holes, they are also
called the \defin{locally non-rotating observers}  \cite{Barde70b} or \defin{zero-angular-momentum observers 
(ZAMO)} \cite{ThornM82} . 
Thanks to relation (\ref{e:ein:def_N}), the proper time $\tau$ of a ZAMO is
related to $t$ by
\be
  d\tau = N \, dt , 
\ee
hence the name \emph{lapse function} given to $N$. 
The 4-acceleration of a ZAMO is the vector 
$\vv{a} := \wnab_{\vv{n}}\vv{n}$. This vector is tangent to $\Sigma_t$ and is related to the spatial gradient of the lapse function by (see e.g. \S~3.3.3 of \cite{Gourg07a}
for details)
\be
  \vv{a} = \vv{D} \ln N = \vv{\gamma}(\vv{\nabla} \ln N) . 
\ee
In general $\vv{a}\not=0$, so that the ZAMO's are \emph{accelerated} observers (they ``feel'' the gravitational ``force''). On the other side, they are \emph{non-rotating}: being orthogonal to the hypersurfaces $\Sigma_t$, their worldlines form a congruence without any twist. Physically this means that the ZAMO's do not ``feel'' any centrifugal nor Coriolis force.

Let us express the energy-momentum tensor $\w{T}$ in terms of the energy $E$ density, the momentum density $\w{p}$ and the stress tensor $\w{S}$, these three quantities being measured
by the ZAMO: setting $\vv{u} = \vv{n}$ and $\w{\bot} = \vv{\gamma}$ in Eqs.~(\ref{e:ein:E_Tuu}), (\ref{e:ein:pa_T}) and 
(\ref{e:ein:Sab_T}), we get
\bea
  E & = & T_{\mu\nu} n^\mu n^\nu \label{e:ein:E_euler} \\
  p_\alpha & = & - T_{\mu\nu} \, u^\nu \gamma^\mu_{\ \, \alpha}  \label{e:ein:p_euler} \\
    S_{\alpha\beta} & = &  T_{\mu\nu}  \gamma^\mu_{\ \, \alpha} \gamma^\nu_{\ \, \beta} . 
	\label{e:ein:S_euler}
\eea
The associated 3+1 decomposition of the energy-momentum tensor is then:
\be \label{e:ein:T_3p1}
  T_{\alpha\beta} = E \, n_\alpha \, n_\beta + p_\alpha \, n_\beta 
  + n_\alpha \, p_\beta + S_{\alpha\beta} . 
\ee

\begin{figure}
\centerline{\includegraphics[width=0.7\textwidth]{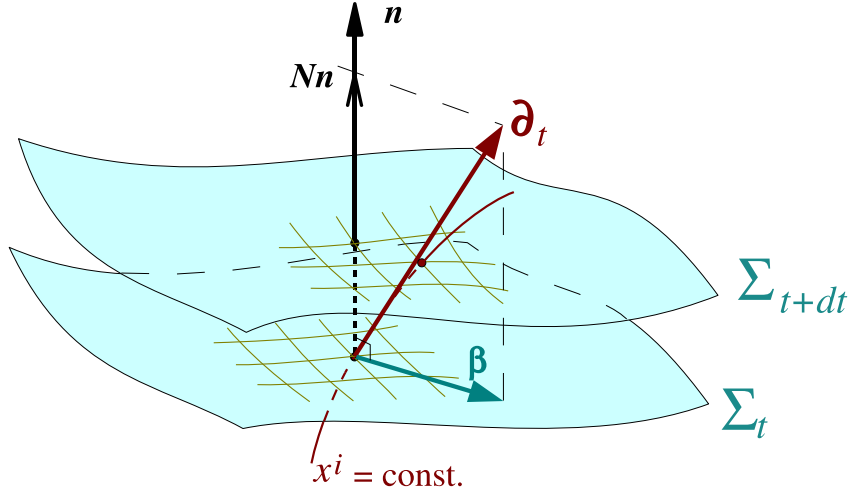}}
\caption[]{\label{f:ein:shift} \small
Coordinates $(x^i)$ on the hypersurfaces $\Sigma_t$:
each line $x^i={\rm const}$ cuts across the foliation 
$(\Sigma_t)_{t\in\R}$ and defines the time vector $\vpar_t$ 
and the shift vector $\w{\beta}$ of the
spacetime coordinate system $(x^\alpha)=(t,x^i)$.
}
\end{figure}

\subsection{Adapted coordinates and shift vector}

A coordinate system $(x^\alpha)$ is said to be \defin{adapted to the foliation}
$(\Sigma_t)_{t\in\R}$ iff $x^0 = t$. Then the triplet\footnote{Latin indices
($i,j,...$)
run in $\{1,2,3\}$, whereas Greek indices ($\alpha,\beta,...$) run in 
$\{0,1,2,3\}$.} $(x^i)=(x^1,x^2,x^3)$ constitutes a coordinate system on each hypersurface $\Sigma_t$, which we may call \defin{spatial coordinates}. 
Given such a coordinate system, we may decompose the natural basis vector 
$\vpar_t$ into a part along $\vv{n}$ and a part tangent to $\Sigma_t$:
\be \label{e:ein:par_t_N_beta}
  \encadre{ \vpar_t = N \vv{n} + \vv{\beta} } \qquad\mbox{with}\quad
  \vv{n}\cdot\vv{\beta} = 0 . 
\ee
The spacelike vector $\vv{\beta}$ is called the \defin{shift vector}. Indeed it measures
the shift of the lines of constant spatial coordinates with respect to the normal to the hypersurfaces $\Sigma_t$ (cf. Fig.~\ref{f:ein:shift}). 
The fact that the coefficient of $\vv{n}$ in Eq.~(\ref{e:ein:par_t_N_beta}) is the lapse function
$N$ is an immediate consequence of $\langle \dd t , \vpar_t\rangle = 1$ [Eq.~(\ref{e:ein:dual_basis})] and
relation (\ref{e:ein:def_N}). As for any vector tangent to $\Sigma_t$, the shift vector has no component along $\vpar_t$:
\be
    \vv{\beta} = \beta^i \, \vpar_i . 
\ee
Equation~(\ref{e:ein:par_t_N_beta}) leads then to the following expression for the 
components of $\vv{n}$: 
\be  \label{e:ein:comp_n}
  n^\alpha = \left( \frac{1}{N}, - \frac{\beta^1}{N}, - \frac{\beta^2}{N}, - \frac{\beta^3}{N}
  \right) .
\ee
The covariant components (i.e. the components of the 1-form associated with $\vv{n}$ by metric
duality) are given by Eq.~(\ref{e:ein:def_N}) and (\ref{e:ein:def_comp_form}): 
\be \label{e:ein:comp_n_cov}
  n_\alpha = (-N,0,0,0) . 
\ee
\begin{remark}
It is immediate to check on (\ref{e:ein:comp_n})-(\ref{e:ein:comp_n_cov}) that $\vv{n}$ is
a timelike unit vector: $\vv{n}\cdot\vv{n} = n_\mu n^\mu = (-N) (1/N) = -1$. 
\end{remark}

The components $(g_{\alpha\beta})$ of the spacetime metric are expressible in terms of the components 
$(\gamma_{ij})$ of the induced metric in $\Sigma_t$, the components of the shift vector and the lapse function: 
\be \label{e:ein:metric3p1}
	\encadre{	g_{\alpha\beta} \, dx^\alpha\, dx^\beta
	= - N^2 dt^2 + \gamma_{ij} (dx^i + \beta^i dt)
		(dx^j + \beta^j dt) } .  
\ee

\subsection{Extrinsic curvature} \label{s:ein:extr_curv}

The \defin{intrinsic curvature} of the hypersurface $\Sigma_t$ equipped with the induced metric $\w{\gamma}$ is given by the Riemann tensor of the Levi-Civita connection $\w{D}$
associated with $\w{\gamma}$ (cf. \S~\ref{s:ein:ee}). 
On the other side, the \emph{extrinsic curvature} describes the way $\Sigma_t$ is embedded into the 
spacetime $(\M,\w{g})$. It is measurable by the variation of the normal unit vector $\vv{n}$ as one moves on $\Sigma_t$. More precisely, the \defin{extrinsic curvature tensor} $\w{K}$ is 
the bilinear form defined on $\Sigma_t$ by
\be
  \forall (\vv{u},\vv{v})\in\T_p(\Sigma_t)\times\T_p(\Sigma_t),\quad
  \w{K}(\vv{u},\vv{v}) := - \vv{u} \cdot\wnab_{\vv{v}} \, \vv{n} . 
\ee
It can be shown (see e.g. \S~2.3.4 of \cite{Gourg07a}) that, as a consequence of $\vv{n}$ being 
hypersurface-orthogonal, the bilinear form $\w{K}$ is symmetric (this result is known as
the \defin{Weingarten property}).

The components of $\w{K}$ with respect to the coordinates $(x^i)$ in $\Sigma_t$ are expressible
in terms of the time derivative of the induced metric $\w{\gamma}$ according to 
\be \label{e:ein:Kij_dergam_ij}
  K_{ij} = - \frac{1}{2N} \left( \der{\gamma_{ij}}{t}
    - \Liec{\beta} \gamma_{ij} \right) = 
\frac{1}{2N} \left( - \der{\gamma_{ij}}{t}
  + \beta^k \der{\gamma_{ij}}{x^k} + \gamma_{kj} \der{\beta^k}{x^i} + 
  \gamma_{ik} \der{\beta^k}{x^j} \right) . 
\ee
$\Liec{\beta} \gamma_{ij}$ stands for the components of the Lie derivative of $\w{\gamma}$
along the vector field $\vv{\beta}$ (cf. Appendix~\ref{s:lie}) and the second equality
results from the 3-dimensional version of Eq.~(\ref{e:lie:der_bilin}).
The trace of $\w{K}$ with respect to the metric $\w{\gamma}$ is connected to the covariant divergence of the unit normal to $\Sigma_t$ :
\be  \label{e:ein:trace_K}
    K := \gamma^{ij} K_{ij} = - \nabla_\mu n^\mu . 
\ee

\subsection{3+1 Einstein equations}

Projecting the Einstein equation (\ref{e:ein:ee}) (i) twice onto $\Sigma_t$, (ii) twice along $\vv{n}$ and (iii) once on $\Sigma_t$ and once along $\vv{n}$, one gets respectively the following
equations \cite{Alcub08,BaumgS10,Gourg07a} :
\bea 
 & & \encadre{  \der{K_{ij} }{t} - \Liec{\beta} K_{ij}  
	= - D_i D_j N + N\left\{
	{}^3 R_{ij} + K K_{ij} -2 K_{ik} K^k_{\ \, j} 
	+ 4\pi \left[ (S-E) \gamma_{ij} - 2 S_{ij} \right] \right\} }\nonumber \\
	& & \label{e:ein:Einstein_PDE1} \\
 & & \encadre{ {}^3 R + K^2 - K_{ij} K^{ij} = 16\pi E } \label{e:ein:Einstein_PDE2}\\
 & & \encadre{ D_j K^j_{\ \, i} - D_i K = 8\pi  p_i } . \label{e:ein:Einstein_PDE3}
\eea
In this system, $E$, $p_i$ and $S_{ij}$ are the matter quantities relative to the 
ZAMO and are defined respectively by Eqs.~(\ref{e:ein:E_euler})-(\ref{e:ein:S_euler}). The scalar $S$ is the trace of $\w{S}$ with respect to the metric $\w{\gamma}$: 
$S = \gamma^{ij} S_{ij}$. 
The covariant derivatives $D_i$ can be expressed in terms
of partial derivatives with respect to the spatial coordinates $(x^i)$
by means of the Christoffel symbols ${}^3 \Gamma^i_{\ \, jk}$
of $\w{D}$ associated with $(x^i)$:
\bea
  & & D_i D_j N = \dder{N}{x^i}{x^j} - {}^3 \Gamma^k_{\ \, ij}
	\der{N}{x^k} ,  \\
  &&	D_j K^j_{\ \, i} = \der{K^j_{\ \, i}}{x^j}
	+ {}^3 \Gamma^j_{\ \, jk} K^k_{\ \, i} 
	- {}^3 \Gamma^k_{\ \, ji} K^j_{\ \, k} , \\
  && D_i K = \der{K}{x^i} .
\eea
The $\Liec{\beta} K_{ij}$ are the components of the Lie derivative of the tensor $\w{K}$ along the vector
$\vv{\beta}$ (cf. Appendix~\ref{s:lie}); according to
the 3-dimensional version of formula~(\ref{e:lie:der_bilin}), they can be expressed 
in terms of partial derivatives with respect to the 
spatial coordinates $(x^i)$ :
\be \label{e:ein:Lie_beta_K}
   \Liec{\beta} K_{ij} = \beta^k \der{K_{ij}}{x^k} 
	+ K_{kj} \der{ \beta^k}{x^i} + K_{ik} \der{\beta^k}{x^j} . 
\ee
The Ricci tensor and scalar curvature of $\w{\gamma}$ are expressible 
according to the 3-dimensional analog of formula (\ref{e:ein:Ricci}):
\bea
  & & {}^3 R_{ij} = \der{\, {}^3 \Gamma^k_{\ \, ij}}{x^k} - \der{\, {}^3 \Gamma^k_{\ \, ik}}{x^j}
	+ {}^3 \Gamma^k_{\ \, ij} {}^3 \Gamma^l_{\ \, kl}
	- {}^3 \Gamma^l_{\ \, ik} {}^3 \Gamma^k_{\ \, lj}  \label{e:sym:3Ricci} \\
  & & {}^3 R = \gamma^{ij} \, {}^3 R_{ij} . \label{e:sym:3Ricci_scal} 
\eea
Finally, let us recall that $K_{ij}$ is related to the derivatives of $\gamma_{ij}$
by Eq.~(\ref{e:ein:Kij_dergam_ij}). 

Equations~(\ref{e:ein:Einstein_PDE2}) and (\ref{e:ein:Einstein_PDE3}), which 
do not contain any second order time derivative of $\gamma_{ij}$, are called 
respectively the \defin{Hamiltonian constraint} and the \defin{momentum constraint}. 

The 3+1 Einstein equations are at the basis of the formulation of general relativity as a Cauchy problem and are much employed in numerical relativity (for reviews see e.g. 
\cite{ChoquY80,Gourg07a,Alcub08}).

%
%

\chapter{Stationary and axisymmetric spacetimes} \label{s:sym}


\minitoc
\vspace{1cm}

Having reviewed general relativity in Chap.~\ref{s:ein}, we focus now on spacetimes that
possess two symmetries: stationarity and axisymmetry, in view of considering rotating
stars in Chap.~\ref{s:eer}. 


\section{Stationary and axisymmetric spacetimes} \label{s:sym:def_sym}

\subsection{Definitions} \label{s:sym:defin}

\subsubsection{Group action on spacetime}

\begin{figure}
\centerline{\includegraphics[height=0.3\textheight]{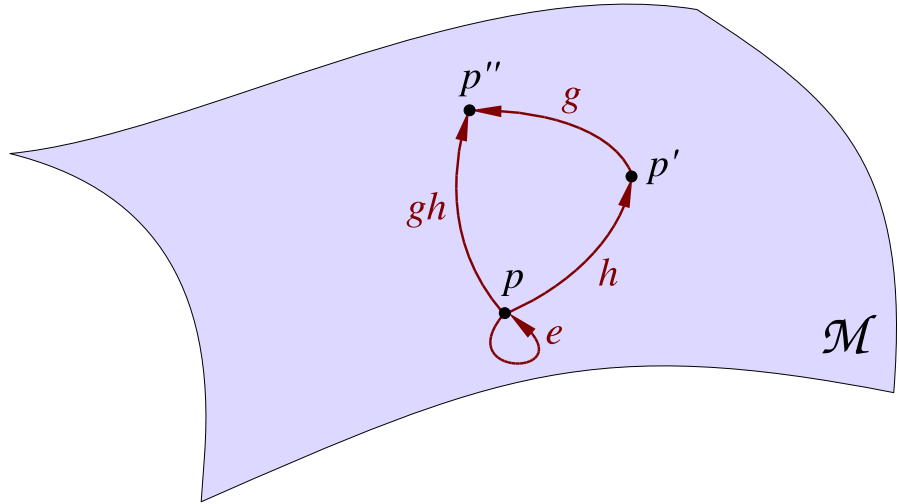}}
\caption{\label{f:sym:action} \small
Group action on the spacetime manifold $\M$.}
\end{figure}

Symmetries of spacetime are described in a coordinate-independent way by means of 
a (symmetry) group acting on the spacetime manifold $\M$. 
Through this action, each transformation belonging to the group displaces points within $\M$ and one demands that the metric $\w{g}$ is invariant under such displacement. 
More precisely, given 
a group $G$, a \defin{group action} of $G$ on $\M$ is an application\footnote{Do no confuse the generic element $g$ of the group $G$ with the metric tensor $\w{g}$.} 
\be
	\begin{array}{rccl}
	\Phi: & G\times \M & \longrightarrow & \M \\
		& (g,p) & \longmapsto & \Phi(g,p) =: g(p) 
	\end{array}
\ee
such that 
\begin{itemize}
\item $\forall (g,h) \in G^2,\  \forall p\in\M,\  g(h(p)) = gh(p)$, where $gh$ denotes the product of $g$ by $h$ according to the group law of
$G$ (cf. Fig.~\ref{f:sym:action}); 
\item if $e$ is the identity element of $G$, then
$\forall p\in \M,\  e(p) = p$ . 
\end{itemize}
The \defin{orbit} of a point $p\in\M$ is the set $\{g(p),\ g\in G\}\subset\M$, i.e. the set of points which are connected to $p$ by some group transformation. 

\begin{figure}
\centerline{\includegraphics[height=0.3\textheight]{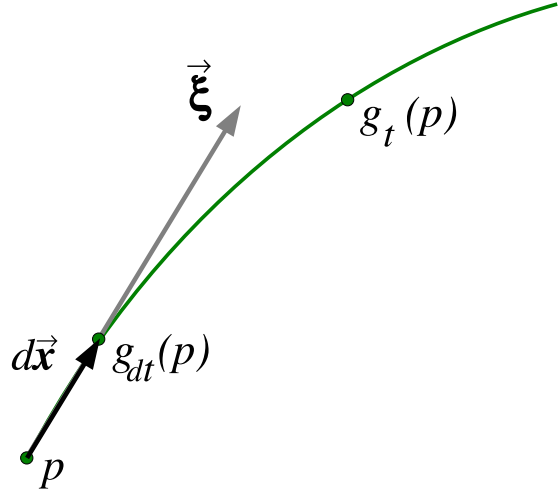}}
\caption{\label{f:sym:orbite} \small
Orbit of a point $p$ under the action of a one-dimensional Lie group, the elements $g_t$ of which being labelled by the parameter $t\in\R$. The vector $\vxi = d\vv{x}/dt$ is the group 
generator associated with this parameter.}
\end{figure}

An important class of group actions are those for which $G$ is a one-dimensional 
\emph{Lie group} (i.e. a ``continuous'' group). Then around $e$, the elements of $G$ can be labelled by a parameter $t\in \R$, such that $g_{t=0} = e$. 
The orbit of a given point $p\in\M$ under the group action is then either $\{p\}$ (case where $p$ is fixed point of the
group action) or a one-dimensional curve of $\M$. In the latter case, 
$t$ is then a natural parameter along the curve (cf. Fig.~\ref{f:sym:orbite}). The tangent vector 
corresponding to that parameter is called the \defin{generator of the symmetry group associated with the $t$ parametrization}. It is given by
\be
    \vxi = \frac{d\vv{x}}{dt} , 
\ee
where $d\vv{x}$ is the infinitesimal vector connecting the point $p$ to the point
$g_{dt}(p)$ (cf. \S~\ref{s:ein:gr_space_time} and Fig.~\ref{f:sym:orbite}). The action of $G$ on $\M$ in any infinitesimal neighbourhood
of $p$ amounts then to translations along the infinitesimal vector $dt\, \vxi$. 

\subsubsection{Stationarity}

A spacetime $(\M,\w{g})$ is said to be \defin{stationary} iff 
there exists a group action on $\M$ with the following properties:
\begin{enumerate}
\item the group $G$ is isomorphic to $(\R,+)$, i.e. the group of unidimensional translations; 
\item the orbits are timelike curves in $\M$; 
\item the metric is invariant under the group action, which is translated by
\be \label{e:ein:Lie_xi_g}
      \encadre{ \Lie{\vxi} \, \w{g} = 0 }, 
\ee
where $\vxi$ is the generator of $G$ associated with some parameter $t$ of $G$
and $\Lie{\vxi} \, \w{g}$ denotes the \emph{Lie derivative} of $\w{g}$ along the vector field
$\vxi$. The Lie derivative measures the variation of $\w{g}$ along the field lines of $\vxi$, i.e. the variation under the group action, and is defined in Appendix~\ref{s:lie}.
\end{enumerate}
\begin{remark}[1]
The property (\ref{e:ein:Lie_xi_g}) expresses the fact that if two vectors $\vv{u}$
and $\vv{v}$ are invariant under transport along the field lines of $\vxi$, then their
scalar product $\vv{u}\cdot\vv{v}$ is also invariant. Indeed, thanks to the Leibniz rule:
\[
  \wnab_{\vxi} \, (\vv{u}\cdot\vv{v}) = \Lie{\vxi} (\vv{u}\cdot\vv{v}) = 
  \Lie{\vxi} \left[ \w{g}(\vv{u},\vv{v}) \right] = 
  ( \Lie{\vxi} \w{g} ) (\vv{u},\vv{v})
  + \w{g} ( \Lie{\vxi} \vv{u}, \vv{v} ) 
  + \w{g} (\vv{u},  \Lie{\vxi} \vv{v} ) , 
\]
so that if $\Lie{\vxi} \vv{u}=0$ and $\Lie{\vxi} \vv{v}=0$ hold, 
$\Lie{\vxi} \w{g} = 0$
implies $\wnab_{\vxi} \, (\vv{u}\cdot\vv{v}) = 0$.
\end{remark}
\begin{remark}[2]
The property (\ref{e:ein:Lie_xi_g}) is independent of the choice of the generator $\vxi$, i.e. of the parametrization $t$ of $G$. Indeed, under a change of parametrization 
$t\mapsto t'$, inducing a change of generator $\vxi\mapsto \vxi'$, the following scaling law holds: 
\[
  \Lie{\vxi'} \w{g} = \frac{dt}{dt'} \, \Lie{\vxi} \, \w{g} .
\]
\end{remark}
Expressing the Lie derivative $\Lie{\vxi} \, \w{g}$ according to Eq.~(\ref{e:lie:der_bilin_nab}) 
and using  $\nabla_\mu g_{\alpha\beta}=0$ [property (\ref{e:nabla_g_zero})] 
as well as $\xi_\alpha = g_{\alpha\mu} \xi^\mu$
shows that the condition~(\ref{e:ein:Lie_xi_g}) is equivalent to 
\be \label{e:sym:Killing}
  \encadre{ \nabla_\alpha \xi_\beta + \nabla_\beta \xi_\alpha = 0 } . 
\ee
This equation is known as \defin{Killing equation}\footnote{Named after the German mathematician
Wilhelm Killing (1847-1923).}. Accordingly, the symmetry generator $\vxi$ is called a 
\defin{Killing vector}.

For an asymptotically flat spacetime, as we are considering here, we can determine the
Killing field uniquely $\vxi$ by demanding that far from the central object 
$\vxi\cdot\vxi \longrightarrow -1$. Consequently, the associated parameter $t$ coincides with the proper time of the asymptotically inertial observer at rest with respect to the central source. 
In the following we shall employ only that Killing vector. 

\subsubsection{Staticity} 

A spacetime $(\M,\w{g})$ is said to be \defin{static} iff
\begin{enumerate}
\item it is stationary; 
\item the Killing vector field $\vxi$ is orthogonal to a family of hypersurfaces.  
\end{enumerate}

\begin{remark}
Broadly speaking, condition 1 means that nothing depends on time and condition 2 that there is 
``no motion'' in spacetime. This will be made clear below for the specific case of rotating stars: we will show that condition 2 implies that the star is not rotating. 
\end{remark}

\subsubsection{Axisymmetry}

A spacetime $(\M,\w{g})$ is said to be \defin{axisymmetric} iff 
there exists a group action on $\M$ with the following properties:
\begin{enumerate}
\item the group $G$ is isomorphic to $\mathrm{SO}(2)$, i.e. the group of rotations in the
plane; 
\item the metric is invariant under the group action:
\be \label{e:sym:Lie_chi_g}
      \encadre{ \Lie{\vchi} \, \w{g} = 0 }, 
\ee
$\vchi$ being the generator of $G$ associated with some parameter $\varphi$ of $G$
and $\Lie{\vxi} \, \w{g}$ denoting the \emph{Lie derivative} of $\w{g}$ along the vector field $\vchi$;  
\item the set of fixed points under the action of $G$ is a 2-dimensional surface of $\M$, which
we will denote by $\Delta$.
\end{enumerate}
Carter \cite{Carte70} has shown that for asymptotically flat spacetimes, which are our main concern here, the first and second properties in the above definition imply the third one. Moreover, he has also shown that $\Delta$ is necessarily a \defin{timelike} 2-surface, i.e. the metric induced on it by $\w{g}$ has the signature $(-,+)$. $\Delta$ is called the \defin{rotation axis}. 
\begin{remark}
The 2-dimensional and timelike characters of the rotation axis can be understood by considering $\Delta$ as the time development (history) of the ``standard'' one-dimensional rotation axis in a 3-dimensional space. 
\end{remark}
From the very definition of a generator of a symmetry group, the vector field $\vchi$
must vanish on the rotation axis (otherwise, the latter would not be a set of 
fixed points):
\be \label{e:sym:chi_zero_axis}
  \encadre{ \left. \vchi \right| _\Delta = 0 }.
\ee
 In addition, 
as for $\vxi$, the condition (\ref{e:sym:Lie_chi_g}) is equivalent to demanding that 
$\vchi$ be a Killing vector: 
\be
  \encadre{ \nabla_\alpha \chi_\beta + \nabla_\beta \chi_\alpha = 0 } . 
\ee

Given a $\mathrm{SO}(2)$ group action, we can determine uniquely the Killing vector $\vchi$
by demanding that the associated parameter $\ph$ takes its values in $[0,2\pi[$. In the following, we shall always employ that Killing vector.

\begin{figure}
\centerline{\includegraphics[height=0.3\textheight]{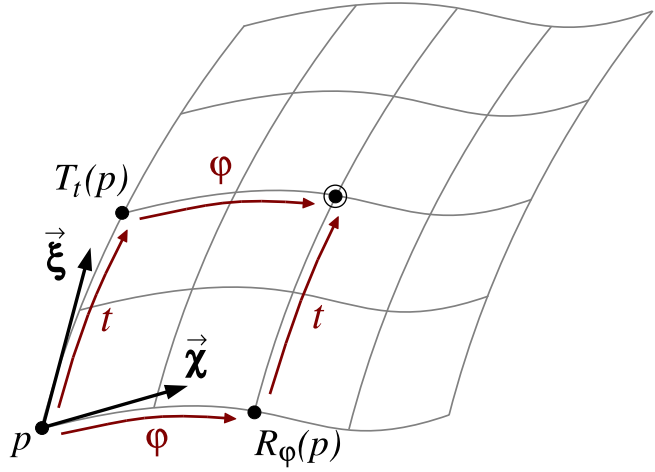}}
\caption{\label{f:sym:commut_coord} \small
Commutativity of the stationary and axisymmetric actions.}
\end{figure}

\subsection{Stationarity and axisymmetry}

We consider a spacetime $(\M,\w{g})$ that is both stationary and axisymmetric. 
Carter has shown in 1970 \cite{Carte70} that no generality is lost by considering that the stationary and axisymmetric actions commute. 
In other words, the spacetime $(\M,\w{g})$ is invariant under the action  of the \emph{Abelian} group $\R\times \mathrm{SO}(2)$, and not only under the actions of 
$\R$ and $\mathrm{SO}(2)$ separately. 
Saying that the stationary and axisymmetric actions \defin{commute} means that starting
from any point $p\in\M$, moving to the point $q = T_t(p)$ under the action of an element $T_t$ of the group $(\R,+)$ and then displacing $q$ via an element $R_\varphi$ of $\mathrm{SO}(2)$ yields 
the same point as when performing the displacements in the reverse order
(cf. Fig.~\ref{f:sym:commut_coord}):
\be \label{e:sym:actions_commute}
    \forall p\in\M,\ \forall t\in\R,\ \forall \varphi\in[0,2\pi[,\quad 
  R_\varphi(T_t(p)) = T_t(R_\varphi(p)) . 
\ee
This property can be translated in terms of the Killing vectors
$\vxi$ and $\vchi$ associated 
respectively with the parameter $t$ of $(\R,+)$ and the parameter 
$\varphi$ of $\mathrm{SO}(2)$. Indeed (\ref{e:sym:actions_commute})
is equivalent to the vanishing of their commutator: 
\be
   \encadre{ [\vxi,\vchi] = 0 }, 
\ee
the \defin{commutator} being the vector whose components are
\be
  [\vxi,\vchi]^\alpha = \xi^\mu \der{\chi^\alpha}{x^\mu}
    - \chi^\mu \der{\xi^\alpha}{x^\mu}
    = \xi^\mu \nabla_\mu \chi^\alpha
    - \chi^\mu \nabla_\mu \xi^\alpha.
\ee

An important consequence of the above commutation property is that the parameters
$t$ and $\varphi$ labelling the two symmetry groups $(\R,+)$ and $\mathrm{SO}(2)$
can be chosen as coordinates on the spacetime manifold $\M$.
Indeed, having set the coordinates $(t,\varphi)=(0,0)$ to a given point $p$, 
(\ref{e:sym:actions_commute}) implies that we can unambiguously attribute the coordinates
$(t,\varphi)$ to the point obtained from $p$ by a time translation of parameter $t$ and a rotation of parameter $\varphi$, whatever the order of these two transformations 
(cf. Fig.~\ref{f:sym:commut_coord}).

\begin{figure}
\centerline{\includegraphics[height=0.3\textheight]{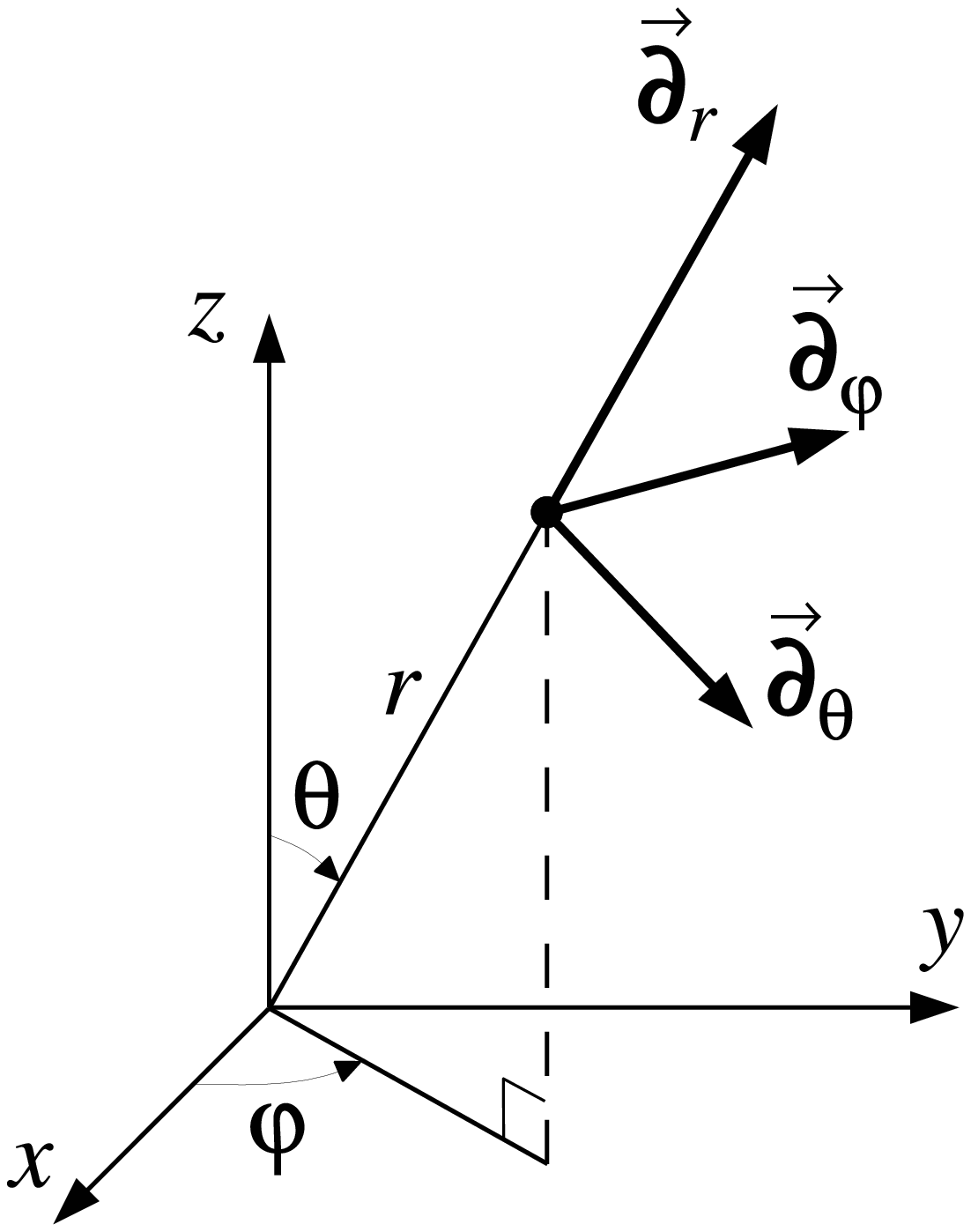}}
\caption{\label{f:sym:coord_spher} \small
Spherical coordinates $(r,\theta,\ph)$ and the associated natural basis 
$(\vpar_r,\vpar_\theta,\vpar_\ph)$.}
\end{figure}

Let us complete $(t,\ph)$ by two other coordinates
$(x^1,x^2)=(r,\theta)$ to get a full coordinate system on $\M$:
\be \label{e:sym:coord_adapt}
  \encadre{ (x^\alpha)  = (t,r,\theta,\ph) } .
\ee
$(r,\theta)$ are chosen so that the spatial coordinates $(r,\theta,\ph)$ are of \emph{spherical type}: 
$r\in[0,+\infty[$, $\theta\in[0,\pi]$ and $\theta=0$ or $\theta=\pi$ on the rotation axis 
$\Delta$ (cf. Fig.~\ref{f:sym:coord_spher}).  
\begin{remark}
An alternative choice would have been $(x^1,x^2)=(\rho,z)$ such that 
the spatial coordinates $(\rho,z,\ph)$ are of \emph{cylindrical type}:
$\rho\in[0,+\infty[$, $z\in\R$ and $\rho=0$ on $\Delta$. The relation between
the two types of coordinates is $\rho = r\sin\theta$ and $z=r\cos\theta$. 
\end{remark}

By construction, the coordinate system (\ref{e:sym:coord_adapt}) is such that the first and fourth
vectors of the associated natural basis (cf. \S~\ref{s:ein:gr_space_time})
are the two Killing vectors: 
\be \label{e:sym:adapated_coord}
    \vpar_t = \vxi \qquad\mbox{and}\qquad \vpar_\ph = \vchi . 
\ee
Consequently, in terms of tensor components in the coordinates $(x^\alpha)$, the Lie derivatives with respect to $\vxi$ (resp. $\vchi$) reduce to partial derivatives with 
respect to $t$ (resp. $\ph$) [cf. Eq.~(\ref{e:Lie_adapted})]. In particular, the stationary and axisymmetry conditions (\ref{e:ein:Lie_xi_g}) and (\ref{e:sym:Lie_chi_g}) are respectively equivalent to 
\be \label{e:sym:ignorables}
  \encadre{ \der{g_{\alpha\beta}}{t} = 0 } 
  \qquad\mbox{and}\qquad
  \encadre{ \der{g_{\alpha\beta}}{\ph} = 0 } .
\ee
For this reason, $t$ and $\ph$ are called \defin{ignorable coordinates}. 
The four coordinates $(x^\alpha)$ are said to be \defin{adapted coordinates} to the spacetime symmetries. 

\begin{figure}
\centerline{\includegraphics[height=0.3\textheight]{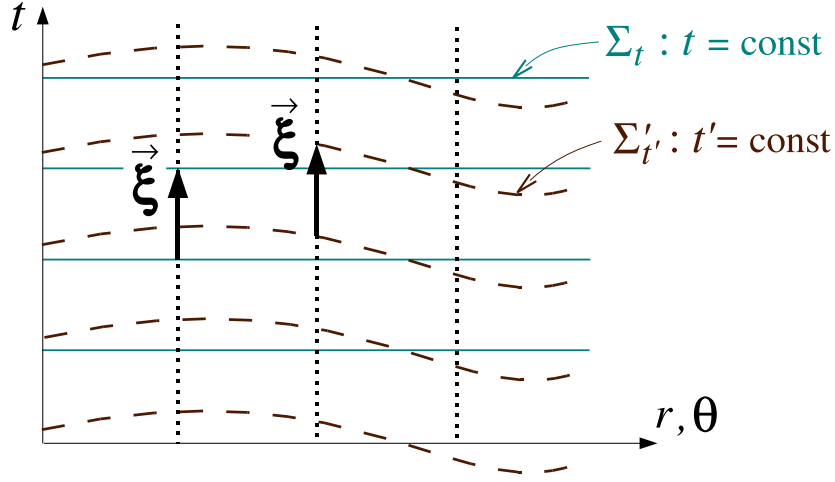}}
\caption{\label{f:sym:change_t} \small
Change of coordinate $t\mapsto t'$ according to (\ref{e:sym:chg_coord}).
The solid lines represent the hypersurfaces
of constant $t$ ($\Sigma_t$) and the dashed lines those of constant $t'$
($\Sigma'_{t'}$). The dotted lines are the orbits of the stationarity action (generator $\vxi$).}
\end{figure}

There is some freedom in choosing the coordinates $(x^\alpha)$ with the above properties.
Indeed, any change of coordinates of the form
\be \label{e:sym:chg_coord}
  \left\{ \begin{array}{lcl}
  t' & = & t + T(r,\theta) \\
  r' & = & R(r,\theta) \\
  \theta' & = & \Theta(r,\theta) \\
  \ph' & = & \ph + \Phi(r,\theta) \ , 
  \end{array} \right. 
\ee
where $T$, $R$, $\Theta$ and $\Phi$ are arbitrary (smooth) functions\footnote{Of course, to preserve 
the spherical type of the spatial coordinates, the functions $R$ and $\Theta$ have to fulfill certain
properties.} of $(r,\theta)$,
lead to another adapted coordinate system: 
\be
  \vpar_{t'} = \vxi 
  \qquad\mbox{and}\qquad \vpar_{\ph'} = \vchi . 
\ee
It is important to realize that in the above equation, $\vxi$ and $\vchi$ 
are the same vectors than in Eq.~(\ref{e:sym:adapated_coord}). 
The change of coordinate $t\mapsto t'$ according to (\ref{e:sym:chg_coord}) is merely a reparametrization of each orbit of the stationarity action --- reparametrization which leaves invariant the associated tangent vector $\vxi$ (cf. Fig.~\ref{f:sym:change_t}). A similar thing can be said for $\ph\mapsto\ph'$. 

\begin{remark}
As illustrated in Fig.~\ref{f:sym:change_t}, stationarity by itself does not introduce a privileged 3+1 slicing
of spacetime (cf. \S~\ref{s:ein:3p1}), for the change of $t$ coordinate according to 
(\ref{e:sym:chg_coord}) changes the hypersurfaces of constant $t$.  
\end{remark}

\section{Circular stationary and axisymmetric spacetimes} 

\subsection{Orthogonal transitivity} \label{s:sym:ortho_trans}

Regarding the components of the metric tensor, the properties (\ref{e:sym:ignorables}) are a priori the only ones implied by the spacetime symmetries (stationarity and axisymmetry). 
There is however a wide subclass of stationary and axisymmetric spacetimes in which, in addition to 
(\ref{e:sym:ignorables}), we are allowed to 
set to zero five metric components $g_{\alpha\beta}$: the so-called \emph{circular spacetimes}. Furthermore, these spacetimes are much relevant for astrophysics. 

Let us first remark that having $g_{\alpha\beta}=0$ for some value of $(\alpha,\beta)$ 
is an \emph{orthogonality} condition: that of
the vectors $\vpar_\alpha$ and $\vpar_\beta$, i.e. of the lines
$\{x^{\alpha'}=\mathrm{const},\ \alpha'\not=\alpha\}$ and
$\{x^{\beta'}=\mathrm{const},\ \beta'\not=\beta\}$. 
Next, we note that, in the present context, there are privileged 2-dimensional surfaces in spacetime: the orbits of the $\R\times\mathrm{SO}(2)$ group action. They are called the \defin{surfaces of transitivity} and 
denoted $\Sp_A$, where $A$ is some label. 
The two Killing vectors $\vxi$ and $\vchi$ are everywhere
tangent to these surfaces and, except on the rotation axis, they form a basis of the tangent space 
to $\Sp_A$ at each point. Once adapted coordinates $(x^\alpha)=(t,r,\theta,\ph)$ are chosen, 
the surfaces of transitivity can be labelled by the value of $(r,\theta)$ since both $r$ and $\theta$
are fixed on each of these surfaces: 
\be
  \Sp_A = \Sp_{r\theta} = \{ p\in\M \,/\  x^1(p) = r,\; x^2(p) = \theta \} . 
\ee
\begin{remark}
$r\theta$ is a mere label for the surfaces $\Sp_{r\theta}$; the latter depend only on the symmetry group $\R\times\mathrm{SO}(2)$ and not on the choice of the $(r,\theta)$ coordinates. A coordinate
change $(r,\theta)\mapsto (r',\theta')$ according to (\ref{e:sym:chg_coord}) will simply result in 
a relabelling of the surfaces $\Sp_{r\theta}$. 
\end{remark}

Given the family of 2-surfaces $(\Sp_{r\theta})$, one may ask if there exists another family of 2-surfaces, 
$(\mathcal{M}_B)$ say, which are everywhere orthogonal to $\Sp_{r\theta}$. If this is the case, 
the $\R\times\mathrm{SO}(2)$ group action is said to be \defin{orthogonally transitive} \cite{Carte69}
and the spacetime $(\M,\w{g})$ to be \defin{circular}. 
One may choose the coordinates $(r,\theta)$ to lie in $\mathcal{M}_B$, i.e. the label $B$ to be
$t\ph$ since both $t$ and $\ph$ are then fixed on each of these surfaces:
\be \label{e:sym:merid_surf}
  \mathcal{M}_B = \mathcal{M}_{t\ph} = \{ p\in\M \,/\  x^0(p) = t,\; x^3(p) = \ph \} . 
\ee
The 2-surfaces $\mathcal{M}_B$ are called the \defin{meridional surfaces}. 
In case of orthogonal transitivity, the following metric components are identically zero, reflecting 
the orthogonality between $\Sp_{r\theta}$ and $\mathcal{M}_{t\ph}$:
\be \label{e:sym:gab_circular}
  g_{t r} = 0,\quad g_{t\theta}=0,\quad g_{\ph r} = 0, \quad g_{\ph\theta} = 0 . 
\ee
The following theorem states under which conditions this is guaranteed: 
\begin{quote}
\textbf{Generalized Papapetrou theorem:} a stationary (Killing vector $\vxi$)
and axisymmetric (Killing vector $\vchi$) spacetime ruled by the Einstein equation 
is circular iff the energy-momentum tensor $\w{T}$ obeys to 
\bea
    \xi^\mu T_\mu^{\ \, [\alpha} \xi^\beta \chi^{\gamma]} &=& 0 \label{e:sym:PapapC1} \\ 
    \chi^\mu T_\mu^{\ \, [\alpha} \xi^\beta \chi^{\gamma]} &=& 0 , \label{e:sym:PapapC2} 
\eea
where the square brackets denote a full antisymmetrization. 
\end{quote}
This theorem has been demonstrated in the case of vacuum solutions ($\w{T}=0$) by Papapetrou (1966)
\cite{Papap66} and extended to the non-vacuum case by Kundt \& Tr\"umper (1966)
\cite{KundtT66}  and Carter (1969) \cite{Carte69,Carte73}
(see also \S~7.1 of Wald's textbook \cite{Wald84}, \S~7.2.1 of Straumann's one \cite{Strau04}, or \S~19.2 of \cite{StephKMHH03}).  
Note that Eqs.~(\ref{e:sym:PapapC1})-(\ref{e:sym:PapapC2}) are equivalent to 
\bea
  & & \vv{V}\in \mathrm{Vect}(\vxi,\vchi) \\
  & & \vv{W}\in \mathrm{Vect}(\vxi,\vchi) ,
\eea
where $\vv{V}$ (resp. $\vv{W}$) is the vector whose components are
$V^\alpha = \xi^\mu T_\mu^{\ \, \alpha}$ (resp. $W^\alpha = \chi^\mu T_\mu^{\ \, \alpha}$)
and $\mathrm{Vect}(\vxi,\vchi)$ denotes the vector plane generated by $\vxi$ and $\vchi$. 
 
In the important case of a perfect fluid source, the energy-momentum tensor takes the form
(\ref{e:ein:T_perfect_fluid}), so that
\bea 
 \xi^\mu T_\mu^{\ \, [\alpha} \xi^\beta \chi^{\gamma]} & = & 
  (\vep+p) \xi^\mu u_\mu \, u^{[\alpha} \xi^\beta \chi^{\gamma]} 
  + p \, \xi^\mu \delta_\mu^{\ \, [\alpha} \xi^\beta \chi^{\gamma]} \nonumber \\
  & = & (\vep+p) \xi^\mu u_\mu \, u^{[\alpha} \xi^\beta \chi^{\gamma]} 
  + p \, \underbrace{\xi^{[\alpha} \xi^\beta \chi^{\gamma]}}_{0} \nonumber \\
  & = & (\vep+p) \xi^\mu u_\mu \, u^{[\alpha} \xi^\beta \chi^{\gamma]}  . \nonumber 
\eea
Similarly,
\[
  \chi^\mu T_\mu^{\ \, [\alpha} \xi^\beta \chi^{\gamma]} = (\vep+p) \chi^\mu u_\mu \, u^{[\alpha} \xi^\beta \chi^{\gamma]} .
\]
Since $\vxi$ and $\vv{u}$ are both timelike, we have $\xi^\mu u_\mu\not=0$.
The circularity conditions
(\ref{e:sym:PapapC1})-(\ref{e:sym:PapapC2}) are therefore equivalent to 
\be
   u^{[\alpha} \xi^\beta \chi^{\gamma]} = 0 , 
\ee
i.e. to 
\be
  \encadre{ \vv{u} \in \mathrm{Vect}(\vxi,\vchi) } . 
\ee
Taking into account that $\vxi = \vpar_t$ and $\vchi=\vpar_\ph$, the above condition is equivalent
to $u^r = 0$ and $u^\theta = 0$, or 
\be \label{e:sym:u_circular}
  \encadre{ \vv{u} = u^t \left( \vxi + \Omega \vchi \right) } , 
\ee
with 
\be
  \encadre{ \Omega := \frac{u^\ph}{u^t} = \frac{d\ph}{dt} } . 
\ee
The 4-velocity (\ref{e:sym:u_circular}) describes a pure \defin{circular motion} of the fluid around the 
rotation axis, hence the qualifier \emph{circular} given to spacetimes that obeys the orthogonal 
transitivity property. In such case, there is no fluid motion in the meridional surfaces, i.e. no convection. 
Stationary and axisymmetric spacetimes that are not circular have been studied in Refs.~\cite{GourgB93,BirklSM11}. 

\[
\fbox{In all what follows, we limit ourselves to circular spacetimes.}
\]

\subsection{Quasi-isotropic coordinates} \label{s:sym:QI}

In a stationary and axisymmetric spacetime $(\M,\w{g})$ that is circular, 
we may use adapted coordinates 
$(x^\alpha)=(t,r,\theta,\ph)$ such that $(r,\theta)$ span the 2-surfaces $\mathcal{M}_{t\ph}$ 
which are orthogonal to the surfaces of transitivity $\Sp_{r\theta}$, leading to the vanishing of the 
metric components listed in (\ref{e:sym:gab_circular}). 
Moreover, we can always choose the coordinates $(r,\theta)$ in each 2-surface $\mathcal{M}_{t\ph}$
such that the metric induced by $\w{g}$ takes the form\footnote{Indices $a$ and $b$ take their values
in $\{1,2\}$.}
\be \label{e:sym:gab_2D}
  g_{ab} \, dx^a dx^b = A^2 (dr^2 + r^2 d\theta^2) , 
\ee 
where $A = A(r,\theta)$. Indeed, $dr^2 + r^2 d\theta^2$ is the line element of the 2-dimensional 
flat metric in polar coordinates and, in dimension 2, all the metrics are 
\defin{conformally related}, meaning that they differ only by a scalar factor $A^2$ as in 
(\ref{e:sym:gab_2D}). 
\begin{remark}
This is no longer true in dimension 3 or higher: in general
one cannot write the metric line element as a conformal factor times the flat one by a mere choice of
coordinates. 
\end{remark}
Note that the choice of coordinates $(r,\theta)$ leading to (\ref{e:sym:gab_2D}) is equivalent to 
the two conditions:
\be \label{e:sym:gab_QI}
   g_{r\theta} = 0 \qquad\mbox{and}\qquad
   g_{\theta\theta} = r^2 g_{rr} . 
\ee
The coordinates $(t,r,\theta,\ph)$ with the above choice for $(r,\theta)$ are called 
\defin{quasi-isotropic coordinates (QI)}. The related cylindrical coordinates $(t,\rho,z,\ph)$, 
with $\rho := r\sin\theta$ and $z:=r\cos\theta$, are called \defin{Lewis-Papapetrou coordinates}, 
from the work of Lewis (1932) \cite{Lewis32} and Papapetrou (1966) \cite{Papap66}. 

Let us define the scalar function $\omega=\omega(r,\theta)$ as minus the scalar product of the two Killing vectors 
$\vxi$ and $\vchi$, normalized by the scalar square of $\vchi$:
\be \label{e:sym:def_omega}
	\encadre{ \omega := - \frac{\vxi\cdot\vchi}{\vchi\cdot\vchi} } .
\ee
As we will see below, the minus sign ensures that for a rotating star, $\omega \geq 0$
(cf. Fig.~\ref{f:sym:prof_omega}). 
Since $g_{t\ph} = \vxi\cdot\vchi$ and $g_{\ph\ph} = \vchi\cdot\vchi$, we may write
\be \label{e:sym:gtp_omega}
	g_{t\ph} = - \omega g_{\ph\ph} . 
\ee

Besides, let us introduce the following function of $(r,\theta)$:
\be \label{e:sym:def_B2}
  B^2 := \frac{g_{\ph\ph}}{r^2\sin^2\theta} 
\ee
Collecting relations (\ref{e:sym:gab_circular}), (\ref{e:sym:gab_2D}), (\ref{e:sym:gtp_omega}) and (\ref{e:sym:def_B2}), 
we may write the components of the metric tensor in the form
\be \label{e:sym:metricQI}
  \encadre{ g_{\alpha\beta} \, dx^\alpha\, dx^\beta
	= - N^2 dt^2 + A^2 (dr^2 + r^2 d\theta^2)
  + B^2 r^2 \sin^2 \theta (d\ph - \omega dt)^2 } , 
\ee
where $N$, $A$, $B$ and $\omega$ are four functions of $(r,\theta)$:
\be \label{e:sym:liste4}
  \encadre{N=N(r,\theta)},\quad
  \encadre{A=A(r,\theta)},\quad
  \encadre{B=B(r,\theta)},\quad
  \encadre{\omega=\omega(r,\theta)} .  
\ee
Equivalently, in matrix form:
\be \label{e:sym:metric_down}
	g_{\alpha\beta} = \left( \begin{array}{cccc}
	- N^2 + B^2 \omega^2 r^2 \sin^2\theta & 0 & 0 
		& - \omega B^2 r^2 \sin^2\theta \\
	0 & A^2 & 0 & 0 \\
	0 & 0 & A^2 r^2 & 0 \\
	- \omega B^2 r^2 \sin^2\theta & 0 & 0 & B^2 r^2 \sin^2 \theta 
	\end{array} \right) .
\ee
The inverse of this matrix is 
\be \label{e:sym:metric_up}
	g^{\alpha\beta} =  \left( \begin{array}{cccc}
	- N^{-2} & 0 & 0 & - \omega / N^2 \\
	0 & A^{-2} & 0 & 0 \\
	0 & 0 & A^{-2} r^{-2} & 0 \\
	- \omega / N^2 & 0 & 0 & (B r \sin\theta)^{-2} - \omega^2 / N^2 
	\end{array} \right) .
\ee
To prove it, it suffices to check that relation (\ref{e:ein:inv_metric}) holds.

\begin{figure}
\centerline{\includegraphics[height=0.3\textheight]{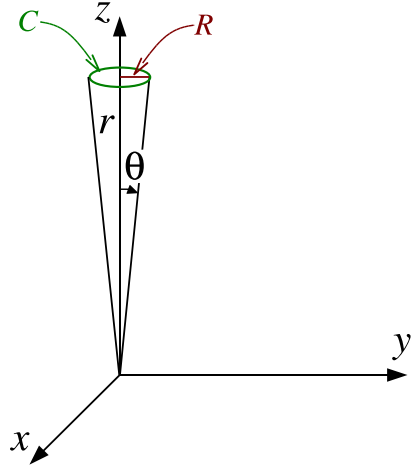}}
\caption{\label{f:sym:flat_axis} \small
Radius $R$ and circumference $C$ of a small circle around the rotation axis
($\theta\ll 1$).}
\end{figure}

The local flatness of spacetime implies that the metric functions $A$ and $B$ must
coincide on the rotation axis:
\be \label{e:sym:A_B_axis}
  \left. A \right| _\Delta = \left. B \right| _\Delta . 
\ee
Indeed, let us consider a small circle around
the rotation axis at a fixed value of both $t$ and $\theta$, with
$\theta \ll 1$, and centered on the point of coordinates
$(x^\alpha) = (t,r,0,0)$ (cf. Fig.~\ref{f:sym:flat_axis}). 
According to the line element (\ref{e:sym:metricQI}), the metric length of its radius
is $R\simeq A(r,0) \, r \theta$, whereas its metric circumference is
$C \simeq B(r/\cos\theta,\theta) \, r \sin\theta \times 2\pi \simeq 2\pi B(r,0) \, r\theta$. 
The local flatness hypothesis implies that $C = 2\pi R$. Would this relation not hold, a conical singularity would be present on the rotation axis. From the above values of $C$ and $R$, 
we get $A(r,0) = B(r,0)$, i.e. the property (\ref{e:sym:A_B_axis}) for the half part of 
$\Delta$ corresponding to $\theta=0$. The demonstration for the second part ($\theta=\pi$)
is similar. 

\subsection{Link with the 3+1 formalism}

In terms of the 3+1 formalism introduced in \S~\ref{s:ein:3p1}, the comparison of
(\ref{e:sym:metricQI}) with (\ref{e:ein:metric3p1}) leads immediately to (i) the identification of
$N$ as the lapse function, (ii) the following components of the shift vector:
\be \label{e:sym:beta_omega}
	\encadre{\beta^i = (0,0,-\omega)} , 
\ee
and (iii) the following expression of the induced metric in the hypersurfaces $\Sigma_t$ : 
\be \label{e:sym:3metQI}
	\encadre{ \gamma_{ij} \, dx^i dx^j = 
	A^2 (dr^2 + r^2 d\theta^2)
  + B^2 r^2 \sin^2 \theta \, d\ph^2 } . 
\ee
Therefore 
\be \label{e:sym:gam_dd}
  \encadre{ \gamma_{ij} = \mathrm{diag}\left(A^2,\;  A^2 r^2,\;  B^2 r^2 \sin^2\theta \right)} 
\ee
and
\be \label{e:sym:gam_uu}
  \encadre{ \gamma^{ij} = \mathrm{diag}\left(\frac{1}{A^2},\; \frac{1}{A^2 r^2},\;  
  \frac{1}{B^2 r^2 \sin^2\theta} \right)}.
\ee
\begin{remark}
Comparing with (\ref{e:sym:metric_down})-(\ref{e:sym:metric_up}), note that $g_{ij} = \gamma_{ij}$ but $g^{ij} \not= \gamma^{ij}$. 
\end{remark}

Moreover, since in the present case $\vpar_t = \vxi$ and 
$\vv{\beta} = -\omega \vpar_\ph = - \omega \vchi$, relation (\ref{e:ein:par_t_N_beta})
becomes
\be \label{e:sym:xi_n_chi}
	\encadre{ \vxi = N \vv{n}  - \omega \vchi } . 
\ee
In particular, if $\omega=0$, then $\vxi$ is colinear to $\vv{n}$ and hence orthogonal to the 
hypersurfaces $\Sigma_t$. According to the definition given in \S~\ref{s:sym:defin}, this
means that for $\omega=0$, the spacetime is static. The reverse is true, 
assuming that in case of staticity, the axisymmetric action takes place in the surfaces
orthogonal to $\vxi$. We may then state
\be \label{e:sym:omega_zero}
  \encadre{ \omega = 0 \iff (\M,\w{g}) \ \mbox{static} } .
\ee 

Let us evaluate the extrinsic curvature tensor $\w{K}$ of the hypersurfaces $\Sigma_t$
(cf. \S~\ref{s:ein:extr_curv}). Since $\dert{\gamma_{ij}}{t} = 0$ and 
$\beta^i=0$ except for $\beta^\ph=-\omega$, Eq.~(\ref{e:ein:Kij_dergam_ij}) leads to 
\[
K_{ij} = 
\frac{1}{2N} \bigg(
  -\omega \underbrace{\der{\gamma_{ij}}{\ph}}_{0} - \gamma_{\ph j} \der{\omega}{x^i} -
  \gamma_{i\ph} \der{\omega}{x^j} \bigg) . 
\]
Since $(\gamma_{ij})$ is diagonal [Eq.~(\ref{e:sym:gam_dd})] and $\dert{\omega}{\ph}=0$, 
we conclude that all the components
of $\w{K}$ vanish, except for
\be \label{e:sym:Krph}
	\encadre{ K_{r\ph} = K_{\ph r} = - \frac{B^2 r^2\sin^2\theta}{2N} \der{\omega}{r} } , 
\ee
\be \label{e:sym:Kthph}
	\encadre{ K_{\theta\ph} = K_{\ph\theta} = - \frac{B^2 r^2\sin^2\theta}{2N} \der{\omega}{\theta} } . 
\ee
In particular, the trace of $\w{K}$ defined by (\ref{e:ein:trace_K}) vanishes identically:
\be
  K := \gamma^{ij} K_{ij} = \frac{1}{A^2} \underbrace{K_{rr}}_{0} 
  + \frac{1}{A^2 r^2} \underbrace{K_{\theta\theta}}_{0}
  + \frac{1}{B^2 r^2 \sin^2\theta} \underbrace{K_{\ph\ph}}_{0} = 0 .
\ee
This is equivalent to the vanishing of the divergence of the unit normal $\vv{n}$ to  $\Sigma_t$ [cf. Eq.~(\ref{e:ein:trace_K})]. $\Sigma_t$ is then of 
\emph{maximal volume} with respect to nearby hypersurfaces (see e.g. \S~9.2.2 of Ref.~\cite{Gourg07a} for more details). This is fully analogous to the concept of
\emph{minimal surfaces} in a 3-dimensional Euclidean space (the change from \emph{minimal}
to \emph{maximal} being due to the change of metric signature). For this reason, the 
foliation $(\Sigma_t)$ with $K=0$ is called a \defin{maximal slicing}. 

The quadratic term $K_{ij} K^{ij}$ which appears in the Hamiltonian constraint
(\ref{e:ein:Einstein_PDE2}) is
\bea
	K_{ij} K^{ij} & = & K_{r\ph} K^{r\ph} + K_{\ph r} K^{\ph r} + K_{\theta\ph} K^{\theta\ph} 
	+ K_{\ph \theta} K^{\ph \theta}
	= 2  K_{r\ph} K^{r\ph} + 2 K_{\theta\ph} K^{\theta\ph} \nonumber \\
	& = & 2 \gamma^{rr} \gamma^{\ph\ph} (K_{r\ph})^2
	+ 2 \gamma^{\theta\theta} \gamma^{\ph\ph} (K_{\theta\ph})^2 , \nonumber
\eea
where to get the last line we used the fact that $\gamma^{ij}$ is diagonal. 
We thus get 
\be \label{e:sym:kcar}
	K_{ij} K^{ij} = \frac{B^2 r^2 \sin^2\theta}{2 A^2 N^2} 
	\, \partial \omega \partial\omega , 
\ee
with the following short-hand notation for any scalar fields $u$
and $v$ :
\be \label{s:sym:def_der_der}
	\encadre{ \partial u \partial v := \der{u}{r} \der{v}{r} 
	+ \frac{1}{r^2} \der{u}{\theta} \der{v}{\theta} } . 
\ee

%
%

\chapter{Einstein equations for rotating stars} \label{s:eer}


\minitoc
\vspace{1cm}

\section{General framework}

We focus at present on the case of a single rotating star in equilibrium. The corresponding 
spacetime $(\M,\w{g})$ is then stationary and asymptotically flat. 
In addition, it is reasonable to assume that the spacetime is axisymmetric. 
This has been shown to be necessary for viscous and heat-conducting fluids by Lindblom (1976) 
\cite{Lindb76}. For perfect fluids, the general argument is that a rotating non-axisymmetric body
will emit gravitational radiation, and therefore cannot be stationary. In this respect the situation is different from the Newtonian one, where stationary non-axisymmetric rotating configurations do exist, the 
best known example being that of the Jacobi ellipsoids. 

In addition of being axisymmetric, we will assume that the spacetime is circular (cf. \S~\ref{s:sym:ortho_trans}), since this is relevant 
for a perfect fluid rotating about the axis of symmetry [cf. Eq.~(\ref{e:sym:u_circular})].  
We may then use the  
QI coordinates $(x^\alpha)=(t,r,\theta,\ph)$ introduced in \S~\ref{s:sym:QI}. 
Consequently, the metric tensor is
fully specified by the four functions $N$, $A$, $B$ and $\omega$ listed in
(\ref{e:sym:liste4}). In this chapter we derive the partial differential equations that these functions have to fulfill in order for the metric to obey the Einstein equation (\ref{e:ein:ee}), as well as 
the equations of equilibrium for a perfect fluid. We shall also discuss the numerical method of resolution. 

\section{Einstein equations in QI coordinates}

\subsection{Derivation}

We shall derive the equations within the framework of the 3+1 formalism presented 
in \S~\ref{s:ein:3p1}.
Let us first consider the trace of the 3+1 Einstein equation (\ref{e:ein:Einstein_PDE1})
in which we make use of the Hamiltonian constraint (\ref{e:ein:Einstein_PDE2}) 
to replace ${}^3 R + K^2$ by $16\pi E + K_{ij} K^{ij}$, to get
\be  \label{e:sym:evol_trK}
    \der{K}{t} - \beta^i \der{K}{x^i} = 
  - D_i D^i N + N \left[ 4\pi (E+S) + K_{ij} K^{ij} \right] .  
\ee
In the present context, $K=0$ and $K_{ij} K^{ij}$ is given by 
Eq.~(\ref{e:sym:kcar}). The Laplacian of $N$ can be evaluated via 
the standard formula: 
\be
  D_i D^i N = \frac{1}{\sqrt{\gamma}} \der{}{x^i} \left( \sqrt{\gamma} \, \gamma^{ij}
  \der{N}{x^j} \right) ,  
\ee
where $\sqrt{\gamma}$ is the square root of the determinant of the components $\gamma_{ij}$
of the metric $\w{\gamma}$. From expression (\ref{e:sym:gam_dd}), we get
\be \label{e:eer:sqrt_gam}
  \sqrt{\gamma} = A^2 B r^2 \sin\theta .
\ee
Taking into account expression (\ref{e:sym:gam_uu}) for $\gamma^{ij}$, 
we conclude that (\ref{e:sym:evol_trK}) is equivalent to 
\be \label{e:sym:eqN_prov}
  \dderr{N}{r} + \frac{2}{r} \der{N}{r} + \frac{1}{r^2} \dderr{N}{\theta}
  + \frac{1}{r^2\tan\theta} \der{N}{\theta}
  = 4\pi A^2 N (E+S) + \frac{B^2r^2 \sin^2\theta }{2N} \partial\omega \partial \omega
  - \partial\ln B \partial N . 
\ee

Let us now consider the Hamiltonian constraint (\ref{e:ein:Einstein_PDE2}).
On the right-hand side, we have $K^2=0$ and $K_{ij} K^{ij}$ given by Eq.~(\ref{e:sym:kcar}).
There remains to compute the Ricci scalar ${}^3 R$. This can be done by means of 
formulas (\ref{e:sym:3Ricci})-(\ref{e:sym:3Ricci_scal}). One gets
\bea
  {}^3 R &=& -\frac{2}{A^2} \Bigg\{ \frac{1}{A} 
  \left[ \dderr{A}{r} + \frac{1}{r} \der{A}{r}
  - \frac{1}{A} \left( \der{A}{r} \right) ^2 
  + \frac{1}{r^2} \dderr{A}{\theta} - \frac{1}{r^2 A} \left( \der{A}{\theta} \right) ^2 
   \right] \nonumber \\
  &  & \qquad + \frac{1}{B} \left( \dderr{B}{r} + \frac{3}{r} \der{B}{r}
  + \frac{1}{r^2} \dderr{B}{\theta} + \frac{2}{r^2\tan\theta} \der{B}{\theta} \right)
    \Bigg\} .
\eea
The Hamiltonian constraint (\ref{e:ein:Einstein_PDE2}) thus becomes
\bea
  & & \frac{1}{A} 
  \left[ \dderr{A}{r} + \frac{1}{r} \der{A}{r}
  - \frac{1}{A} \left( \der{A}{r} \right) ^2 
  + \frac{1}{r^2} \dderr{A}{\theta} - \frac{1}{r^2 A} \left( \der{A}{\theta} \right) ^2 
   \right]  + \frac{B^2 r^2 \sin^2\theta}{4 N^2} 
	\, \partial \omega \partial\omega \nonumber \\
  & & + \frac{1}{B} \left( \dderr{B}{r} + \frac{3}{r} \der{B}{r}
  + \frac{1}{r^2} \dderr{B}{\theta} + \frac{2}{r^2\tan\theta} \der{B}{\theta} \right)
  \ = \ - 8\pi A^2 E . \label{e:sym:Ham_constr}
\eea

Moving to the momentum constraint (\ref{e:ein:Einstein_PDE3}), 
we first notice that $D_i K=0$. 
We may use the standard formula
to evaluate the divergence of a symmetric tensor (such as $\w{K}$):
\be
  D_j K^j_{\ \, i} = \frac{1}{\sqrt{\gamma}}
  \der{}{x^j} \left( \sqrt{\gamma}  K^j_{\ \, i} \right)
  - \frac{1}{2} \der{\gamma_{jk}}{x^i} K^{jk} . 
\ee
In the present case, the last term always vanishes since $K^{jk}$ is non-zero only for
non-diagonal terms ($jk$=$r\ph$, $\ph r$, $\theta\ph$ or $\ph\theta$) and $\gamma_{jk}$
is diagonal [Eq.~(\ref{e:sym:gam_dd})]. Since in addition $\dert{}{\ph} = 0$, 
we are left with
\be \label{e:sym:div_Kij}
  D_j K^j_{\ \, i} = \frac{1}{\sqrt{\gamma}} \left[ 
  \der{}{r} \left( \sqrt{\gamma}  K^r_{\ \, i} \right)
  + \der{}{\theta} \left( \sqrt{\gamma}  K^\theta_{\ \, i} \right) \right] . 
\ee
Now, $K^r_{\ \, r} = K^\theta_{\ \, r} = 0$ and 
$K^r_{\ \, \theta} = K^\theta_{\ \, \theta} = 0$. We thus conclude that the first two components of the momentum constraint (\ref{e:ein:Einstein_PDE3}) reduce to 
\be
  p_r = 0 \qquad \mbox{and}\qquad p_\theta = 0 . 
\ee
For the third component, we use Eq.~(\ref{e:sym:div_Kij}) for $i=\ph$, taking into 
account that $K^r_{\ \, \ph} = \gamma^{rj} K_{j\ph} = \gamma^{rr} K_{r\ph} = A^{-2} K_{r\ph}$
and $K^\theta_{\ \, \ph} = \gamma^{\theta j} K_{j\ph} = \gamma^{\theta\theta} 
K_{\theta\ph} = A^{-2} r^{-2} K_{\theta\ph}$ with the values 
(\ref{e:sym:Krph})-(\ref{e:sym:Kthph}) for $K_{r\ph}$ and $K_{\theta\ph}$.
We thus obtain
\be \label{e:sym:eq_shift_prov}
  \frac{N}{B^3} \left[ \frac{\sin\theta}{r^3} \der{}{r} \left( r^4 \frac{B^3}{N}
  \der{\omega}{r} \right) + \frac{1}{r\sin^2\theta} 
  \der{}{\theta} \left( \sin^3\theta \frac{B^3}{N}
  \der{\omega}{\theta} \right)
\right] = - 16\pi \frac{N A^2}{B^2} \frac{p_\ph}{r\sin\theta} . 
\ee

Finally, let us consider the $\ph\ph$ component of the 3+1 Einstein equation
(\ref{e:ein:Einstein_PDE1}). We have $K_{\ph\ph}=0$ and, according to Eq.~(\ref{e:ein:Lie_beta_K}),
\[
  \Liec{\beta} K_{\ph\ph} = \beta^k \der{}{x^k} \underbrace{K_{\ph\ph}}_{0}
	+ K_{k\ph} \underbrace{\der{ \beta^k}{\ph}}_{0} 
    + K_{\ph k} \underbrace{\der{\beta^k}{\ph}}_{0} = 0 . 
\]
Thus the left-hand side of the $\ph\ph$ component of Eq.~(\ref{e:ein:Einstein_PDE1}) vanishes identically. 
On the right-hand side, we have $D_j N = \dert{N}{x^j}$,
\bea
  D_\ph D_\ph N & = &\der{}{\ph} \underbrace{\der{N}{\ph}}_{0} - {}^3\Gamma^k_{\ \, \ph\ph} \der{N}{x^k}
    = - {}^3\Gamma^r_{\ \, \ph\ph} \der{N}{r} - {}^3\Gamma^\theta_{\ \, \ph\ph} \der{N}{\theta}
    \nonumber \\
    & = & r^2\sin^2\theta \, \frac{B^2}{A^2} 
    \left[ \left( \frac{1}{B} \der{B}{r} + \frac{1}{r} \right) \der{N}{r}
  + \frac{1}{r^2} \left( \frac{1}{B} \der{B}{\theta} + \frac{1}{\tan\theta} \right)
  \der{N}{\theta} \right] , \nonumber 
\eea
\[
   {}^3 R_{\ph\ph} = - r^2 \sin^2\theta \frac{B}{A^2}
    \left( \dderr{B}{r} + \frac{3}{r} \der{B}{r}
  + \frac{1}{r^2} \dderr{B}{\theta} + \frac{2}{r^2\tan\theta} \der{B}{\theta} \right) . 
\]
and 
\[
	K_{\ph k} K^k_{\ \, \ph} = K_{\ph r} K^r_{\ \, \ph} + K_{\ph \theta} K^\theta_{\ \, \ph} = \frac{1}{A^2} (K_{\ph r})^2 +
	\frac{1}{A^2 r^2} (K_{\ph\theta})^2
	= \frac{B^4 r^4\sin^4\theta}{4A^2 N^2} \, \partial\omega\partial \omega . 
\]
In addition, the matter term is 
\[
  (S-E)\gamma_{\ph\ph} - 2 S_{\ph\ph} = 
    (S^r_{\ \, r} + S^\theta_{\ \, \theta} + S^\ph_{\ \, \ph} - E) \gamma_{\ph\ph}
  - 2 \gamma_{\ph\ph} S^\ph_{\ \, \ph}
  = r^2\sin^2\theta \, B^2 (S^r_{\ \, r} + S^\theta_{\ \, \theta} - S^\ph_{\ \, \ph} - E) . 
\]
Accordingly, after division by $r^2\sin^2\theta\, N B^2/A^2$,
the $\ph\ph$ component of the 3+1 Einstein equation
(\ref{e:ein:Einstein_PDE1}) writes
\bea
	& & \frac{1}{B} \left( \dderr{B}{r} + \frac{3}{r} \der{B}{r}
  + \frac{1}{r^2} \dderr{B}{\theta} + \frac{2}{r^2\tan\theta} \der{B}{\theta} \right)
	+ \frac{1}{BN}\partial B \partial N 
	+ \frac{B^2 r^2\sin^2\theta}{2 N^2} \, \partial\omega\partial \omega 
	\nonumber \\
  && \qquad + \frac{1}{r N} \left(\der{N}{r} + \frac{1}{r\tan\theta} \der{N}{\theta} \right)
	\  = \  4\pi A^2 \left( S^r_{\ \, r} + S^\theta_{\ \, \theta} - S^\ph_{\ \, \ph} - E
		\right) . 
	\label{e:sym:eq_B_prov}
\eea

At this stage, we have four equations, (\ref{e:sym:eqN_prov}), (\ref{e:sym:Ham_constr})
(\ref{e:sym:eq_shift_prov}) and (\ref{e:sym:eq_B_prov}), for the four unknowns
$N$, $A$, $B$ and $\omega$. We construct linear combinations of these
equations in which classical elliptic operators appear. First, multiplying Eq.~(\ref{e:sym:eqN_prov}) by $B$, Eq.~(\ref{e:sym:eq_B_prov}) by $N B$ and adding the two
yields
\be
    \dderr{}{r}(NB) + \frac{3}{r} \der{}{r}(NB)
	+ \frac{1}{r^2} \dderr{}{\theta}(NB)
	+ \frac{2}{r^2\tan\theta} \der{}{\theta}(NB) 
 = 8\pi N A^2 B (S^r_{\ \, r} + S^\theta_{\ \, \theta} ) .  \label{e:sym:eq_NB_prov}
\ee 
Second, dividing Eq.~(\ref{e:sym:eqN_prov}) by $N$, adding Eq.~(\ref{e:sym:Ham_constr})
and subtracting Eq.~(\ref{e:sym:eq_B_prov}) yields
\bea
	& & \frac{1}{N} \dderr{N}{r} + \frac{1}{rN}\der{N}{r}
	+ \frac{1}{r^2N}\dderr{N}{\theta}
	+ \frac{1}{A} \dderr{A}{r} + \frac{1}{rA}\der{A}{r}
	+ \frac{1}{r^2A}\dderr{A}{\theta} \nonumber \\
	& & \qquad \qquad \qquad 
	= 8\pi A^2 S^\ph_{\ \, \ph} 
	+ \frac{1}{A^2} \partial A \partial A 
	+ \frac{3 B^2 r^2\sin^2 \theta}{4N^2} \, \partial \omega\partial \omega . 
		\label{e:sym:eq_AN_prov}
\eea
Reorganizing slightly Eqs.~(\ref{e:sym:eqN_prov}), (\ref{e:sym:eq_shift_prov}), 
(\ref{e:sym:eq_NB_prov}) and (\ref{e:sym:eq_AN_prov}), we obtain the final system:
\be \label{e:sym:eq_N}
 \encadre{ \Delta_3 \nu = 4\pi A^2(E+S)
	+ \frac{B^2 r^2\sin^2 \theta}{2N^2} \, \partial \omega\partial \omega
	- \partial \nu \partial(\nu + \ln B) }
\ee
\be \label{e:sym:eq_omeg}
  \encadre{ \tilde \Delta_3 (\omega r \sin\theta)
	= - 16\pi \frac{N A^2}{B^2} \frac{p_\ph}{r\sin\theta}
	+ r\sin\theta  \, \partial\omega \partial(\nu - 3\ln B ) }
\ee
\be \label{e:sym:eq_NB}
   \encadre{ \Delta_2 \left[ (NB-1) r\sin\theta \right]
	= 8\pi N A^2 B r\sin\theta (S^r_{\ \, r} + S^\theta_{\ \, \theta} ) }  
\ee
\be \label{e:sym:eq_lnApnu}
  \encadre{ \Delta_2 (\ln A + \nu) = 8\pi A^2 S^\ph_{\ \, \ph} 
 + \frac{3 B^2 r^2\sin^2 \theta}{4 N^2} \, \partial \omega\partial \omega
	- \partial\nu  \partial \nu } , 
\ee
where the following abbreviations have been introduced:
\bea
	&  & \nu := \ln N \\ 
	& & \Delta_2 := \dderr{}{r} + \frac{1}{r}\der{}{r}
	+ \frac{1}{r^2}\dderr{}{\theta} \\
	& & \Delta_3 := \dderr{}{r} + \frac{2}{r}\der{}{r}
	+ \frac{1}{r^2}\dderr{}{\theta} + \frac{1}{r^2\tan\theta} \der{}{\theta} \\
 	& & \tilde\Delta_3 := \Delta_3 - \frac{1}{r^2\sin^2\theta} . 
\eea
Terms of the form $\partial\nu\partial \nu$ have been defined 
by (\ref{s:sym:def_der_der}). 

The operator $\Delta_2$ is nothing but the Laplacian in a 2-dimensional flat space spanned by the polar coordinates $(r,\theta)$, whereas $\Delta_3$ is the Laplacian in a 
3-dimensional flat space, taking into account the axisymmetry ($\dert{}{\ph}=0$).
A deeper understanding of why these operators naturally occur in the problem
is provided by the (2+1)+1 formalism developed in Ref.~\cite{GourgB93}. 

\subsection{Boundary conditions}

Equations~(\ref{e:sym:eq_N})-(\ref{e:sym:eq_lnApnu}) forms a system of elliptic partial differential equations. It must be supplemented by a set of boundary conditions. Those are provided by the asymptotic flatness assumption: for $r\rightarrow +\infty$, the metric tensor tends towards Minkowski metric $\w{\eta}$, whose components in spherical coordinates are 
\be
 \eta_{\alpha\beta} \, dx^\alpha\, dx^\beta
	= - dt^2 + dr^2 + r^2 d\theta^2
  + r^2 \sin^2 \theta \, d\ph^2 .
\ee
Comparing with (\ref{e:sym:metricQI}), we get immediately the boundary conditions:
\be \label{e:sym:BC}
  \left\{ \begin{array}{lcl}
  N & \longrightarrow & 1 \\
  A & \longrightarrow & 1 \\
  B & \longrightarrow & 1 \\
  \omega & \longrightarrow & 0 
  \end{array} \right.
  \qquad \mbox{when} \quad
  r \longrightarrow + \infty . 
\ee

\begin{remark}
Spatial infinity is the only place where exact boundary conditions can be set, because
the right-hand sides of 
Eqs.~(\ref{e:sym:eq_N})-(\ref{e:sym:eq_lnApnu}) have non-compact support, except
for (\ref{e:sym:eq_NB}). This contrasts with the Newtonian case, where the basic
equation is Poisson equation for the gravitational potential $\Phi$
[Eq.~(\ref{e:sym:Poisson_Newt}) below]: 
\be
  \Delta_3 \Phi = 4\pi G\rho .
\ee
For a star, the right-hand side, involving the mass density $\rho$, has clearly compact support. The general solution outside the star is then known in advance: it is the 
(axisymmetric) harmonic function 
\be \label{e:eer:Phi_harm}
  \Phi(r,\theta) = \sum_{\ell=0}^\infty \alpha_\ell 
  \frac{P_\ell(\cos\theta)}{r^{\ell+1}} ,
\ee
where $P_\ell$ is the Legendre polynomial of degree $\ell$.
In this case, one can set the boundary conditions for a finite value of $r$ and perform
some matching of $\Phi$ and $\dert{\Phi}{r}$ to determine the coefficients $\alpha_\ell$. 
\end{remark}

From the properties of the operators $\Delta_3$ and $\tilde\Delta_3$ and the
boundary conditions (\ref{e:sym:BC}), one can infer the following
asymptotic behavior of the functions $N$ and $\omega$:
\be \label{e:eer:N_asympt}
	\encadre{ N = 1 + \frac{K_0}{r} + O\left(\frac{1}{r^2} \right) },
\ee
\be \label{e:eer:omega_asympt}
	\encadre{ \omega = \frac{K_1}{r^3} + O\left(\frac{1}{r^4} \right) },
\ee
where $K_0$ and $K_1$ are two constants. In Chap.~\ref{s:glo}, we will show that 
\be \label{e:eer:K0_K1}
	K_0 = - M \qquad\mbox{and}\qquad K_1 = 2J , 
\ee
$M$ and $J$ being respectively the mass of the star and its angular momentum
[cf. Eqs.~(\ref{e:glo:N_asympt}) and (\ref{e:glo:omega_asympt})]. 

\subsection{Case of a perfect fluid}

Since we are interested in rotating stars, let us write the matter source terms in
the system (\ref{e:sym:eq_N})-(\ref{e:sym:eq_lnApnu}) for the case of 
a perfect fluid [cf. Eq.~(\ref{e:ein:T_perfect_fluid})].
Due to the circularity hypothesis, the fluid 4-velocity 
is necessarily of the form (\ref{e:sym:u_circular}):
$\vv{u} = u^t ( \vxi + \Omega \vchi )$. We may use 
Eq.~(\ref{e:sym:xi_n_chi}) to express $\vxi$ in terms of the unit timelike vector $\vv{n}$
(the 4-velocity of the ZAMO)
and $\omega \vchi$. We thus get
\be \label{e:sym:u_decomp_n_chi}
	\vv{u} = (N u^t) \left[ \vv{n} + \frac{1}{N} (\Omega - \omega) \vchi \right] . 
\ee
Now, according to Eq.~(\ref{e:ein:comp_n_cov}), $\vv{n}\cdot\vchi = n_\mu \chi^\mu = n_\ph = 0$. 
Therefore, Eq.~(\ref{e:sym:u_decomp_n_chi}) constitutes an orthogonal decomposition of 
the fluid 4-velocity with respect to the ZAMO 4-velocity $\vv{n}$:
\be \label{e:sym:u_GnU}
	\encadre{ \vv{u} = \Gamma \left( \vv{n} + \vv{U} \right) }, 
\ee
with 
\be \label{e:sym:Gam_U}
	\encadre{\Gamma = N u^t} \qquad \mbox{and}\qquad 
	\encadre{\vv{U} =  \frac{1}{N} (\Omega - \omega) \vchi } . 
\ee
$\Gamma$ is the \defin{Lorentz factor} of the fluid with respect to the ZAMO and the vector
$\vv{U}$ is the fluid \defin{velocity} (3-velocity) with respect to the ZAMO
(see e.g. Ref.~\cite{Gourg06} for details). 
$\vv{U}$ is a spacelike vector, tangent to the hypersurfaces $\Sigma_t$. 
Let us define 
\be \label{e:sym:U_Omeg_omeg}
	\encadre{ U := \frac{B}{N}(\Omega-\omega) r\sin\theta }. 
\ee
Since $\vchi\cdot\vchi = B^2 r^2\sin^2\theta$, we have
\be \label{e:sym:U2}
	\vv{U}\cdot\vv{U} = U^2 . 
\ee
From $\vv{n}\cdot\vv{n}=-1$ and $\vv{n}\cdot\vv{U}=0$, Eq.~(\ref{e:sym:u_GnU}) yields the
geometric interpretation of the Lorentz factor as (minus) the scalar product of the ZAMO 
4-velocity with the fluid one: 
\be \label{e:eer:Gam_nu}
  \encadre{\Gamma = - \vv{n}\cdot\vv{u} } . 
\ee
The normalization condition 
$\vv{u}\cdot\vv{u}=-1$ expressed in terms of (\ref{e:sym:u_GnU}) leads to the standard relation 
\be \label{e:sym:Gam_U2}
	\encadre{ \Gamma = \left( 1 - U^2 \right) ^{-1/2} } . 
\ee

The quantities $E$, $p_\ph$, $S^r_{\ \, r}$, $S^\theta_{\ \, \theta}$ and
$S^\ph_{\ \, \ph}$ which appear in the right-hand of Eqs.~(\ref{e:sym:eq_N})-(\ref{e:sym:eq_lnApnu}) are computed by means of 
Eqs.~(\ref{e:ein:E_euler})-(\ref{e:ein:S_euler}), using the form 
(\ref{e:ein:T_perfect_fluid}) of the energy momentum tensor. 
We get, using (\ref{e:sym:u_GnU}), 
\[
  E = T_{\mu\nu} n^\mu n^\nu = (\vep+p) \underbrace{u_\mu n^\mu}_{-\Gamma}
  \underbrace{u_\nu n^\nu}_{-\Gamma} + p \underbrace{g_{\mu\nu} n^\mu n^\nu}_{-1}
  = \Gamma^2 (\vep+p) - p , 
\]
\[
  p_\ph = - T_{\mu\nu} n^\nu \gamma^\mu_{\ \, \ph}
  = - (\vep+p) \underbrace{u_\mu \gamma^\mu_{\ \, \ph}}_{\Gamma U_\ph}
    \underbrace{u_\nu n^\nu}_{-\Gamma} + p 
  \underbrace{g_{\mu\nu} n^\nu  \gamma^\mu_{\ \, \ph}}_{0} 
  = \Gamma^2 (\vep+p) U_\ph , 
\]
\[
  S^\alpha_{\ \, \beta} = T_{\mu\nu} \gamma^{\mu \alpha} \gamma^\nu_{\ \, \beta} 
  = (\vep + p) \underbrace{u_\mu \gamma^{\mu \alpha}}_{\Gamma U^\alpha} 
    \underbrace{u_\nu  \gamma^\nu_{\ \, \beta}}_{\Gamma U_\beta}
  + p \underbrace{g_{\mu\nu} \gamma^{\mu \alpha} \gamma^\nu_{\ \, \beta}}_{\gamma^\alpha_{\ \, \nu} \gamma^\nu_{\ \, \beta}} = \Gamma^2 (\vep+p) U^\alpha U_\beta
  + p \, \gamma^\alpha_{\ \, \beta}  . 
\]
Now, 
from Eq.~(\ref{e:sym:Gam_U}), $U^\alpha = (0,0,0,U^\ph)$, and from 
Eq.~(\ref{e:sym:U2}), $U_\ph U^\ph = U^2$ and 
$U_\ph = B r \sin\theta\, U$. 
Besides $\gamma^r_{\ \, r} = \gamma^\theta_{\ \, \theta} 
  = \gamma^\ph_{\ \, \ph} = 1$. 
Accordingly, the above results can be expressed as
\be \label{e:sym:E_G2_ep}
  \encadre{ E = \Gamma^2 (\vep+p) - p } , 
\ee
\be \label{e:sym:p_ph_U}
  \encadre{ p_\ph = B (E+p) U r\sin\theta },
\ee
\be \label{e:sym:S_fluid_parfait}
  \encadre{ S^r_{\ \, r} = p },\quad
  \encadre{S^\theta_{\ \, \theta} = p},\quad
  \encadre{ S^\ph_{\ \, \ph} = p + (E+p) U^2 }, \quad 
  \encadre{S = 3p + (E+p) U^2 } . 
\ee
From the above expressions, the term $E+S$ in the right-hand side of Eq.~(\ref{e:sym:eq_N}) can be written
\be \label{e:sym:EpS_fluid}
  E+S = \frac{1+U^2}{1-U^2} (\vep+p) + 2 p . 
\ee

\subsection{Newtonian limit}

In the non-relativistic limit, $\nu$ tends to the Newtonian gravitational potential $\Phi$
(divided by $c^2$):
\be \label{e:sym:lim_Newt_nu}
    \encadre{ \nu \simeq \frac{\Phi}{c^2} } _{\mbox{\footnotesize\  Newt.}}
\ee
Moreover, Eq.~(\ref{e:sym:U_Omeg_omeg}) reduces to 
\be \label{e:sym:U_Newt}
	\encadre{ U = \Omega r\sin\theta } _{\mbox{\footnotesize\  Newt.}}
\ee
and $U^2\ll 1$,  $p\ll \vep$ and $\vep \simeq \rho c^2$, where $\rho$ is the 
mass density. Accordingly, (\ref{e:sym:EpS_fluid}) gives
\be
  E+S \simeq \rho c^2 . 
\ee
Since in addition $A\simeq 1$ and all the quadratic terms involving gradient of the 
metric potentials tend to zero, the first equation of the Einstein system
(\ref{e:sym:eq_N})-(\ref{e:sym:eq_lnApnu}) reduces to (after restoring the $G$ and $c$'s)
\be \label{e:sym:Poisson_Newt}
  \encadre{ \Delta_3 \Phi = 4\pi G \rho } _{\mbox{\footnotesize\  Newt.}}
\ee
The other three equations become trivial, of the type ``$0=0$''. 
We conclude that, at the non-relativistic limit, the Einstein equations (\ref{e:sym:eq_N})-(\ref{e:sym:eq_lnApnu}) reduce to the 
Poisson equation\index{Poisson equation} (\ref{e:sym:Poisson_Newt}), which is the basic equation for Newtonian gravity. 

\subsection{Historical note} \label{s:sym:hist}

The Einstein equations for stationary axisymmetric bodies were written quite early. In 1917,
Weyl \cite{Weyl1917} treated the special case of \emph{static} bodies, i.e. with
orthogonal Killing vectors
$\vxi$ and $\vchi$ , or equivalently with $\omega=0$  [cf. Eq.~(\ref{e:sym:omega_zero})]. 
The effect of rotation was first taken into account by Lense and Thirring in 1918 \cite{LenseT1918},
who derived the first order correction induced by rotation in the metric outside a spherical body. 
The first full formulation of the problem of axisymmetric rotating matter was obtained by
Lanczos in 1924 \cite{Lancz1924} and van Stockum in 1937 \cite{Stock37}. They wrote
the Einstein equations in Lewis-Papapetrou coordinates (cf. \S~\ref{s:sym:QI}) 
and obtained an exact solution
describing dust (i.e. pressureless fluid) rotating rigidly about some axis. However this solution does not correspond to a star since the dust fills the entire spacetime, with an increasing
density away from the rotation axis. In particular the solution is not asymptotically flat. 
Rotating stars have been first considered by Hartle and Sharp (1967) \cite{HartlS67}, who formulated a variational principle for rotating fluid bodies in general relativity. Hartle (1967)
\cite{Hartl67} and Sedrakian \& Chubaryan (1968) \cite{SedraC68} wrote Einstein equations for a slowly rotating star, in coordinates
different from the QI ones\footnote{Namely coordinates for which 
$g_{\theta\theta} = g_{\ph\ph}/\sin^2\theta$ instead of (\ref{e:sym:gab_QI}).}. 
For rapidly rotating stars, a system of elliptic equations similar to (\ref{e:sym:eq_N})-(\ref{e:sym:eq_lnApnu}) was presented (in integral form) by Bonazzola \& Machio (1971) \cite{BonazM71}. It differs from (\ref{e:sym:eq_N})-(\ref{e:sym:eq_lnApnu})
by the use of $\ln g_{tt}$ instead of $\ln N = \nu$. 
Bardeen \& Wagoner (1971) \cite{BardeW71} presented a system composed of three elliptic
equations equivalent to Eqs.~(\ref{e:sym:eq_N}), (\ref{e:sym:eq_omeg}) and 
(\ref{e:sym:eq_NB}), supplemented by a first order equation of the form
\be \label{e:sym:eq_BW}
	\der{}{\theta}(\ln A + \nu) = \mathcal{S}(r,\theta) , 
\ee
where $\mathcal{S}(r,\theta)$ is a complicated expression containing first and second derivatives
of $\nu$, $A$, $B$ and $\omega$ and no matter term [see Eq.~(4d) of Ref.~\cite{ButteI76}
(9 lines !) or Eq.~(II.16) of \cite{BardeW71}]. Equation~(\ref{e:sym:eq_BW}) results from
the properties
$R_{rr} = R_{\theta\theta}/r^2$ and $R_{r\theta}=0$ of the Ricci tensor. 
Note that these are verified
only for an isotropic energy-momentum tensor, such as the perfect fluid one 
[Eq.~(\ref{e:ein:T_perfect_fluid})]. For anisotropic ones (like that of an electromagnetic field), Eq.~(\ref{e:sym:eq_BW}) should be modified so as to include the anisotropic part 
of the energy-momentum tensor on its right-hand side. 
The full system of the four elliptic equations (\ref{e:sym:eq_N})-(\ref{e:sym:eq_lnApnu}) has been first written by 
Bonazzola, Gourgoulhon,  Salgado \& Marck (1993)\footnote{In the original 
article \cite{BonazGSM93}, the functions $\tilde A = (AB)^{1/4}$ and $\tilde B = (B/A)^{1/2}$
were employed instead of $A$ and $B$. The equations with the present values of $A$ and $B$
are given in Ref.~\cite{GourgHLPBM99}.} \cite{BonazGSM93}.


\section{Spherical symmetry limit}

\subsection{Taking the limit}

In spherical symmetry, the metric components $g_{\alpha\beta}$ simplify significantly. 
Indeed, we have seen that $A$ and $B$ must be equal on the rotation axis 
[Eq.~(\ref{e:sym:A_B_axis})]. But, if the star is spherically symmetric, there is no privileged axis. Therefore, we can extend (\ref{e:sym:A_B_axis}) to all space: 
\be \label{e:sym:A_B_spher}
	\encadre{ A = B } _{\mbox{\footnotesize\  spher.  sym.}}
\ee
Besides, to comply with spherical symmetry, the parameter $\Omega$ in the decomposition 
(\ref{e:sym:u_circular}) of the fluid 4-velocity must vanish. Otherwise, there would exist 
a privileged direction around which the fluid is rotating. Hence we have 
\be \label{e:eer:ang_vel_zero}
	\encadre{ \Omega = 0 } _{\mbox{\footnotesize\  spher.  sym.}}
\ee
and $\vv{u}$ is colinear to the Killing vector $\vxi$.
From Eqs.~(\ref{e:sym:p_ph_U}) and (\ref{e:sym:U_Omeg_omeg}) with $\Omega=0$, we have $p_\ph \propto \omega$. 
Equation~(\ref{e:sym:eq_omeg}) is then
a linear equation in $\omega$. It admits a unique solution which is 
well-behaved in all space:
\be \label{e:sym:omeg_zero}
	\encadre{ \omega = 0 } _{\mbox{\footnotesize\  spher.  sym.}}
\ee
According to (\ref{e:sym:omega_zero}), this implies that the spacetime is static.

Properties (\ref{e:eer:ang_vel_zero}) and (\ref{e:sym:omeg_zero}), once inserted in 
Eq.~(\ref{e:sym:U_Omeg_omeg}) leads to the vanishing of the fluid velocity with respect to the
ZAMO:
\be \label{e:sym:U_zero}
	\encadre{ U = 0 } _{\mbox{\footnotesize\  spher.  sym.}}
\ee

Thanks to (\ref{e:sym:A_B_spher}) and (\ref{e:sym:omeg_zero}), the spacetime metric
(\ref{e:sym:metricQI}) becomes
\be  \label{e:sym:gab_spher}
  \encadre{ g_{\alpha\beta} \, dx^\alpha\, dx^\beta
	= - N(r)^2 dt^2 + A(r)^2 (dr^2 + r^2 d\theta^2 +  r^2 \sin^2 \theta \, d\ph^2 )
	} _{\mbox{\footnotesize\  spher.  sym.}} . 
\ee
The coordinates $(t,r,\theta,\ph)$ are in this case truly \emph{isotropic}, the spatial
metric being conformal to a flat one. This justify the name \emph{quasi-isotropic}
given to these coordinates in the general case. 

Since $U=0$ implies $\Gamma=1$, the Einstein equations 
(\ref{e:sym:eq_N})-(\ref{e:sym:eq_lnApnu}) with a perfect fluid source and spherical
symmetry reduce to 
\be \label{e:sym:eq_N_spher}
 \encadre{ \frac{d^2\nu}{dr^2} + \frac{2}{r} \frac{d\nu}{dr} = 4\pi A^2 (\vep+3p)
	- \frac{d\nu}{dr}  \frac{d}{dr}(\nu + \ln A) } _{\mbox{\footnotesize\  spher.  sym.}}
\ee
\be \label{e:sym:eq_NB_spher}
   \encadre{ \frac{d^2}{dr^2}(NA) + \frac{3}{r} \frac{d}{dr}(NA)
	= 16\pi N A^3  p }  _{\mbox{\footnotesize\  spher.  sym.}}
\ee
\be \label{e:sym:eq_lnApnu_spher}
  \encadre{ \frac{d^2}{dr^2} (\ln A + \nu) + \frac{1}{r} \frac{d}{dr}
	(\ln A + \nu) = 8\pi A^2 p
	- \left(\frac{d\nu}{dr}\right) ^2 } _{\mbox{\footnotesize\  spher.  sym.}}  
\ee
We have omitted Eq.~(\ref{e:sym:eq_omeg}) since, as seen above, it is trivially
satisfied in spherical symmetry. Besides, we have used the equivalent form
(\ref{e:sym:eq_NB_prov}) of Eq.~(\ref{e:sym:eq_NB}). 

\subsection{Link with the Tolman-Oppenheimer-Volkoff system}

The reader familiar with static models of neutron stars might have been surprised that the spherical 
limit of the Einstein equations (\ref{e:sym:eq_N})-(\ref{e:sym:eq_lnApnu}) does not give
the familiar \defin{Tolman-Oppenheimer-Volkoff system} (TOV). Indeed, the TOV system is 
(see e.g. \cite{HaensPY07,Hartl03}):
\bea
&& \frac{dm}{d\rr} = 4\pi \rr^2 \, \vep  \label{e:sym:TOV1} \\
&& \frac{d\nu}{d\rr} = \left( 1 - \frac{2 m(\rr)}{\rr} \right) ^{-1}
	\left( \frac{m(\rr)}{\rr^2} + 4\pi  \rr p \right) \\
&& \frac{dp}{d\rr} = - ( \vep + p) \frac{d\nu}{d\rr}  , \label{e:sym:TOV3}
\eea
which is pretty different from the system 
(\ref{e:sym:eq_N_spher})-(\ref{e:sym:eq_lnApnu_spher}). In particular, the latter is
second order, whereas (\ref{e:sym:TOV1})-(\ref{e:sym:TOV3}) is clearly first order. 
However, the reason of the discrepancy is already apparent in the notations used in 
(\ref{e:sym:TOV1})-(\ref{e:sym:TOV3}) : the radial coordinate is not the same in the two
system: in (\ref{e:sym:eq_N_spher})-(\ref{e:sym:eq_lnApnu_spher}), it the isotropic 
coordinate $r$, whereas in the TOV system, it is the \defin{areal radius} $\rr$. 
With this radial coordinate, the metric components have the form
\be \label{e:sym:gab_areal}
  {\bar g}_{\alpha\beta} \, d{\bar x}^\alpha\, d{\bar x}^\beta
	= - N(\rr)^2 dt^2 + \left( 1 - \frac{2 m(\rr)}{\rr} \right) ^{-1} d\rr^2 + \rr^2 d\theta^2 +  \rr^2 \sin^2 \theta \, d\ph^2 .
\ee
This form is different from the isotropic one, given by (\ref{e:sym:gab_spher}).
Note however that the coordinates ${\bar x}^\alpha=(t,\rr,\theta,\ph)$ differ from the
isotropic coordinates $x^\alpha=(t,r,\theta,\ph)$ only in the choice of $r$. 
From (\ref{e:sym:gab_areal}), it is clear that the area of a 2-sphere defined
by $t=\mathrm{const}$ and $\rr=\mathrm{const}$ is given by
\be
  \mathcal{A} = 4\pi \rr^2 , 
\ee
hence the qualifier \emph{areal} given to the coordinate $\rr$. 
On the contrary, we see from (\ref{e:sym:gab_spher}) that the area of a 2-sphere defined
by $t=\mathrm{const}$ and $r=\mathrm{const}$ is
\be
  \mathcal{A} = 4\pi A(r)^2 r^2 ,
\ee
so that there is no simple relation between the area and $r$. 

Outside the star, the solution is the Schwarzschild metric (see \S~\ref{s:sym:outside}) and
the coordinates $({\bar x}^\alpha)$ are simply the standard 
\emph{Schwarzschild coordinates}. In particular, outside the star $m(\rr) = M =
\mathrm{const}$, where $M$ is the gravitational mass of the star and 
$N(\rr)^2  = (1-2M/\rr)$, so that we recognize in (\ref{e:sym:gab_areal}) the standard 
expression of Schwarzschild solution. The expression of this solution in terms of the isotropic 
coordinates $(x^\alpha)$ used here is derived in \S~\ref{s:sym:outside}.
The relation between the two radial coordinates is given by
\be \label{e:eer:rr_r}
  \rr = r \left( 1 + \frac{M}{2r} \right) ^2  \qquad \mbox{(outside the star)} . 
\ee
Far from the star, in the weak field region ($M/r \ll 1$), we have of course $\rr\simeq r$. 

There exist two coordinate systems that extend the TOV areal coordinates
to the rotating case: 
\begin{itemize}
\item the coordinates employed by Hartle \& Thorne \cite{Hartl67,HartlT68} 
and Sedrakyan \& Chubaryan \cite{SedraC68} to compute stellar models in the slow rotation approximation (metric expanded to the order $\Omega^2$); 
\item the so-called \emph{radial gauge} introduced by Bardeen \& Piran \cite{BardeP83}; however
the entire hypersurface $\Sigma_t$ cannot be covered in a regular way by these coordinates,
as shown in Appendix~A of Ref.~\cite{BonazGSM93}. 
\end{itemize}

\subsection{Solution outside the star} \label{s:sym:outside}

Outside the star, $\vep=0$ and $p=0$. Equation (\ref{e:sym:eq_NB_spher}) reduces then
to
\[
	\frac{d^2}{dr^2}(NA) + \frac{3}{r} \frac{d}{dr}(NA) = 
	\frac{1}{r^3} \frac{d}{dr} \left[ r^3 \frac{d}{dr}(NA) \right] = 0 , 
\]
from which we deduce immediately that 
\[
	\frac{d}{dr}(NA) = \frac{2 a}{r^3} , \quad\mbox{with}\ a = \mathrm{const} . 
\]
The factor $2$ is chosen for later convenience. 
The above equation is easily integrated, taken into account the boundary conditions
(\ref{e:sym:BC}): 
\be \label{e:sym:NA_outside}
	NA = 1 - \frac{a}{r^2} . 
\ee
Besides, subtracting Eq.~(\ref{e:sym:eq_N_spher}) from Eq.~(\ref{e:sym:eq_lnApnu_spher}), 
both with $\vep=0$ and $p=0$, we get the following equation outside the star: 
\be \label{e:sym:eq_A_outside_prov}
	\frac{d^2}{dr^2} \ln A + \frac{1}{r} \frac{d}{dr} \ln A
	- \frac{1}{r}  \frac{d\nu}{dr} = \frac{d\nu}{dr} \frac{d}{dr}\ln A . 
\ee
Now Eq.~(\ref{e:sym:NA_outside}) gives, since $\nu=\ln N$, 
\[
	\frac{d\nu}{dr} =  - \frac{d}{dr}\ln A 
	+ \frac{2 a}{r(r^2-a)} . 
\]
Inserting this relation into Eq.~(\ref{e:sym:eq_A_outside_prov}), we get
\be
	\frac{d^2 A}{dr^2} 
	+ \frac{2(r^2 -2a)}{r(r^2-a)} \frac{dA}{dr}
	- \frac{2 a}{r^2(r^2-a)} A = 0 . 
\ee
The reader can check easily that the unique solution of this linear ordinary differential equation which satisfies to the boundary condition (\ref{e:sym:BC}) requires $a\geq 0$ and is
\be \label{e:eer:A_sym_spher}
	A = \left( 1 + \frac{\sqrt{a}}{r} \right) ^2 . 
\ee
Relation (\ref{e:sym:NA_outside}) yields then
\be
	N = \left( 1 - \frac{\sqrt{a}}{r} \right) \left( 1 + \frac{\sqrt{a}}{r} \right) ^{-1} . 
\ee
The constant $\sqrt{a}$ is actually half the gravitational mass of the star.
Indeed the asymptotic expansion for $r\rightarrow\infty$ of the above expression is
\[
  N \simeq \left( 1 - \frac{\sqrt{a}}{r} \right) \times
  \left( 1 - \frac{\sqrt{a}}{r} \right) \simeq 1 - \frac{2\sqrt{a}}{r} .
\]
Comparing with Eqs.~(\ref{e:eer:N_asympt}) and (\ref{e:eer:K0_K1}), we get
\be
	\sqrt{a} = \frac{M}{2} . 
\ee
Replacing $a$, $A$ and $N$ by the above values in (\ref{e:sym:gab_spher}),
we obtain the following form of the metric outside the star: 
\be \label{e:eer:Schwarz_QI}
	   g_{\alpha\beta} \, dx^\alpha\, dx^\beta  = - \left( 
    \frac{1 - \frac{M}{2r}}{ 1 + \frac{M}{2r}} \right) ^2
         dt^2 
    + \left( 1 + \frac{M}{2r} \right) ^4 \left[ d{r}^2 
    + {r}^2 (d\theta^2 + \sin^2\theta \, d\ph^2) \right]  .
\ee
We recognize the \defin{Schwarzschild metric}\index{Schwarzschild!metric} 
expressed in isotropic coordinates. 
Actually, we have recovered the \defin{Birkhoff theorem}\index{Birkhoff theorem}\footnote{More precisely, we have established Birkhoff theorem under the hypothesis of stationarity; however this hypothesis is not necessary.}: With spherical symmetry, the only solution 
of Einstein equation outside the central body is the Schwarzschild solution. 

\begin{remark}
There is no equivalent of the Birkhoff theorem for axisymmetric rotating bodies: 
for black holes, 
the generalization of the Schwarzschild metric beyond spherical symmetry is the \emph{Kerr metric}\index{Kerr!metric}, and the generic metric outside a rotating star 
is \emph{not} the Kerr metric. Moreover, no fluid star has been found to be a source for the
Kerr metric (see e.g. \cite{MartiMR08}). 
The only matter source for the Kerr metric found so far is made of two 
counterrotating thin disks of collisionless particles \cite{BicakL93}.
It has also been shown that uniformly rotating fluid bodies are finitely separated from 
Kerr solutions, except extreme Kerr \cite{Meine06}; this means that one cannot have a quasiequilibrium transition from a star to a black hole.   
\end{remark}

\section{Fluid motion}

Having discussed the gravitational field equations (Einstein equation), let us now turn to 
the equation governing the equilibrium of the fluid, i.e. Eq.~(\ref{e:ein:divT_zero}). 

\subsection{Equation of motion at zero temperature}

We consider a perfect fluid at zero temperature, which is a very good approximation for
a neutron star, except immediately after its birth. 
The case of finite temperature has been treated by Goussard, Haensel \& Zdunik (1997) \cite{GoussHZ97}. 
The energy-momentum tensor has the form
(\ref{e:ein:T_perfect_fluid}) and, thanks to the zero temperature hypothesis, the equation of state (EOS) can be written as
\bea	
	\vep &  =  & \vep(n_{\rm b}) \\
	p & = & p(n_{\rm b}) , 
\eea
where $n_{\rm b}$ is the baryon number density in the fluid frame. 

The equations of motion are the energy-momentum conservation law (\ref{e:ein:divT_zero}) : 
\be \label{e:sym:T_perfect_fluid}
	\nabla^\mu T_{\alpha\mu} = 0 
\ee
and the baryon number conservation law: 
\be \label{e:eer:cons_baryon}
	 \nabla_\mu (n_{\rm b} u^\mu) = 0 . 
\ee
Inserting the perfect fluid form (\ref{e:ein:T_perfect_fluid}) into Eq.~(\ref{e:sym:T_perfect_fluid}), expanding and projecting orthogonally to the fluid 4-velocity
$\vv{u}$ [via the projector $\w{\bot}$ given by (\ref{e:ein:comp_ortho_proj})], 
we get the \emph{relativistic Euler equation}: 
\be \label{e:sym:Euler}
	(\vep + p) u^\mu \nabla_\mu u_\alpha +
	(\delta^\mu_{\ \, \alpha} + u^\mu u_\alpha) \nabla_\mu p = 0 . 
\ee
Now the Gibbs-Duhem relation at zero temperature states that
\be \label{e:sym:Gibbs-Duhem}
	dp = n_{\rm b}\, d\mu , 
\ee
where $\mu$ is the baryon 
chemical potential, $\mu := d\vep/dn_{\rm b}$. Moreover, thanks to 
the first law of Thermodynamics at zero temperature (see e.g. Ref.~\cite{Gourg06} for details), $\mu$ is equal to the enthalpy per baryon $h$ defined by
\be \label{e:sym:def_h}
	\encadre{ h := \frac{\vep+p}{n_{\rm b}} } . 
\ee
Thus we may rewrite (\ref{e:sym:Gibbs-Duhem}) as $dp = n_{\rm b}\, dh$, hence
\[
	\nabla_\alpha p = n_{\rm b} \nabla_\alpha h . 
\]
Accordingly, Eq.~(\ref{e:sym:Euler}) becomes, after division by $n_{\rm b}$, 
\[
	h u^\mu \nabla_\mu u_\alpha +
	(\delta^\mu_{\ \, \alpha} + u^\mu u_\alpha) \nabla_\mu h = 0 , 
\]
which can be written in the compact form
\be \label{e:sym:fluid_eom}
	\encadre{ u^\mu \nabla_\mu (h u_\alpha) + \nabla_\alpha h = 0 } . 
\ee
Thanks to the properties $u^\mu u_\mu =-1$ and $u^\mu \nabla_\alpha u_\mu = 0$ (the latter being a consequence of the former), Eq.~(\ref{e:sym:fluid_eom}) can be rewritten in the equivalent form
\be \label{e:eer:fluid_canon}
 \encadre{ u^\mu \left[ \nabla_\mu (h u_\alpha) -  \nabla_\alpha (h u_\mu) \right] = 0 }.
\ee
This makes appear the antisymmetric bilinear form 
$\Omega_{\alpha\beta} := \nabla_\alpha (h u_\beta) -  \nabla_\alpha (h u_\beta)$, called the 
\defin{vorticity 2-form}. 
The fluid equation of motion (\ref{e:eer:fluid_canon}) has been
popularized by Lichnerowicz \cite{Lichn55,Lichn67} and Carter \cite{Carte79}, when treating relativistic hydrodynamics by means of Cartan's exterior calculus (see e.g. Ref.~\cite{Gourg06} for an introduction).  

Equation~(\ref{e:sym:fluid_eom}) [or (\ref{e:eer:fluid_canon})] is more useful than the original Euler equation (\ref{e:sym:Euler}). 
It leads easily to the relativistic generalization of the classical Bernoulli 
theorem (\S~\ref{s:sym:Bernoulli}) and to a first integral of motion for a rotating star
(\S~\ref{s:sym:first_integ}). 

\subsection{Bernoulli theorem} \label{s:sym:Bernoulli}

Equation~(\ref{e:sym:fluid_eom}) is an identity between 1-forms. Applying these 1-forms to the 
stationarity Killing vector $\vxi$ (i.e. contracting (\ref{e:sym:fluid_eom}) with $\xi^\alpha$), 
we get successively
\bea
	& & \xi^\nu u^\mu \nabla_\mu (h u_\nu) + \underbrace{\xi^\nu \nabla_\nu h}_{0} = 0 ,
		\nonumber \\
	& & u^\mu \nabla_\mu (h u_\nu \xi^\nu) - h u_\nu u^\mu \nabla_\mu \xi^\nu = 0 , \nonumber \\
	& & u^\mu \nabla_\mu (h u_\nu \xi^\nu) - h 
	\underbrace{u ^\mu u^\nu \nabla_\mu \xi_\nu}_{0} = 0  , \nonumber \\
    & & u^\mu \nabla_\mu (h u_\nu \xi^\nu) = 0  . \nonumber 
\eea
Note that $\xi^\nu \nabla_\nu h=0$ stems from the stationarity of the fluid flow
and $u ^\mu u^\nu \nabla_\mu \xi_\nu =0$ from Killing equation (\ref{e:sym:Killing}). We thus have
\be \label{e:sym:Bernoulli}
   \encadre{ \wnab_{\vv{u}} (h \vv{u}\cdot\vxi) = 0 } . 
\ee
In other words, the scalar quantity $h \vv{u}\cdot\vxi$ is constant along any given fluid line. This results constitutes a relativistic generalization of the classical
Bernoulli theorem. To show it, let us first recast Eq.~(\ref{e:sym:Bernoulli}) in an alternative
form. Combining 
Eq.~(\ref{e:sym:u_GnU}) with Eq.~(\ref{e:sym:xi_n_chi}), we get
(using $\vv{n}\cdot\vv{n}=-1$, $\vv{n}\cdot\vchi=0$ and $\vv{n}\cdot\vv{U}=0$)
\be \label{e:eer:h_u_xi}
  h \vv{u}\cdot\vxi = h \Gamma (\vv{n} + \vv{U})\cdot(N\vv{n} - \omega \vchi)
	= - h \Gamma N \left( 1 + \frac{\omega}{N} \vchi \cdot \vv{U} \right) . 
\ee
Accordingly, by taking the logarithm of
$-h \vv{u}\cdot\vxi$, we obtain that property (\ref{e:sym:Bernoulli}) is  equivalent to
\be \label{e:sym:Bernoulli_const}
	\encadre{ H + \nu + \ln \Gamma  + \ln \left( 1 + \frac{\omega}{N} \vchi \cdot \vv{U} \right) 
  = \mbox{const along a fluid line}}, 
\ee
where we have introduced the \defin{log-enthalpy}
\be \label{e:eer:def_H}
	\encadre{ H := \ln \left(\frac{h}{m_{\rm b}} \right) }, 
\ee
$m_{\rm b}$ being the mean baryon mass : $m_{\rm b} \simeq 1.66\times 10^{-27} {\; \rm kg}$. 
Defining the fluid internal energy density by
\be \label{e:eer:def_eps_int}
	\vep_{\rm int} := \vep - m_{\rm b} n_{\rm b} 
\ee
we have [cf. (\ref{e:sym:def_h})]
\be
	H = \ln \left( 1 + \frac{\vep_{\rm int} + p}{m_{\rm b} n_{\rm b}} \right) .
\ee
At the Newtonian limit,
\[
  \frac{\vep_{\rm int} + p}{m_{\rm b} n_{\rm b}} \ll 1 , 
\]
and $m_{\rm b} n_{\rm b} \simeq \rho$ (the mass density), so that $H$ tends towards the (non-relativistic) specific enthalpy:
\be
  \encadre{ H \simeq \frac{\vep_{\rm int} + p}{\rho} }  _{\mbox{\footnotesize\  Newt.}} .
\ee
Besides, thanks to (\ref{e:sym:Gam_U2}) and $U^2 \ll 1$, we have
$\ln\Gamma \simeq U^2 / 2$. Moreover we have already seen that $\nu$ tends towards the Newtonian gravitational potential $\Phi$ [Eq.~(\ref{e:sym:lim_Newt_nu})]. 
In addition, at the Newtonian limit, $\omega=0$.
Consequently, the Newtonian limit of (\ref{e:sym:Bernoulli_const}) is 
\be \label{e:sym:Bernoulli_Newt}
  \encadre{ H + \Phi + \frac{U^2}{2}  = \mbox{const along a fluid line} } _{\mbox{\footnotesize\  Newt.}} ,
\ee
which is nothing but the classical Bernoulli theorem.

The general relativistic Bernoulli theorem (\ref{e:sym:Bernoulli}) has been first derived by
Lichnerowicz in 1940 \cite{Lichn40,Lichn41}. The special relativistic version had been obtained 
previously by Synge in 1937 \cite{Synge37}. 

\begin{remark}
That a symmetry (here stationarity, via the Killing vector $\vxi$) gives rise to 
a conserved quantity [Eq.~(\ref{e:sym:Bernoulli_const})] is of course a manifestation of the 
\emph{Noether theorem}. We will encounter another one in \S~\ref{s:glo:redshift_gal}.
\end{remark}

To establish (\ref{e:sym:Bernoulli}) we have used nothing but the fact that the flow obeys
the symmetry and $\vxi$ is a Killing vector. For a flow that is axisymmetric, we thus have the
analogous property
\be 
   \encadre{ \wnab_{\vv{u}} (h \vv{u}\cdot\vchi) = 0 } . 
\ee
The quantity $h \vv{u}\cdot\vchi$ is thus conserved along the fluid lines.
Thanks to (\ref{e:sym:u_GnU}) and $\vv{n}\cdot\vchi=0$, we may write it as
\be \label{e:eer:h_u_chi}
  h \vv{u}\cdot\vchi = h \Gamma \vchi\cdot \vv{U} = h \Gamma B^2 r^2 \sin^2 \theta \, U^\ph .
\ee
At the Newtonian limit, $h\rightarrow 1$, $\Gamma\rightarrow 1$, $B \rightarrow 1$
and $U^\ph \simeq \dot\ph$, so that the conserved quantity along each fluid line is 
the specific angular momentum about the rotation axis, $r^2 \sin^2\theta\, \dot\ph$.

\begin{remark}
It is well known that, along a timelike geodesic, the conserved quantities associated with the Killing vectors $\vxi$ and $\vchi$ are $\vv{u}\cdot\vxi$ and
$\vv{u}\cdot\vchi$ \cite{MisneTW73,Wald84}. Here we have instead $h\vv{u}\cdot\vxi$ and
$h\vv{u}\cdot\vchi$, for the fluid lines are not geodesics, except when $p=0$. In the latter case, $h$ reduces to a constant (typically $m_ {\rm b}$, cf. Eq.~(\ref{e:sym:def_h}) and Eq.~(\ref{e:eer:def_eps_int}) with $\vep_{\rm int}=0$)
and we recover the classical result. 
\end{remark}

\subsection{First integral of motion} \label{s:sym:first_integ}

In the derivation of the Bernoulli theorem, we have assumed a general stationary fluid.
Let us now focus to the case of a rotating star, namely of a circular fluid motion around the rotation axis\footnote{For more general motions see \S~VI.E of \cite{GourgMUE11}.}. The fluid 4-velocity takes then the form
(\ref{e:sym:u_circular}): 
\be  \label{e:sym:u_k}
  \encadre{ \vv{u} = u^t \vv{k} }, 
\ee
with 
\be \label{e:sym:def_k}
  \encadre{ \vv{k} := \vxi + \Omega \vchi }.
\ee
Note that if $\Omega$ is a constant, $\vv{k}$ is a Killing vector: it trivially satisfies Killing 
equation since $\vxi$ and $\vchi$ both do. $\vv{u}$ is then colinear to a Killing vector and  
one says that the star is 
undergoing \defin{rigid rotation}. 
If $\Omega$ is not constant, the star is said to be in \defin{differential rotation}. 

To derive the first integral of motion, it is convenient to start from the fluid equation of motion in the Carter-Lichnerowicz form (\ref{e:eer:fluid_canon}).
Thanks to the symmetry of the Christoffel symbols (\ref{e:ein:Christoffel}) with 
respect to their last two indices, we may replace the
covariant derivatives by partial ones in Eq.~(\ref{e:eer:fluid_canon}): 
\be \label{e:eer:fluid_canon_par}
  u^\mu \left[ \der{}{x^\mu} (h u_\alpha) -  \der{}{x^\alpha} (h u_\mu) \right] = 0 . 
\ee
Assuming a circular motion, i.e. $\vv{u}$ of the form (\ref{e:sym:u_k}), we have
$u^\mu = (u^t,0,0,u^t\Omega)$, so that 
\[
  u^\mu \der{}{x^\mu} (h u_\alpha) = u^t \left[ \der{}{t}(h u_\alpha) + \Omega \der{}{\ph}(h u_\alpha) \right] = 0 . 
\]
Hence Eq.~(\ref{e:eer:fluid_canon_par}) reduces to (after division by $u^t$)
\[
  \der{}{x^\alpha} (h u_t) + \Omega \der{}{x^\alpha} (h u_\ph) = 0 , 
\]
or equivalently
\[
  \der{}{x^\alpha} [h (u_t + \Omega u_\ph)] - h u_\ph \der{\Omega}{x^\alpha} = 0 . 
\]
Noticing that $u_t = \vv{u}\cdot\vxi$ and $u_\ph = \vv{u}\cdot\vchi$, we
may rewrite the above result as
\be
  \wnab( h \vv{u}\cdot\vv{k}) - h \vv{u}\cdot\vchi \, \wnab \Omega = 0 . 
\ee
Dividing by $h \vv{u}\cdot\vv{k}$, we get 
\be \label{e:sym:grad_first_integ}
  \encadre{ 
    \wnab \ln(- h \vv{u}\cdot\vv{k})
  - \frac{\vv{u}\cdot\vchi}{\vv{u}\cdot\vv{k}} \, \wnab \Omega = 0} . 
\ee
\begin{remark}
The minus sign in the logarithm is justified by the fact, that since $\vv{u}$
and $\vv{k}$ are both future directed timelike vectors, $\vv{u}\cdot\vv{k} <0$.
\end{remark}
The integrability condition of Eq.~(\ref{e:sym:grad_first_integ}) is either 
(i) $\Omega = \mathrm{const}$ (rigid rotation) or (ii)
the factor in front of $\wnab\Omega$ is a function of $\Omega$:
\be \label{e:sym:def_F_omega}
  - \frac{\vv{u}\cdot\vchi}{\vv{u}\cdot\vv{k}} = F(\Omega) . 
\ee
We shall discuss these two cases below.
But before proceeding, let us give an equivalent form of Eq.~(\ref{e:sym:grad_first_integ}), based on the \defin{specific angular momentum}
\be \label{e:eer:def_ell}
  \encadre{ \ell := - \frac{\vv{u}\cdot\vchi}{\vv{u}\cdot\vxi} = - \frac{u_\ph}{u_t} } . 
\ee
This quantity is (minus) the quotient of the two Bernoulli quantities discussed in 
\S~\ref{s:sym:Bernoulli}. It is thus conserved along any fluid line in a stationary and axisymmetric spacetime, irrespective of the fluid motion. Its name stems from 
the fact that $\ell$ is the ``angular momentum'' (per unit mass) $h\vv{u}\cdot\vchi$ 
divided by the ``energy'' (per unit mass) $-h\vv{u}\cdot\vxi$. 
By combining (\ref{e:eer:h_u_xi}) with (\ref{e:eer:h_u_chi}) and
$U^\ph = (\Omega-\omega)/N$ [Eq.~(\ref{e:sym:Gam_U})], we get 
\be
  \ell = \frac{B^2 (\Omega-\omega)  r^2\sin^2\theta}{N^2 + B^2 
	\omega (\Omega-\omega) r^2\sin^2\theta } .
\ee
At the Newtonian limit, $B\rightarrow 1$, $N\rightarrow 1$, $\omega \rightarrow 0$ and
$\ell \simeq \Omega r^2\sin^2\theta$. 

Since $\vv{u}\cdot\vv{k} = u_t( 1 - \Omega \ell) $, we may rewrite the equation of motion 
(\ref{e:sym:grad_first_integ}) as
\be \label{e:eer:eom_ut_ell_prov}
    \wnab \ln[- h u_t( 1 - \Omega \ell) ]
  + \frac{\ell}{1 - \Omega \ell}\, \wnab \Omega = 0 . 
\ee
Besides, we have $u_\mu u^\mu = u_t u^t + u_\ph u^\ph = u^t(u_t + \Omega u_\ph)
  = u^t u_t ( 1 - \Omega \ell)$, so that the normalization $u_\mu u^\mu = -1$ leads
to
\[
  u_t ( 1 - \Omega \ell) = - \frac{1}{u^t} . 
\]
Therefore, Eq.~(\ref{e:eer:eom_ut_ell_prov}) can be recast as
\be \label{e:eer:eom_ut_ell}
  \encadre{ \wnab \ln \left( \frac{h}{u^t} \right) 
  + \frac{\ell}{1 - \Omega \ell}\, \wnab \Omega = 0 }. 
\ee
This is a variant of the equation of motion (\ref{e:sym:grad_first_integ}).

\subsubsection{Rigid rotation}

If $\Omega = \mathrm{const}$, then $\wnab \Omega = 0$ and
Eq.~(\ref{e:sym:grad_first_integ}) leads immediately to the first integral of motion
\be \label{e:sym:int_prem_rigid_h}
   \ln( -h \vv{u}\cdot\vv{k} ) = \mathrm{const}  . 
\ee
Thanks to (\ref{e:sym:u_k}) and (\ref{e:sym:Gam_U}), we have
\be \label{e:sym:uck}
	\vv{u}\cdot\vv{k} = {\underbrace{(u^t)}_{\Gamma/N}}^{-1} 
	\, \underbrace{\vv{u}\cdot\vv{u}}_{-1} = - \frac{N}{\Gamma} , 
\ee
so that the first integral (\ref{e:sym:int_prem_rigid_h})
becomes  
\be \label{e:sym:int_prem_rigid}
	\encadre{ H + \nu -\ln \Gamma = \mathrm{const} } . 
\ee
This first integral of motion has been obtained first in 1965 by Boyer \cite{Boyer65} for incompressible fluids\footnote{More precisely, the first integral obtained by Boyer is (within our notations)
$N^2 (1-U^2) = \mathrm{const} / h^2$. Taking the square root and the logarithm, we obtain
(\ref{e:sym:int_prem_rigid}).}
 and generalized to compressible fluids by Hartle \& Sharp (1965)
\cite{HartlS65} and Boyer \& Lindquist (1966) \cite{BoyerL66b}. 
Note the difference of sign in front of $\ln\Gamma$ with respect to the Bernoulli theorem
(\ref{e:sym:Bernoulli_const}). 
\begin{remark}
Another important difference is that (\ref{e:sym:Bernoulli_const}) provides a constant along each fluid line, but this constant
may vary from one fluid line to another one, whereas (\ref{e:sym:int_prem_rigid})
involves a constant throughout the entire star. However, (\ref{e:sym:int_prem_rigid}) requires the fluid to be in a pure circular motion, while the Bernoulli theorem (\ref{e:sym:Bernoulli_const}) is valid for any stationary flow. 
\end{remark}

The Newtonian limit of (\ref{e:sym:int_prem_rigid}) is readily obtained
since $\nu\simeq \Phi$ and $\ln\Gamma \simeq U^2/2$ (cf. \S~\ref{s:sym:Bernoulli}); 
it reads
\be
  \encadre{ H + \Phi - \frac{U^2}{2}  = \mathrm{const} } _{\mbox{\footnotesize\  Newt.}} . 
\ee
Again notice the change of sign in front of $U^2/2$ with respect to the Bernoulli 
expression (\ref{e:sym:Bernoulli_Newt}). Notice also that 
$U^2 = \Omega^2 r^2\sin^2\theta$ [Eq.~(\ref{e:sym:U_Newt})] and that
\be
	\Phi_{\rm tot} := \Phi - \frac{1}{2} \Omega^2 r^2\sin^2\theta 
\ee
is the total potential (i.e. that generating the gravitational force and the centrifugal one) in the frame rotating at the angular velocity $\Omega$ about the rotation axis. 

\subsubsection{Differential rotation}

Equation~(\ref{e:sym:grad_first_integ}) gives, 
thanks to (\ref{e:eer:def_H}), (\ref{e:sym:def_F_omega}) and (\ref{e:sym:uck}), 
\[
	\wnab \left( H + \nu - \ln\Gamma \right)
	+ F(\Omega) \wnab \Omega = 0 . 
\]
We deduce immediately the first integral of motion :
\be \label{e:sym:int_prem_diff}
	\encadre{ H + \nu -\ln \Gamma  + \int_0^\Omega F(\Omega') d\Omega' = \mathrm{const} } . 
\ee
\begin{remark}
The lower boundary in the above integral has been set to zero but it can be changed to any value: this will simply change the value of the constant in the right-hand side.  
\end{remark}
Let us make explicit Eq.~(\ref{e:sym:def_F_omega}), 
by means of respectively Eqs.~(\ref{e:sym:uck}), (\ref{e:sym:u_GnU}), 
(\ref{e:sym:Gam_U}), (\ref{e:sym:Gam_U2}) and (\ref{e:sym:U_Omeg_omeg}):
\bea
 F(\Omega) & = & - \frac{\vv{u}\cdot\vchi}{\vv{u}\cdot\vv{k}} 
	= \frac{\Gamma}{N} \Gamma (\vv{n} + \vv{U})\cdot\vchi 
	= \frac{\Gamma^2}{N} \vv{U}\cdot\vchi
	= \frac{\Gamma^2}{N^2} (\Omega-\omega) \vchi\cdot\vchi \nonumber \\
	& = & \frac{1}{N^2(1-U^2)} (\Omega-\omega) B^2 r^2\sin^2\theta 
	= \frac{B^2 (\Omega-\omega) r^2\sin^2\theta}{N^2 [ 1 -(B/N)^2 
	(\Omega-\omega)^2 r^2\sin^2\theta]} . \nonumber 
\eea
Hence 
\be \label{e:sym:F_omega}
	\encadre{ F(\Omega) = \frac{B^2 (\Omega-\omega)  r^2\sin^2\theta}{N^2 - B^2 
	(\Omega-\omega)^2 r^2\sin^2\theta} } . 
\ee
At the Newtonian limit, this relation reduces to 
\be \label{e:sym:F_omega_Newt}
	\encadre{ F(\Omega) = \Omega r^2 \sin^2 \theta} _{\mbox{\footnotesize\  Newt.}}
\ee
and (\ref{e:sym:int_prem_diff}) to
\be
  \encadre{ H + \Phi - \frac{U^2}{2} + \int_0^\Omega F(\Omega') d\Omega' 
	= \mathrm{const} } _{\mbox{\footnotesize\  Newt.}} . 
\ee
Equation (\ref{e:sym:F_omega_Newt}) implies that $\Omega$ must be a function of the 
distance from the rotation axis:
\be
	\encadre{ \Omega = \Omega(r\sin\theta) } _{\mbox{\footnotesize\  Newt.}} . 
\ee
We thus recover a well known result for Newtonian differentially rotating stars, 
the so-called \defin{Poincaré-Wavre theorem} (see e.g. \S~3.2.1 of Ref.~\cite{Tasso00}). 
In the general relativistic case, once some choice of the function $F$ is made, one must
solve Eq.~(\ref{e:sym:F_omega}) in $\Omega$ to get the value of $\Omega$ at each point. 
The first integral (\ref{e:sym:int_prem_diff}) has been first written\footnote{in the 
equivalent form $N^2 (1-U^2) = [1+\Gamma(\Omega)]^2/h^2$, the link between the 
functions $\Gamma(\Omega)$ and $F(\Omega)$ being $\ln[1+\Gamma(\Omega)] = - \int_0^\Omega F(\Omega') d\Omega'$} by Boyer (1966) \cite{Boyer66}.  The relation (\ref{e:sym:F_omega}) has been
presented by Bonazzola \& Machio (1971) \cite{BonazM71} (see also the work 
by Abramowicz (1971, 1974) \cite{Abram71,Abram74}).

From the alternative form (\ref{e:eer:eom_ut_ell}) of the equation of fluid motion, 
we get, as an integrability condition, that $\ell / (1- \Omega\ell)$ must be a
function of $\Omega$, i.e. that $\ell$ itselft must be a function of $\Omega$: 
\be
  \encadre{ \ell = \ell(\Omega) }. 
\ee
The relation between this function and $F$ is obtained by 
comparing (\ref{e:eer:eom_ut_ell_prov}) with (\ref{e:sym:grad_first_integ}):
\be
  F(\Omega) = \frac{\ell(\Omega)}{1 - \Omega \ell(\Omega)} 
  \iff \ell(\Omega) = \frac{F(\Omega)}{1 + \Omega F(\Omega)} .  
\ee

\subsection{Stellar surface and maximum rotation velocity}

Let us consider a rigidly rotating star and rewrite the first integral of motion 
(\ref{e:sym:int_prem_rigid}) as
\be \label{e:eer:int_prem_Hc}
    H = H_{\rm c} + \nu_{\rm c} - \nu + \ln\Gamma , 
\ee
where $H_{\rm c}$ and $\nu_{\rm c}$ are the values of $H$ and $\nu$ at $r=0$ [we have
$\ln\Gamma_{\rm c} = 0$ since $U=0$ at $r=0$ by Eq.~(\ref{e:sym:U_Omeg_omeg})]. 
The surface of the star is defined by the vanishing of the pressure:
\be \label{e:eer:def_surf}
   \encadre{ \mbox{stellar surface:}\quad p = 0 }, 
\ee
which expresses the local equilibrium with respect to the ambient vacuum. 
Let $H_0$ be the value of $H$ corresponding to $p=0$ (usually, $H_0 = 0$).
As a solution of the Poisson-like equation (\ref{e:sym:eq_N}), $\nu$ is an increasing function 
of $r$, from $\nu_{\rm c} < 0$ to $\nu=0$ at $r=+\infty$.  
For instance, at the Newtonian limit and in spherical symmetry,
$\nu = -M/r$ outside the star. Therefore $-\nu$ is a decreasing function of $r$ and if
$\ln\Gamma =0$ (no rotation), (\ref{e:eer:int_prem_Hc}) implies that $H$ is a decreasing function of
$r$ as well. Provided that $H_{\rm c} > H_0$, it always reaches $H_0$ at some point, thereby defining the surface of the star. 
If the star is rotating, $\ln\Gamma$ is an increasing function of $r$. It takes its maximum 
value in the equatorial plane $\theta=\pi/2$. For instance, for a Newtonian star, 
$\ln\Gamma = (\Omega^2/2) r^2 \sin^2\theta$. If $\Omega$ is too large, it could be that 
$\ln\Gamma$ compensate the decay of $-\nu$ in such a way that $H$ remains always larger than $H_0$. 
Therefore, for a given value of $H_{\rm c}$, there exists a maximum rotation speed. 
Physically this corresponds to the fluid particles
reaching the orbital Keplerian velocity at the stellar equator. For this reason, the maximal
value of $\Omega$ is called the \defin{Keplerian limit} and denoted $\Omega_{\rm K}$.   

\section{Numerical resolution}

Having established all the relevant equations in the preceding sections, let us now 
discuss briefly their numerical resolution to get a model of a rotating star. 

\subsection{The self-consistent-field method} \label{s:eer:SCF}

To be specific, we consider rigidly rotating stars: $\Omega = \mathrm{const}$, but the following
discussion can be adapted to take into account differential rotation. 
A standard algorithm to numerically construct a rotating stellar model is the following one.
Choose 
\begin{itemize}
\item a barotropic EOS, of the form
\be \label{e:sym:EOS_H}
  \vep = \vep(H) \qquad\mbox{and}\qquad  p = p(H), 
\ee
assuming that $H=0$ corresponds to $p=0$;
\item some central value $H_{\rm c}$ of the log-enthalpy $H$;
\item some value for the constant angular velocity $\Omega$.
\end{itemize}
In addition assume some values for the four metric functions
$N$, $\omega$, $A$ and $B$. These may be crude values, such as the flat spacetime ones:
$N=A=B=1$ and $\omega=0$. 
Next, initialize $U$ to zero and the enthalpy to a crude profile, like
$H = H_{\rm c} (1-r^2/R^2)$ where $R$ is some radius chosen, also crudely, around the expected radius of the final model.  
Then
\begin{enumerate}
\item Via the EOS (\ref{e:sym:EOS_H}), evaluate $\vep$ and $p$. The surface of the star is then defined by $p=0$.
\item From $U$, $\Gamma$, $\vep$ and $p$, compute via Eqs.~(\ref{e:sym:E_G2_ep})-(\ref{e:sym:S_fluid_parfait}) the matter sources 
$E$, $p_\ph$, $S$, $S^r_{\ \, r}$, $S^\theta_{\ \, \theta}$, $S^\ph_{\ \, \ph}$ that appear in the
right-hand side of Einstein equations.
\item Solve the Einstein equations (\ref{e:sym:eq_N})-(\ref{e:sym:eq_lnApnu})
as flat Poisson-like equations, using the current value of 
$\nu$, $\omega$, $A$ and $B$ on their right-hand sides, 
thereby getting new values of $N=\exp \nu$, 
$\omega$, $A$ and $B$.
\item Compute the velocity field from Eq.~(\ref{e:sym:U_Omeg_omeg}):
$U = (B/N)(\Omega-\omega) r\sin\theta$; evaluate the Lorentz factor via
Eq.~(\ref{e:sym:Gam_U2}): $\Gamma = (1-U^2)^{-1/2}$.
\item Use the first integral of fluid motion in the form
(\ref{e:eer:int_prem_Hc}):
\[
  H = H_{\rm c} + \nu_{\rm c} - \nu +\ln \Gamma 
\]
to compute $H$ in all space. Go to step 1. 
\end{enumerate}
In practice, this method leads to a unique solution for a given value
of the input parameters $(H_{\rm c},\Omega)$ and a fixed EOS. 

The above method is called the \defin{self-consistent-field method} (SCF) and was 
introduced for computing Newtonian rotating stars by Ostriker \& Mark (1968) \cite{OstriM68}
(see Ref.~\cite{Rieut06} for some historical account). The first application of this 
method to compute rapidly rotating stellar models in full general relativity, along
the lines sketched above is due to Bonazzola \& Maschio (1971) \cite{BonazM71}.

\begin{remark}
The procedure described above is only a sketch of the actual algorithm implemented in a numerical code; the latter is generally more complicated, involving some relaxation, progressive increase of $\Omega$ from null value, convergence to a given mass, etc.  
\end{remark}

The mathematical analysis of the self-consistent-field method applied to the Einstein equations (\ref{e:sym:eq_N})-(\ref{e:sym:eq_lnApnu})
has been performed by Schaudt \& Pfister \cite{SchauP96,Schau98,PfistS00}. They proved the exponential 
convergence of the method and that the exterior and interior Dirichlet problem for relativistic rotating
stars is solvable, under some conditions on the boundary data. 
The full mathematical demonstration of existence of rotating fluid stars in general relativity has been given by Heilig in 1995 \cite{Heili95}, by a method different from the self-consistent-field, and
only for sufficiently small angular velocities $\Omega$. 
Let us stress that, up to now, we do not know any exact solution describing both the interior and the exterior of a rotating star in general relativity (see e.g. \cite{MarsS98,Senov93} for a discussion). 

\subsection{Numerical codes}

The first numerical models of rotating relativistic stars have been computed by Hartle \& Thorne
(1968) \cite{HartlT68} in the approximation of slow rotation
(cf. \S~\ref{s:sym:hist}). The first rapidly rotating models
were obtained by Bonazzola \& Maschio (1971) \cite{BonazM71}
using a formulation very close to that presented
here, namely based on Lewis-Papapetrou coordinates (the cylindrical version of QI coordinates, cf. \S~\ref{s:sym:QI}) and applying the 
self-consistent-field method to a system of equations similar to  (\ref{e:sym:eq_N})-(\ref{e:sym:eq_lnApnu}) (basically using the variable $\sqrt{-g_{tt}}$ instead of $N$). All the numerical codes developed since then have been based on the self-consistent-field method. Almost all make use of QI coordinates, an exception being
the code of Lin \& Novak (2006) \cite{LinN06} based on Dirac gauge.   
In 1972, Wilson \cite{Wilso72} used the Bardeen-Wagoner formulation \cite{BardeW71}
(cf. \S~\ref{s:sym:hist}) to compute models of differentially rotating stars without an explicit EOS. 
In 1974 Bonazzola \& Schneider (1974) \cite{BonazS74} improved Bonazzola 
\& Maschio's code and computed the first models models of rapidly rotating relativistic stars constructed on an explicit EOS (an ideal Fermi gas of neutron). 
In 1975, Butterworth \& Ipser \cite{ButteI75,ButteI76} developed a code based on 
the Bardeen-Wagoner formulation and computed homogeneous rotating bodies \cite{ButteI75,ButteI76} as well as polytropes \cite{Butte76}.
Friedman, Ipser \& Parker
(1986, 1989) \cite{FriedIP86,FriedIP89} applied their method to some realistic (i.e. based onto 
detailed microphysics calculations) EOS. 
This study has been extended by Lattimer et al. (1990) \cite{LattiPMY90} to 
include more EOS. 

In 1989, Komatsu, Eriguchi \&  Hachisu (KEH) \cite{KomatEH89a} developed a new code, also based on the Bardeen-Wagoner formulation, and
applied it to differentially rotating polytropes \cite{KomatEH89b}. 
Their method has been improved by Cook, 
Shapiro \& Teukolsky (CST) (1992,1994) \cite{CookST92,CookST94a}, by introducing the
radial variable $s=r/(r + r_{\rm eq})$, where $r_{\rm eq}$ is the $r$ coordinate of 
the stellar equator, in order that the computational domain extends to spatial 
infinity --- the only place where the boundary conditions are known in advance 
[Eq.~(\ref{e:sym:BC})]. In 1995, Stergioulas \& Friedman \cite{StergF95} developed their own code 
based on the CST scheme, improving the accuracy, leading to the famous public domain code
\texttt{rns} \cite{RNS}.  

In 1993, Bonazzola, Gourgoulhon, Salgado \& Marck (BGSM) \cite{BonazGSM93} developed a new code based 
a formulation very close to that presented here (cf. \S~\ref{s:sym:hist}), using spectral 
methods \cite{GrandN09}, whereas all previous codes employed finite differences. 
This code has been improved in 1998 \cite{BonazGM98,GourgHLPBM99} (by introducing numerical domains adapted to the surface of the star) and incorporated into the \textsc{Lorene} library, to become the public domain code \texttt{nrotstar} \cite{Lorene}. It is presented briefly in Appendix~\ref{s:lor}. 

In 1998, Nozawa et al. \cite{NozawSGE98} have performed a detailed comparison of the KEH code, 
the \texttt{rns} code and the BGSM code, showing that the relative difference between 
\texttt{rns} and BGSM is of the order of $10^{-4}$ (at most $10^{-3}$ in some extreme cases). 
The relative difference between KEH and BGSM is only of the order $10^{-3}$ to $10^{-2}$.
This is due to the finite computational domain used by KEH, implying approximate boundary conditions 
for some finite value of $r$. On the contrary, thanks to some compactification, both \texttt{rns} and BGSM are based on a computational domain that extends to $r=+\infty$ where the exact boundary conditions (\ref{e:sym:BC}) can be imposed. 

A new multi-domain spectral code has been developed by Ansorg, Kleinwaechter \& Meinel (AKM) in 2002 \cite{AnsorKM02,AnsorFKMPS04}. It is extremely accurate, leading to rotating stellar models to machine accuracy (i. e. $10^{-15}$ for 16-digits computations). It is has been used recently to compute
sequences of differentially rotating stars, with a very high degree of differential rotating, leading to toroidal stellar shapes \cite{AnsorGV09}. 	

For a more extended discussion of the numerical codes, see the review article by Stergioulas \cite{Sterg03}.
In particular, a comparison between the AKM, KEH, BGSM, \texttt{rns}
and \texttt{Lorene/rotstar} codes can be found in Table~2 of that article. 

\begin{figure}
\centerline{\includegraphics[height=0.35\textheight]{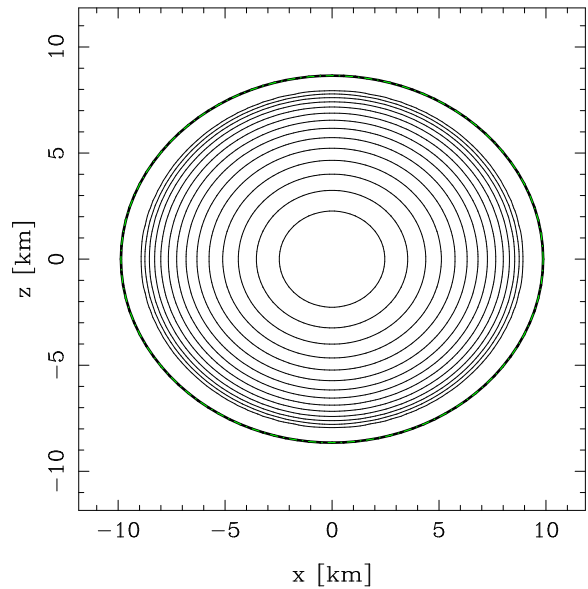}}
\caption{\label{f:sym:coupe_ener} \small
Isocontours of the proper energy density $\vep$ in the meridional plane $\ph=0$ for
the rotating neutron star model discussed in \S~\ref{s:sym:illustr}. The coordinates
$(x,z)$ are defined by $x:=r\sin\theta$ and $z:=r\cos\theta$. The thick solid line marks the surface
of the star.}
\end{figure}

\subsection{An illustrative solution} \label{s:sym:illustr}

In order to provide some view of the various fields in a relativistic rotating star, we present here some plots from a numerical solution\footnote{This solution can be reproduced by using the parameter files stored in the directory \texttt{Lorene/Codes/Nrotstar/Parameters/GR/APR\_1.4Msol\_716Hz}} obtained by the \texttt{Lorene/nrotstar} code
described in Appendix~\ref{s:lor}. 
The solution corresponds to a neutron star of mass $M=1.4\, M_\odot$ rigidly rotating at the frequency $\Omega/(2\pi) = 716 {\rm\; Hz}$. This value has been chosen for it is the highest one among observed neutron stars: it is achieved 
by the pulsar PSR~J1748-2446ad discovered in 2006 \cite{Hesse_al06}. 
The EOS is the following one (see \cite{HaensPY07} for details):
\begin{itemize}
\item \emph{for the core:} the model A18+$\delta v$+UIX* of Akmal, Pandharipande \& Ravenhall (1998) \cite{AkmalPR98}, describing a matter of neutrons, protons, electrons and muons via a Hamiltonian including two-body and three-body interactions, as well as relativistic corrections;
\item \emph{for the inner crust:} the SLy4 model of Douchin \& Haensel \cite{DouchH01};
\item \emph{for the outer crust:} the Haensel \& Pichon (1994) model \cite{HaensP94}, which is based on the experimental masses of neutron rich nuclei.
\end{itemize}

\begin{table}
\centerline{
\begin{tabular}{ll}
\hline \\[-2mm]
Gravitational mass $M$ & $1.400 \, M_\odot$ \\
Baryon mass $M_{\rm b}$ & $1.542 \, M_\odot$ \\
Rotation frequency $\Omega/(2\pi)$ & $716 {\rm\; Hz}$ \\
Central log-enthalpy $H_{\rm c}$ & $0.2262 \, c^2$ \\
Central proper baryon density $n_{\rm b,c}$ & $0.5301 {\rm\; fm}^{-3}$ \\ 
Central proper energy density $\vep_{\rm c}$ &   $5.7838 \; \rho_{\rm nuc} c^2$ \\
Central pressure $p_{\rm c}$ &  $0.8628 \; \rho_{\rm nuc} c^2$ \\
Coordinate equatorial radius $r_{\rm eq}$ & $9.867{\rm\; km}$ \\
Coordinate polar radius $r_{\rm p}$ & $8.649{\rm\; km}$ \\
Axis ratio $r_{\rm p}/r_{\rm eq}$ & $0.8763$ \\
Circumferential equat. radius $R_{\rm circ}$ & $12.08 {\rm\; km}$ \\
Compactness $G M /(c^2 R_{\rm circ})$ & $0.1711$ \\
Angular momentum $J$ &  $0.7238\; G M_\odot^2 / c$ \\
Kerr parameter $c J / (G M^2)$  &  $0.3693$ \\
Moment of inertia $I$ &  $1.417\times 10^{38} {\rm\; kg\, m}^2$ \\
Kinetic energy ratio $T/W$ &  $0.0348$ \\
Velocity at the equator $U_{\rm eq}$ & $0.1967 \, c$ \\
Redshift from equator, forward $z_{\rm eq}^{\rm f}$ & $-0.01362$ \\
Redshift from equator, backward $z_{\rm eq}^{\rm b}$ & $0.5529$ \\
Redshift from pole $z_{\rm p}$  & $0.2618$ \\
\hline
\end{tabular}
}
\caption{\label{t:sym:APR_716Hz} 
Properties of the rotating neutron star model presented in \S~\ref{s:sym:illustr};
\protect\\
$M_\odot := 1.989\times 10^{30} {\rm\; kg}$, $\rho_{\rm nuc} := 1.66\times 10^{17} \; {\rm kg\, m}^{-3}$.}
\end{table}

\begin{figure}
\centerline{\includegraphics[height=0.35\textheight]{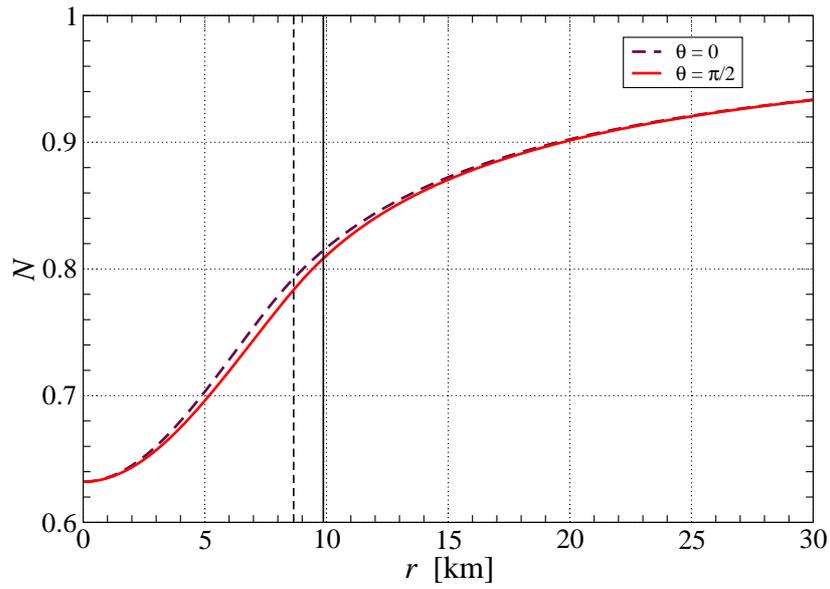}}
\caption{\label{f:sym:prof_n} \small
Profile of the lapse function $N$ in two different directions: $\theta=0$ (rotation axis, dashed line)
and $\theta =\pi/2$ (equatorial plane, solid line). The vertical solid line (resp. dashed line) marks the location of
the stellar surface in the equatorial plane (resp. along the rotation axis).}
\end{figure}

\begin{figure}
\centerline{\includegraphics[height=0.35\textheight]{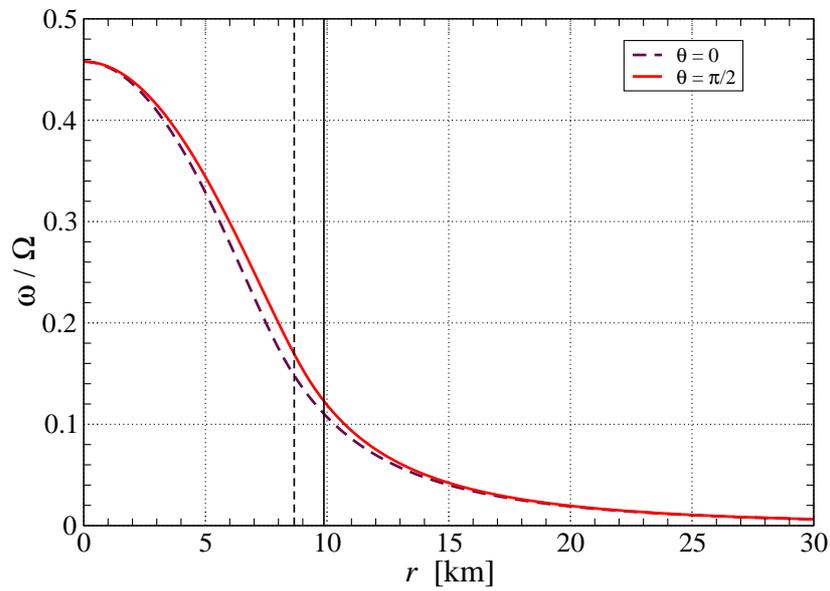}}
\caption{\label{f:sym:prof_omega} \small
Same as Fig.~\ref{f:sym:prof_n} but for the shift vector component $\omega=-\beta^\ph$.}
\end{figure}

\begin{figure}
\centerline{\includegraphics[height=0.35\textheight]{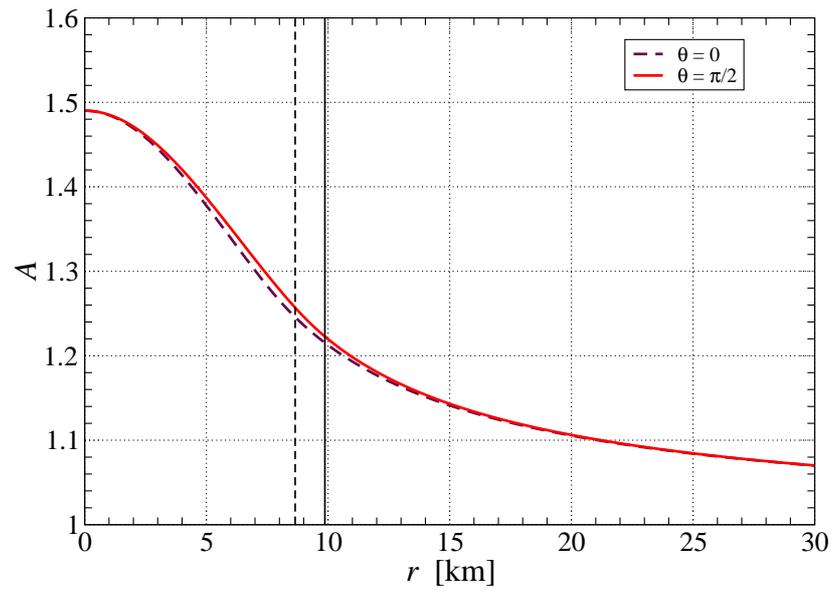}}
\caption{\label{f:sym:prof_a} \small
Same as Fig.~\ref{f:sym:prof_n} but for the metric coefficient $A$.}
\end{figure}

\begin{figure}
\centerline{\includegraphics[height=0.35\textheight]{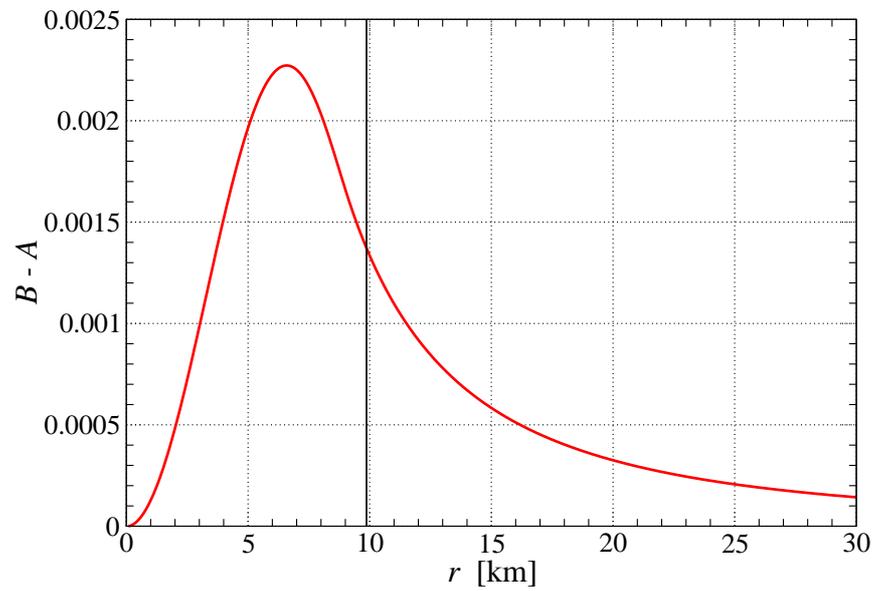}}
\caption{\label{f:sym:prof_bma} \small
Difference between the metric coefficients $B$ and $A$ in the equatorial plane.
The vertical solid line marks the location of
the stellar surface in that plane.}
\end{figure}

A global (coordinate-based) view of the star is provided in Fig.~\ref{f:sym:coupe_ener} and
some of its global properties are listed in Table~\ref{t:sym:APR_716Hz}. All these quantities are properly defined in Chap.~\ref{s:glo}.
The metric functions $N$, $\omega$ and $A$ are displayed in Figs.~\ref{f:sym:prof_n}, \ref{f:sym:prof_omega} and \ref{f:sym:prof_a} respectively.
It appears that the deviations from the flat space values ($N=1$, $\omega=0$
and $A=1$) are maximal at the stellar center. One may also check on these figures the tendency to
fulfill the boundary conditions (\ref{e:sym:BC}) when $r\rightarrow +\infty$. 
Moreover, the faster decay of $\omega$ [in $1/r^3$, cf. Eq.~(\ref{e:eer:omega_asympt})]
with respect to that of $A$ [in $1/r$, cf. (\ref{e:eer:A_sym_spher})]
is clearly apparent.
\begin{remark}
Figures~\ref{f:sym:prof_n} to \ref{f:sym:prof_a} have been truncated at
$r=30 {\rm\; km}$ for graphical convenience, but the computational domain of the
\texttt{Lorene/nrotstar} code extends to $r=+\infty$.
\end{remark}
The metric function $B$ turns out to be very close to $A$: on the rotation axis, it is exactly equal
to $A$ [cf. Eq.~(\ref{e:sym:A_B_axis})] and, in the equatorial plane, the difference $B-A$ is at most $2.3\times 10^{-3}$, as shown in Fig.~\ref{f:sym:prof_bma}.

\begin{figure}
\centerline{\includegraphics[height=0.35\textheight]{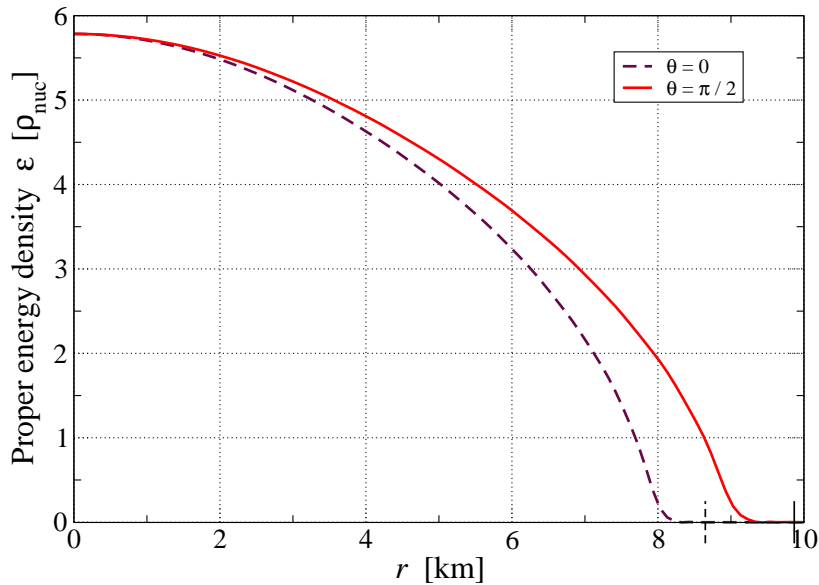}}
\caption{\label{f:sym:prof_ener} \small
Profile of the fluid proper energy density $\vep$ in two different directions: $\theta=0$ (rotation axis, dashed line)
and $\theta =\pi/2$ (equatorial plane, solid line). The small vertical solid line (resp. dashed line) marks the location of
the stellar surface in the equatorial plane (resp. along the rotation axis).
$\rho_{\rm nuc} := 1.66\times 10^{17} \; {\rm kg\, m}^{-3}$.}
\end{figure}

\begin{figure}
\centerline{\includegraphics[height=0.35\textheight]{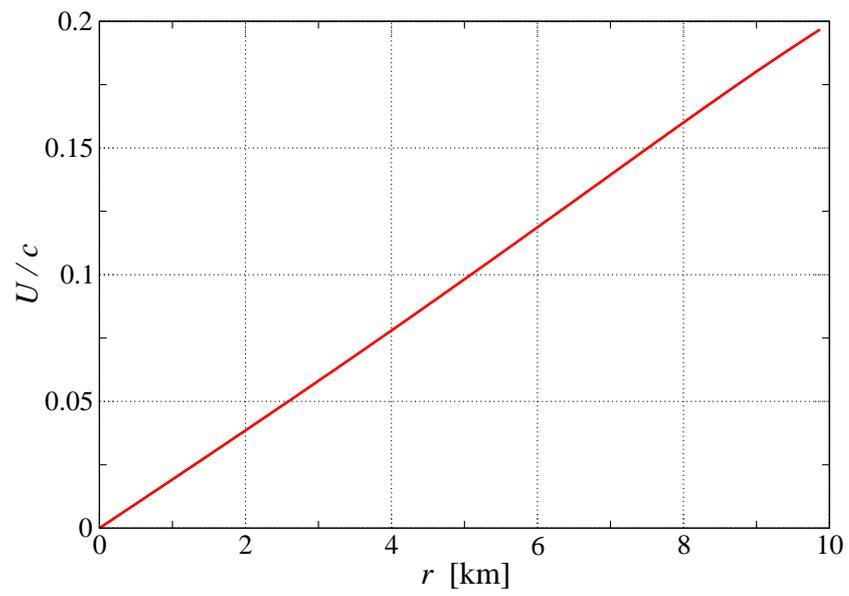}}
\caption{\label{f:sym:prof_u} \small
Profile of the norm $U$ of the fluid velocity with respect to the ZAMO, in the equatorial plane.}
\end{figure}

Two fluid quantities are displayed in Figs.~\ref{f:sym:prof_ener} and \ref{f:sym:prof_u}: the
proper energy density $\vep$ and the fluid velocity $U$ with respect to the ZAMO 
[cf. Eqs.~(\ref{e:sym:U_Omeg_omeg}) and (\ref{e:sym:U2})].
In the first figure, note the presence of the neutron star crust, of width 
$\sim 0.4 {\rm\; km}$ for $\theta=0$ and 
$\sim 0.6 {\rm\; km}$ for $\theta=\pi/2$. The profile of $U$ shown in Fig.~\ref{f:sym:prof_u}
is deceptively simple: it looks very close to the Newtonian profile $\Omega r$
[Eq.~(\ref{e:sym:U_Newt}) with $\theta=\pi/2$],
with one difference however: the mean slope is not $\Omega$ but $\sim 1.3\, \Omega$. 
This slope is approximately constant due to some numerical compensation of the two factors $B/N$ and 
$(1-\omega/\Omega)$ that appear in Eq.~(\ref{e:sym:U_Omeg_omeg}) once rewritten as
$U = (B/N) (1-\omega/\Omega) \, \Omega r$. Indeed $B/N$ is a decreasing function of $r$, taking the values
$2.36$ at $r=0$ and $1.52$ at $r=r_{\rm eq}$. On the contrary, $1-\omega/\Omega$ is an increasing 
function, taking the values $0.54$ at $r=0$ and $0.87$ at $r=r_{\rm eq}$. This results in 
$(B/N) (1-\omega/\Omega)$ taking almost the same values $r=0$ and $r=r_{\rm eq}$
($1.27$ and $1.32$ respectively), hence the straight behavior of the plot of $U(r)$.  
Notice that at the stellar surface, $U$ is a non-negligible fraction of the velocity of light:
$U \simeq 0.2 \, c$.

%
%

\chapter{Global properties of rotating stars} \label{s:glo}


\minitoc
\vspace{1cm}


\section{Total baryon number}

The baryon 4-current is 
\be
  \vv{\jmath}_{\rm b} = n_{\rm b} \vv{u} , 
\ee
where $n_{\rm b}$ is the baryon number density in the fluid frame and $\vv{u}$ is the fluid 4-velocity. For a stellar model computed along the method sketched in \S~\ref{s:eer:SCF}, $n_{\rm b}$ is obtained from the log-enthalpy $H$ and the EOS (\ref{e:sym:EOS_H}) via 
the formula
\be
	n_{\rm b} = \frac{\vep+p}{m_{\rm b}} \exp(-H), 
\ee
which is easily deduced from Eqs.~(\ref{e:eer:def_H}) and (\ref{e:sym:def_h}). 
The baryon 4-current obeys the conservation equation (\ref{e:eer:cons_baryon}):
\be
  \nabla_\mu j_{\rm b}^\mu = 0 . 
\ee
Thanks to this property, it can be shown, via the Gauss-Ostrogradski theorem, that the integral
\be \label{e:glo:def_nbar}
  \encadre{ \mathcal{N} := - \int_{\Sigma_t} \vv{\jmath}_{\rm b} \cdot \vv{n} \sqrt{\gamma} 
  \, dx^1 \, dx^2 \, dx^3 }  
\ee
is independent of the hypersurface $\Sigma_t$, i.e. of the slicing $(\Sigma_t)_{t\in\R}$
of spacetime. It gives the \defin{total number of baryons} in the star. 
Note that, by definition
of the baryon 4-current, the integrand, $n^*_{\rm b} = - \vv{\jmath}_{\rm b} \cdot \vv{n}$, is nothing but the baryon number density as measured by the ZAMO, 
whose 4-velocity is $\vv{n}$. 
Thanks to the relation $\Gamma = - \vv{n}\cdot\vv{u}$ [Eq.~(\ref{e:eer:Gam_nu})], 
we have $n^*_{\rm b} = \Gamma n_{\rm b}$. Choosing $(x^1,x^2,x^3)=(r,\theta,\ph)$, we have
in addition
$\sqrt{\gamma} = A^2 B r^2 \sin\theta$ [Eq.~(\ref{e:eer:sqrt_gam})], so that 
expression (\ref{e:glo:def_nbar}) can be explicited as
\be
  \encadre{\mathcal{N} =  \int_{\Sigma_t} \Gamma n_{\rm b} A^2 B r^2 \sin\theta \, dr\, d\theta\, d\ph } . 
\ee
We may convert $\mathcal{N}$ to a mass, by multiplying it by the mean baryon mass
$m_{\rm b} \simeq 1.66\times 10^{-27} {\; \rm kg}$~:
\be \label{e:glo:def_baryon_mass}
  \encadre{ M_{\rm b} := m_{\rm b} \mathcal{N} } . 
\ee
$M_{\rm b}$ is called the \defin{baryon mass} of the star.

\section{Mass}

Due to the mass-energy equivalence, the mass of the star must involve all forms of energy, including the gravitational field energy. Now the latter is not localizable (as a consequence of the Equivalence Principle): there is no ``energy density'' of the gravitational field.  
This implies that the mass of the star cannot be defined as a integral 
over $\Sigma_t$ of some ``mass density'' as in (\ref{e:glo:def_nbar}), which gives the $\mathcal{N}$  as the integral of the baryon density $n^*_{\rm b}=- \vv{\jmath}_{\rm b} \cdot \vv{n}$ over $\Sigma_t$. 
 
The are two major concepts of mass in general relativity (see Refs.~\cite{Gourg07a}, \cite{Szaba09}, \cite{JaramG10} for more details) : the \emph{ADM mass} which applies to any asymptotically flat spacetime and the \emph{Komar mass} which holds only for stationary spacetimes. 
In the present case, both coincide. We shall examine first the Komar mass definition. 

\subsection{Komar mass} \label{s:glo:Komar_mass}

Given the Killing vector $\vxi$ associated with stationarity, the \defin{Komar mass}
is 
\be \label{e:glo:def_Komar_M}
  \encadre{ M := - \frac{1}{8\pi} \oint_\Sp \nabla^\mu \xi^\mu \, dS_{\mu\nu} } ,  
\ee
where $\Sp$ is any closed 2-surface (sphere) surrounding the star and $dS_{\mu\nu}$ is the area-element 2-form normal to $\Sp$. In the case where $\Sp$ is entirely contained in the hypersurface $\Sigma_t$ and is
defined by $r=\mathrm{const}$, it is spanned by the coordinates $(\theta,\ph)$ and
\be \label{e:glo:dS}
  dS_{\mu\nu} = (s_\mu n_\nu - s_\nu n_\mu) \sqrt{q} \; d\theta \, d\ph ,
\ee
where the $n_\mu$'s are given by (\ref{e:ein:comp_n_cov}), the $s_\mu$'s are the covariant components of the unit normal $\vv{s}$ of $\Sp$ in $(\Sigma_t,\w{\gamma})$ and $q$ is the determinant of the components $q_{ab}$ of the metric 
induced by $\w{\gamma}$ (or equivalently $\w{g}$) on $\Sp$. From the diagonal form
(\ref{e:sym:gam_dd}) of $\w{\gamma}$, we have $\vv{s} = s^r \vpar_r$ and $s^r$ is determined by 
the normalization condition $\vv{s}\cdot\vv{s}=1$. Hence 
\be \label{e:glo:comp_s}
	s^\alpha = (0, A^{-1}, 0, 0) \qquad\mbox{and}\qquad
	s_\alpha = (0, A, 0, 0) . 
\ee
Since $\Sp$ is defined by $\{t=\mathrm{const},\; r=\mathrm{const}\}$, we read on (\ref{e:sym:3metQI})
the following components of the metric $\w{q}$ on $\Sp$:
\be
	q_{ab} \, dx^a \, dx^b = r^2 \left( A^2 d\theta^2 + B^2 \sin^2 \theta \, d\ph^2 \right) . 
\ee
Hence
\be \label{e:glo:sqrt_q}
	\sqrt{q} = AB r^2 \sin\theta . 
\ee

\begin{remark}
The Komar mass as defined by (\ref{e:glo:def_Komar_M}) is the \emph{flux integral} through $\Sp$
of the type $(2,0)$ tensor $\vv{\nabla} \vxi$. 
In dimension 3, there is locally only one direction $\vv{s}$ normal to a given sphere, so that 
instead of the normal 2-form (\ref{e:glo:dS}), one considers the normal 1-form 
$dS_i = s_i \sqrt{q} \; d\theta \, d\ph$;  flux integrals are then defined for vectors, i.e. 
type $(1,0)$ tensors. 
In the 4-dimensional spacetime $(\M,\w{g})$, there exists a whole plane of directions normal to the sphere and the vectors $(\vv{n},\vv{s})$ which appear in (\ref{e:glo:dS}) form a basis of that plane. 
\end{remark}

A priori the quantity $M$ defined by (\ref{e:glo:def_Komar_M}) should depend on the choice of the 
surface $\Sp$. However, owing to the fact that $\vxi$ is a Killing vector, this is not the case, as long as $\Sp$ is located outside the star. This was shown by Komar in 1959 \cite{Komar59}. To prove it, one may use 
the Gauss-Ostrogradski theorem, the Killing equation (\ref{e:sym:Killing})
and the Einstein equation (\ref{e:ein:ee}) to rewrite 
the integral (\ref{e:glo:def_Komar_M}) as\footnote{See e.g. \S~7.6.1 of \cite{Gourg07a} for the detailed demonstration.}
\be \label{e:glo:Komar_M_int}
	M = 2 \int_{\Sigma_t} \left[ \w{T}(\vv{n},\vxi) - \frac{1}{2} T \, \vv{n}\cdot\vxi \right]
	\sqrt{\gamma} \, dr\, d\theta\, d\ph , 
\ee 
where $\w{T}$ is the energy-momentum tensor and $T := g^{\mu\nu} T_{\mu\nu}$ its trace with
respect to $\w{g}$. 
From (\ref{e:glo:Komar_M_int}), it is clear that outside the support of $\w{T}$, i.e. outside
the star, there is no contribution to $M$, hence the independency of $M$ upon the choice of $\Sp$. 

Expressing (\ref{e:glo:def_Komar_M}) in terms of 3+1 quantities, we get (cf. \S~7.6.2 of \cite{Gourg07a} for details), 
\be
	M = \frac{1}{4\pi} \oint_{\Sp} \left( s^i D_i N - K_{ij} s^i \beta^j \right) \sqrt{q} 
		\, d\theta \, d\ph . 
\ee
In the present case, substituting (\ref{e:glo:comp_s}) for $s^i$, 
(\ref{e:sym:Krph})-(\ref{e:sym:Kthph}) for $K_{ij}$, (\ref{e:sym:beta_omega}) for $\beta^j$
and (\ref{e:glo:sqrt_q}) for $\sqrt{q}$, we have
\be
	M = \frac{1}{4\pi} \oint_{\Sp} \left(\der{N}{r}
	 - \frac{B^2 r^2\sin^2\theta}{2N} \, \omega \der{\omega}{r} \right)
	B r^2 \sin\theta \, d\theta \, d\ph . 
\ee
If we let the radius of the sphere extend to $+\infty$, then the term containing $\omega$ does no longer
contribute to the integral, since (\ref{e:eer:omega_asympt}) implies
\[
	r^2 \omega \der{\omega}{r} = O\left(\frac{1}{r^5}\right) . 
\]
Taking into account $B\rightarrow 1$ when $r\rightarrow +\infty$, we may then write
\be
	\encadre{ M = \frac{1}{4\pi} \lim_{\Sp\rightarrow\infty} \oint_{\Sp} \der{N}{r} r^2 \sin\theta \, d\theta \, d\ph },  
\ee
where the limit means that the integral is to be taken on a sphere $\Sp$
of radius $r\rightarrow +\infty$. 
Plugging the asymptotic behavior (\ref{e:eer:N_asympt}) of $N$ into the above formula yields
$K_0 = -M$, hence 
\be \label{e:glo:N_asympt}
	\encadre{ N = 1 - \frac{M}{r} + O\left(\frac{1}{r^2} \right) } . 
\ee

Let us now turn to the volume expression of $M$ provided by formula (\ref{e:glo:Komar_M_int}). Introducing
in it the 3+1 decomposition (\ref{e:ein:T_3p1}) of the energy-momentum tensor, and making use of
$\vxi = N \vv{n} + \vv{\beta}$ [Eq.~(\ref{e:ein:par_t_N_beta})] with $\vpar_t = \vxi$, we get 
\be
	M = \int_{\Sigma_t} \left[ N(E+S) - 2 p_i \beta^i \right]
	\sqrt{\gamma} \, dr\, d\theta\, d\ph . 
\ee
In the present case, taking into account expression (\ref{e:sym:p_ph_U}) for $p_i$, 
(\ref{e:sym:beta_omega}) for $\beta^i$ and (\ref{e:eer:sqrt_gam}) for $\sqrt{\gamma}$, we obtain
\be
	\encadre{ M = \int_{\Sigma_t} \left[ N(E+S) + 2 \omega B (E+p) U r\sin\theta \right]
	A^2 B r^2 \sin\theta \, dr\, d\theta\, d\ph } . 
\ee
At the Newtonian limit, only the term in $E\simeq \rho$ remains in this expression:
\be
	\encadre{ M = \int_{\Sigma_t} \rho r^2 \sin\theta \, dr\, d\theta\, d\ph 
	} _{\mbox{\footnotesize\  Newt.}} . 
\ee
Of course, we recognize the expression of the mass as the integral of the mass density 
$\rho$. 

\subsection{ADM mass}

The \defin{ADM mass} has been introduced by Arnowitt, Deser \& Misner in 1962 \cite{ArnowDM62} for 
any asymptotically flat spacelike hypersurface $\Sigma_t$. 
It is expressible as \cite{Gourg07a,JaramG10}
\be \label{e:glo:M_ADM}
	\encadre{ M_{\rm ADM} = \frac{1}{16\pi}
	\lim_{\Sp\rightarrow\infty}
	\oint_{\Sp} \left[ \Df^j \gamma_{ij} - \Df_i (f^{kl} \gamma_{kl}) \right]
	s^i  \sqrt{q}\,  d\theta\, d\ph }, 
\ee
where (i) $\w{f}$ is a fiducial flat 3-metric on $\Sigma_t$:
$f_{ij} = \mathrm{diag}(1,r^2,r^2\sin^2\theta)$ and $f^{ij} = 
\mathrm{diag}(1,r^{-2},r^{-2}\sin^{-2}\theta)$ and (ii) 
$\w{\Df}$ stands for the connection associated with $\w{f}$.
\begin{remark}
Contrary to the Komar mass definition (\ref{e:glo:def_Komar_M}), the surface integral
in (\ref{e:glo:M_ADM}) has to be taken at infinity. 
Note also that the ADM mass does not require stationarity to be defined.
In non-stationary spacetimes, the Komar mass can still be defined by an integral 
of the form (\ref{e:glo:def_Komar_M}) where $\vxi$ is an asymptotic Killing vector, but then the surface $\Sp$ has to be located at infinity. 
\end{remark}

We leave as an exercise to the reader to show that with the form
(\ref{e:sym:gam_dd}) of $\gamma_{ij}$, the formula (\ref{e:glo:M_ADM}) can be explicited as
\be
  \encadre{ M_{\rm ADM} = -\frac{1}{16\pi} \lim_{\Sp\rightarrow\infty}
	\oint_{\Sp} \left[ \der{}{r} (A^2 + B^2) + \frac{B^2-A^2}{r} \right]
  r^2 \sin\theta \,  d\theta\, d\ph } .
\ee

Although the two definitions look very different,  Beig (1978) \cite{Beig78} and 
Ashtekar \& Magnon-Ashtekar (1979) \cite{AshteM79} have shown that for a stationary spacetime, 
the ADM mass is always equal to the Komar mass, provided that $\Sigma_t$  
is orthogonal to the Killing vector $\vxi$ at spatial
infinity, which is our case here:
\be
  \encadre{M_{\rm ADM} = M} . 
\ee

Therefore, for a stationary rotating star, the Komar mass and ADM mass coincide. This common value,
denoted here by $M$, is sometimes called \defin{gravitational mass}, to distinguish it from
the \emph{baryon mass} defined by (\ref{e:glo:def_baryon_mass}).

\subsection{Binding energy}

The \defin{binding energy} of the star is defined by as the difference between the 
gravitational mass and baryon mass:
\be
  \encadre{ E_{\rm bind} := M - M_{\rm b} }.
\ee
Physically, $|E_{\rm bind}|$ is the energy one should spend to disperse all the particles (baryons) constituting the star to infinity. 
The star is bound iff $E_{\rm bind} < 0$. If $E_{\rm bind} \geq 0$, the star would be
unstable and would explode, releasing the energy $E_{\rm bind}$.

\section{Virial identities}

\subsection{GRV3}

Taking the identity of ADM and Komar mass as a starting point, Gourgoulhon \& Bonazzola (1994) \cite{GourgB94} have derived a formula that generalizes the Newtonian virial theorem 
to general relativity for stationary spacetimes. It is called \defin{GRV3}, for \emph{General Relativistic Virial theorem}, the ``3'' recalling that the integral is a 3-dimensional one. 
For a rotating star, it reads [Eq.~(43) of \cite{GourgB94}]
\bea
  \lefteqn{ \int_{\Sigma_t} \bigg\{ 4\pi S 
  - \frac{1}{A^2} \left( \partial\nu\partial\nu - \frac{1}{2 AB} \partial A \partial B \right) 
  + \frac{1}{2r}\left( \frac{1}{A^2} - \frac{1}{B^2} \right) \bigg[ \frac{1}{A} 
  \left( \der{A}{r} + \frac{1}{r\tan\theta} \der{A}{\theta} \right) } \nonumber \\
  & &  -\frac{1}{2B} 
  \left( \der{B}{r} + \frac{1}{r\tan\theta} \der{B}{\theta} \right) \bigg]
  + \frac{3 B^2 \sin^2\theta}{8 r^2 A^2 N^2} \partial\omega\partial\omega
  \bigg\} A^2 B r^2 \sin\theta \, dr\,d\theta\,d\ph  = 0 , \label{e:glo:GRV3}
\eea
with $S$ given by (\ref{e:sym:S_fluid_parfait}) : 
\[
  S = 3p + (E+p) U^2 .
\]
At the Newtonian limit, $S \simeq 3p + \rho U^2$, $\nu \simeq \Phi / c^2$, $\ln A \simeq -\Phi/c^2$, 
$\ln B \simeq -\Phi/c^2$, $A\simeq 1$, $B\simeq 1$ and $\omega=0$, so that,
after division by $4\pi$, 
(\ref{e:glo:GRV3}) becomes
\be \label{e:glo:GRV3_Newt}
  \encadre{ 2 E_{\rm kin} + 3 P + E_{\rm grav} = 0 } _{\mbox{\footnotesize\  Newt.}}, 
\ee
where 
\bea
  & & E_{\rm kin} = \frac{1}{2} \int_{\Sigma_t} \rho U^2 \, r^2 \sin\theta \, dr\,d\theta\,d\ph \\
  & & P = \int_{\Sigma_t} p \, r^2 \sin\theta \, dr\,d\theta\,d\ph  \\
  & & E_{\rm grav} = - \frac{1}{8\pi G} \int_{\Sigma_t}
    (\vv{\nabla} \Phi)^2 \, r^2 \sin\theta \, dr\,d\theta\,d\ph .
\eea
$E_{\rm kin}$ and $E_{\rm grav}$ are respectively the kinetic energy and the gravitational potential energy of the star, and we recognize in (\ref{e:glo:GRV3_Newt}) the classical virial theorem for stationary configurations (see e.g. Eq.~(2.180) of Ref.~\cite{Tasso00}). 

The GRV3 identity proved to be very useful to check numerical solutions \cite{NozawSGE98,Sterg03}.

\subsection{GRV2}

Another virial identity has been obtained by Bonazzola (1973) \cite{Bonaz73} (see also
Ref.~\cite{BonazGSM93,BonazG94a}). Following \cite{BonazG94a}, it is called GRV2, because contrary
to GRV3, it involves a 2-dimensional integral. 
The derivation of GRV2 is quite straightforward once we consider 
the last equation of the Einstein system 
(\ref{e:sym:eq_N})-(\ref{e:sym:eq_lnApnu}). The operator on the left-hand side of 
(\ref{e:sym:eq_lnApnu})
is the 2-dimensional flat Laplacian $\Delta_2$. It operates in the plane $\R^2$
spanned by the polar coordinates $(r,\theta)$ and its Green
function is
\be
  G(\vec{x},\vec{x}') = \frac{1}{2\pi} \ln \| \vec{x}- \vec{x}' \| ,
\ee
where $\vv{x}$ denotes the vector of $\R^2$ of coordinates $(x^a)=(r,\theta)$
and $\| \vec{x}- \vec{x}' \|$ is the norm of $\vec{x}- \vec{x}'$ with respect to the
flat Euclidean metric of $\R^2$. 
Accordingly, the generic solution of (\ref{e:sym:eq_lnApnu}), assuming its right-hand
side is given, is 
\be \label{e:glo:lnA_nu_asympt}
   \ln A + \nu = \frac{I}{2\pi} \ln r + O\left( \frac{1}{r} \right) ,
\ee
where $I$ is the integral of the right-hand side of (\ref{e:sym:eq_lnApnu}),
\be \label{e:glo:def_sigma}
	\sigma := 8\pi A^2 S^\ph_{\ \, \ph} 
 + \frac{3 B^2 r^2\sin^2 \theta}{4 N^2} \, \partial \omega\partial \omega
	- \partial\nu  \partial \nu ,
\ee
over $\R^2$ with respect to the flat area element $r \, dr\, d\theta$:
\be \label{e:glo:I_int_M}
  I = \int_{r=0}^{r=+\infty} \int_{\theta=0}^{\theta=2\pi} \sigma  \, r \, dr\, d\theta . 
\ee
In the above integral, the coordinate $\theta$ spans the interval $[0,2\pi]$, whereas the 
QI coordinate $\theta$ runs in $[0,\pi]$ only. This is because the meridional surface
$\mathcal{M}_{t\ph}$ covers only half of $\R^2$. To define (\ref{e:glo:I_int_M}), we therefore use the analytical continuation of $\sigma$ from $[0,\pi]$ to $[\pi,2\pi]$. 
More precisely, since the regularity conditions imply that $\sigma$ can be
expanded into a series of $\cos(k\theta)$, the analytical continuation satisfies
\be \label{e:glo:sigma_ana_cont}
	\forall \theta\in [\pi,2\pi], \quad
	\sigma(r,\theta) = \sigma(r,2\pi-\theta) . 
\ee
Now, when $r\rightarrow+\infty$, the asymptotic behavior of $\nu$ is $O(1/r)$
[cf. Eq.~(\ref{e:eer:N_asympt})] and the same things holds for $\ln A$ [see (\ref{e:eer:A_sym_spher}) for the spherically symmetric case]. To have a regular
asymptotically flat solution, we must therefore have $I=0$ in (\ref{e:glo:lnA_nu_asympt}). In view of (\ref{e:glo:def_sigma}), (\ref{e:glo:sigma_ana_cont}) and expression (\ref{e:sym:S_fluid_parfait})
for $S^\ph_{\ \, \ph}$, this implies
\be \label{e:glo:GRV2}
  \encadre{ \int_{r=0}^{r=+\infty} \int_{\theta=0}^{\theta=\pi}
  \left\{ 8\pi A^2  [p + (E+p) U^2]
 + \frac{3 B^2 r^2\sin^2 \theta}{4 N^2} \, \partial \omega\partial \omega
	- \partial\nu  \partial \nu \right\} \, r \, dr\, d\theta = 0 } . 
\ee
This is the \defin{GRV2 identity}. 
Note that the integration domain in $\theta$ has been reduced to $[0,\pi]$ thanks
to the property (\ref{e:glo:sigma_ana_cont}), thanks to which $I$ is twice the integral 
in the left-hand side of (\ref{e:glo:GRV2}). 
As GRV3, it is very useful to check the accuracy of numerical solutions
\cite{NozawSGE98,Sterg03}.

The Newtonian limit of GRV2 is 
\be \label{e:glo:GRV2_Newt}
  \encadre{ \int_{r=0}^{r=+\infty} \int_{\theta=0}^{\theta=\pi}
  \left[ p + \rho U^2 - \frac{1}{8\pi G} (\vv{\nabla} \Phi)^2 \right]
    \, r \, dr\, d\theta = 0 } _{\mbox{\footnotesize\  Newt.}} .
\ee
The identity (\ref{e:glo:GRV2_Newt}) has been exhibited in 1939 by Chandrasekhar \cite{Chand39} in the special
case of spherical symmetry, but it seems that it has not been considered for axisymmetric Newtonian bodies before the work \cite{BonazG94a}. 

\begin{remark}
Although it also involves a 2-dimensional Laplacian, Eq.~(\ref{e:sym:eq_NB})
of the Einstein system does not give birth to a virial identity because of the 
$\sin\theta$ factor on its right-hand side, which leads to 
$\sigma(r,\theta) = -\sigma(r,2\pi-\theta)$ instead of (\ref{e:glo:sigma_ana_cont}), implying
that $I$ is automatically zero. 
\end{remark}

\section{Angular momentum}

The \defin{Komar angular momentum} (or \defin{angular momentum} for brevety)
is defined by a integral similar to that
used to define the Komar mass in \S~\ref{s:glo:Komar_mass} but replacing the
stationarity Killing vector $\vxi$ by the axisymmetry Killing vector $\vchi$: 
\be
\encadre{ J :=  \frac{1}{16\pi} \oint_\Sp \nabla^\mu \chi^\mu \, dS_{\mu\nu} } ,  
\ee
where $\Sp$ is any closed 2-surface (sphere) surrounding the star and $dS_{\mu\nu}$ is the area-element 2-form normal to $\Sp$ given by (\ref{e:glo:dS}).
As for the Komar mass, $J$ is independent of the choice of $\Sp$ provided the star is entirely enclosed in $\Sp$.

In terms of the 3+1 quantities  $J$ is expressible as (cf. \cite{Gourg07a} for details)
\be \label{e:glo:J_surf}
  J = \frac{1}{8\pi} \oint_\Sp K_{ij} s^i \chi^j \sqrt{q} \, d\theta\, d\ph . 
\ee
By means of the Gauss-Ostrogradski theorem and Einstein equation, we can get a volume expression of $J$ similar to (\ref{e:glo:Komar_M_int}), differing only by the 
replacement of $\vxi$ by $\vchi$ and the overall factor $2$ by $-1$. 
Since $\vchi$ is tangent to $\Sigma_t$, we have $\w{T}(\vv{n},\vchi)=-\vv{p}\cdot\vchi$
and $\vv{n}\cdot\vchi=0$, so that the final expression reduces to 
\be
  J = \int_{\Sigma_t} p_i \chi^i \sqrt{\gamma} \, dr\, d\theta\, d\ph . 
\ee
For a perfect fluid rotating star, $p_i \chi^i = p_\ph$ and we get, 
by combining Eqs.~(\ref{e:sym:p_ph_U}) and  (\ref{e:eer:sqrt_gam}),
\be
  \encadre{ J = \int_{\Sigma_t} (E+p) U A^2 B^2 r^3 \sin^2\theta \, dr\, d\theta\, d\ph } . 
\ee
At the Newtonian limit, this expression reduces to 
\be
  \encadre{ J = \int_{\Sigma_t} \rho U r\sin \theta \times r^2 \sin\theta  \, dr\, d\theta\, d\ph  } _{\mbox{\footnotesize\  Newt.}} .
\ee
We recognize the $z$-component of the star total angular momentum. 

Let us make explicit the surface integral expression of $J$, Eq.~(\ref{e:glo:J_surf}), in the
present case. We have from Eq.~(\ref{e:sym:Krph})-(\ref{e:sym:Kthph}) and 
(\ref{e:glo:comp_s}),
\[
	K_{ij} s^i \chi^j = K_{r\ph} s^r = - \frac{B^2 r^2\sin^2\theta}{2AN} \der{\omega}{r}
      \sim - r^2 \sin\theta \der{\omega}{r} \quad \mbox{when}\quad r\rightarrow +\infty. 
\]
Therefore taking the limit $\Sp\rightarrow\infty$ in Eq.~(\ref{e:glo:J_surf}) gives
\[
	J = - \frac{1}{16\pi} \lim_{\Sp\rightarrow\infty}
	\oint_{\Sp} \der{\omega}{r} r^4 \sin^3 \theta \, d\theta \, d\ph . 
\]
Now, from the asymptotic behavior of $\omega$ as given by (\ref{e:eer:omega_asympt}), 
$\dert{\omega}{r} = - 3 K_1 / r^4 + O(1/r^5)$. Hence
\[
	J = \frac{3 K_1}{8} \int_0^\pi \sin^3 \theta \, d\theta 
		= \frac{3 K_1}{8} \int_0^\pi (1-\cos^2\theta)\sin \theta \, d\theta 
		= \frac{K_1}{2} . 
\] 
We may therefore rewrite the asymptotic behavior of $\omega$ as 
\be \label{e:glo:omega_asympt}
	\encadre{ \omega = \frac{2J}{r^3} + O\left(\frac{1}{r^4} \right) } . 
\ee
In particular, we recover the fact that if the star is not rotating ($J=0$), then $\omega = 0$
[cf. (\ref{e:sym:omega_zero})].

Having defined both $J$ and $M$, we can form the dimensionless ratio 
\be
  \encadre{ j := \frac{J}{M^2} } .
\ee
$j$ is called the \defin{Kerr parameter}. For a Kerr black hole of parameters
$(a,M)$, $j=a/M$ and varies between 
$0$ (Schwarzschild black hole) and $1$ (extreme Kerr black hole). 

Next, for a rigidly rotating star, one defines the \defin{moment of inertia} by 
\be
	\encadre{ I := \frac{J}{\Omega} } . 
\ee 

\begin{remark}
Due to the so-called \emph{supertranslation ambiguity}, there is no equivalent of the ADM mass for the angular momentum (see e.g. \cite{JaramG10} for a discussion). 
\end{remark}

\section{Circumferential radius}

The \defin{stellar equator} is the closed line at the surface of the star
[$p=0$, cf. (\ref{e:eer:def_surf})] defined by $t=\mathrm{const}$ and
$\theta = \pi/2$ (equatorial plane). 
It has a constant value of the coordinate $r$, that we have already denoted
by $r_{\rm eq}$. A coordinate-independent characterization of the stellar equator is
the \defin{circumferential radius}: 
\be
  \encadre{ R_{\rm circ} := \frac{\mathscr{C}}{2\pi} }, 
\ee
where $\mathscr{C}$ is the circumference of the star in the equatorial plane, 
i.e. the length of the equator as given by the metric tensor. 
According to the line element (\ref{e:sym:3metQI}), we have
\[
  R_{\rm circ} = \frac{1}{2\pi} \oint_{{r=r_{\rm eq}\atop \theta=\pi/2}}
    ds = \frac{1}{2\pi} \oint_{{r=r_{\rm eq}\atop \theta=\pi/2}}
	\sqrt{B^2 r^2 \sin^2\theta \, d\ph^2}
    = \frac{B(r_{\rm eq},\pi/2) \, r_{\rm eq}}{2\pi}  \int_0^{2\pi} d\ph , 
\]
hence
\be
  \encadre{ R_{\rm circ} = B(r_{\rm eq},\pi/2) \, r_{\rm eq} } .
\ee
Of course, at the Newtonian limit, $B=1$ and $R_{\rm circ} = r_{\rm eq}$.
For a relativistic star, we have $B>1$ (cf.  Fig.~\ref{f:sym:prof_a}, 
taking into account that $B$ is very close to $A$, as shown in
Fig.~\ref{f:sym:prof_bma}), so that $R_{\rm circ} > r_{\rm eq}$. 

\section{Redshifts}

\subsection{General expression of the redshift} \label{s:glo:redshift_gal}

Consider a photon emitted by the fluid at the surface of the star. 
The photon follows a null geodesic $\Li$ of spacetime and is described by its 
4-momentum\footnote{We are using the same letter $p$ as for the matter momentum density with respect to the ZAMO inside the star
[cf. Eq.~(\ref{e:ein:pa_T})], but no confusion should arise.}
$\vv{p}$, which is a future directed null vector:
\be \label{e:glo:p_null}
  \vv{p}\cdot\vv{p} = 0 .
\ee
$\vv{p}$ is tangent to $\Li$ in each point and obeys the geodesic equation
\be \label{e:glo:p_geod}
  \wnab_{\vv{p}}\,  \vv{p} = 0 . 
\ee

For an observer who is receiving the photon, the relative shift in wavelength, called 
\defin{redshift}, is 
\be
  \encadre{ z:= \frac{\lambda_{\rm rec} - \lambda_{\rm em}}{\lambda_{\rm em}} } , 
\ee
where $\lambda_{\rm rec}$ is the photon's wavelength as measured by the observer
(the ``receiver'') and $\lambda_{\rm em}$ is the photon's wavelength in the fluid
frame (the ``emitter''). 
In terms of the photon energy $E=\hbar c / \lambda$, the above formula can be
written
\be
  z = \frac{E_{\rm em}}{E_{\rm rec}} - 1 . 
\ee 
Now, by the very definition of the photon 4-momentum, the photon energy measured by an
observer $\mathcal{O}$ is
\be
  E_{\mathcal{O}} = - \vv{u}_{\mathcal{O}}\cdot \vv{p}  ,  
\ee
where $\vv{u}_{\mathcal{O}}$ is the 4-velocity of $\mathcal{O}$. 
In the present case, $\vv{u}_{\rm em} = \vv{u}$ (the fluid 4-velocity),
whereas, for an observer at infinity and at rest with respect to the star
$\vv{u}_{\rm rec} = \vxi$. Hence
\be \label{e:glo:redshift_gal}
  \encadre{ z = \frac{ \left. \vv{u}\cdot\vv{p} \right| _{\rm em} }{ \left. \vxi\cdot\vv{p} \right| _{\rm rec} } - 1 }, 
\ee 
where the indices `em' and `rec' indicate the location where the scalar products
$\vv{u}\cdot\vv{p}$ and $\vxi\cdot\vv{p}$ are to be taken.

A fundamental property for the scalar product $\vxi\cdot\vv{p}$ is to be conserved along the photon geodesic worldline. To show it, let us take the scalar product of
(\ref{e:glo:p_geod}) by $\vxi$; we have successively
\bea
  & & \xi_\nu p^\mu \nabla_\mu p^\nu = 0 ,\nonumber \\
  & & p^\mu [\nabla_\mu(\xi_\nu p^\nu) - p^\nu \nabla_\mu \xi_\nu] = 0, \nonumber \\
  & & p^\mu \nabla_\mu(\xi_\nu p^\nu) - \underbrace{p^\mu p^\nu \nabla_\mu \xi_\nu}_{0} = 0, \nonumber \\
  & & p^\mu \nabla_\mu(\xi_\nu p^\nu) = 0 , \label{e:glo:nab_xi_p_comp}
\eea
where we have used the Killing equation (\ref{e:sym:Killing}) to set to zero the
term $p^\mu p^\nu \nabla_\mu \xi_\nu$:
\[
  p^\mu p^\nu \nabla_\mu \xi_\nu = \frac{1}{2} \left( 
p^\mu p^\nu \nabla_\mu \xi_\nu + p^\nu p^\mu \nabla_\nu \xi_\mu \right)
  = \frac{1}{2} p^\mu p^\nu (\underbrace{\nabla_\mu \xi_\nu + \nabla_\nu \xi_\mu}_{0})
  = 0 . 
\]
Equation~(\ref{e:glo:nab_xi_p_comp}) can be rewritten as
\be
  \encadre{ \wnab_{\vv{p}} (\vxi \cdot \vv{p}) = 0 } . 
\ee
Since $\vv{p}$ is tangent to the photon worldline, this relation shows that
\be \label{e:glo:cons_xi_p}
  \encadre{ \vxi \cdot \vv{p} = \mbox{const along the photon worldline} } .
\ee
Hence the symmetry encoded by $\vxi$ (stationarity) gives birth to a conserved quantity. This is another manifestation of the Noether theorem (cf. \S~\ref{s:sym:Bernoulli}).
Similarly, we have
\be \label{e:glo:cons_chi_p}
  \encadre{ \vchi \cdot \vv{p} = \mbox{const along the photon worldline} } .
\ee

The conservation laws (\ref{e:glo:cons_xi_p}) and (\ref{e:glo:cons_chi_p}) 
are particularly useful when we specialize $\vv{u}$ in (\ref{e:glo:redshift_gal}) 
to the circular form
(\ref{e:sym:u_circular}):
$\vv{u} = (\Gamma/N) (\vxi + \Omega\vchi)$ (we have used (\ref{e:sym:Gam_U}) to replace
$u^t$ by $\Gamma/N$). We get then 
\[
  \vv{u}\cdot\vv{p} = \frac{\Gamma}{N} \left( \vxi\cdot \vv{p} + \Omega\,  \vchi\cdot\vv{p}\right) ,
\]
so that Eq.~(\ref{e:glo:redshift_gal}) becomes
\be
  z = \left. \frac{\Gamma}{N} \right| _{\rm em} \left( 
  \frac{ \left. \vxi\cdot\vv{p} \right| _{\rm em} }{ \left. \vxi\cdot\vv{p} \right| _{\rm rec} } +  \Omega \, \frac{ \left. \vchi\cdot\vv{p} \right| _{\rm em} }{ \left. \vxi\cdot\vv{p} \right| _{\rm rec} } \right) - 1 .
\ee
Thanks to (\ref{e:glo:cons_xi_p}), $\left. \vxi\cdot\vv{p} \right| _{\rm em} = 
 \left. \vxi\cdot\vv{p} \right| _{\rm rec}$, hence
\be \label{e:glo:redshift_Omega}
  \encadre{ z =  \left. \frac{\Gamma}{N} \right| _{\rm em} \left( 
  1 + \Omega \, \frac{\vchi\cdot\vv{p} }{ \vxi\cdot\vv{p} } \right) - 1} . 
\ee
Note that we have suppressed the indices `em' or `rec' on $\vchi\cdot\vv{p}$ and $\vxi\cdot\vv{p}$
thanks to (\ref{e:glo:cons_xi_p}) and (\ref{e:glo:cons_chi_p}).
\begin{remark}
For differentially rotating stars, $\Omega$ is not a constant and the value to be used
in Eq.~(\ref{e:glo:redshift_Omega}) is the value at the emission point.
\end{remark}

\subsection{Polar redshift}

Let us apply formula (\ref{e:glo:redshift_Omega}) to an emission taking place at one of the poles of the star,
i.e. at one of the two points of intersection of the stellar surface with the rotation axis:
$\theta = 0$ or $\theta=\pi$. 
On the rotation axis $\vchi = 0$ [Eq.~(\ref{e:sym:chi_zero_axis})].
Moreover, $\Gamma = 1$ since $U=0$ on the rotation axis [cf. Eq.~(\ref{e:sym:U_Omeg_omeg}) with $\sin\theta=0$]. Accordingly, formula (\ref{e:glo:redshift_Omega}) simplifies to
\be
  \encadre{ z_{\rm p} = \frac{1}{N}_{\rm p} - 1 } ,
\ee
where the index `p' stands for polar values. 

\subsection{Equatorial redshifts}

Let us now consider the emission from the equator and in a direction tangent to it, i.e. 
an emission from the left or right edge of the star as seen by the distant observer. 
The photon 4-momentum at the emission is then 
\be
  \vv{p} = p^t \vxi + p^\ph \vchi , 
\ee
with $p^\ph > 0$ (resp. $p^\ph < 0$) for an emission in the forward (resp. backward) direction with respect to the fluid motion.
The term in factor of $\Omega$ in formula (\ref{e:glo:redshift_Omega}) is then
\be \label{e:chip_xip_prov}
  \frac{\vchi\cdot\vv{p} }{ \vxi\cdot\vv{p} }
  = \frac{\vchi\cdot\vxi + \alpha \, \vchi\cdot\vchi}{\vxi\cdot\vxi + \alpha \vxi\cdot\vchi}
  =  \frac{g_{t\ph} + \alpha \, g_{\ph\ph}}{g_{tt} + \alpha g_{t\ph}} , 
  \qquad\mbox{with}\quad \alpha := \frac{p^\ph}{p^t} . 
\ee
Note that since $\vv{p}$ is a future directed null vector, we have $p^t > 0$; in particular
$p^t\not=0$, so that $\alpha$ is always well defined. 

Besides, Eq.~(\ref{e:glo:p_null}) writes 
$g_{tt} (p^t)^2 + 2 g_{t\ph} p^t p^\ph + g_{\ph\ph} (p^\ph)^2 = 0$,
i.e.
\be \label{e:glo_equat_alpha}
  g_{\ph\ph} \alpha^2 + 2 g_{t\ph} \alpha + g_{tt} = 0 . 
\ee
From this relation, we can write (\ref{e:chip_xip_prov}) as
\be \label{e:glo:chip_xip}
  \frac{\vchi\cdot\vv{p} }{ \vxi\cdot\vv{p} } = - \frac{1}{\alpha} . 
\ee
Finally, we compute $\alpha$ by solving the second-order equation (\ref{e:glo_equat_alpha}). 
We get two solutions:
\[
  \alpha_\pm = \frac{-g_{t\ph} \pm \sqrt{ g_{t\ph}^2 - g_{tt} g_{\ph\ph}}}{g_{\ph\ph}} 
    = \frac{ B r \omega \pm N}{ Br} , 
\]
where the second equality follows from the expression (\ref{e:sym:metric_down}) of the metric components, specialized at $\theta=\pi/2$ (equatorial plane).
Substituting the above value for $\alpha$ into Eq.~(\ref{e:glo:chip_xip}) 
leads to
\be \label{e:glo:chip_xip_NB}
    \frac{\vchi\cdot\vv{p} }{ \vxi\cdot\vv{p} } = \frac{Br}{\mp N - Br\omega} . 
\ee
Since $\vxi$ and $\vv{p}$ are both future directed, we have always $\vxi\cdot\vv{p} < 0$.
For a emission in the forward direction
$\vchi\cdot\vv{p} = B^2 r^2 (p^\ph -\omega p^t) \geq 0$, so the $-$ sign must selected in the
above expression. On the contrary, the $+$ sign corresponds to an emission in the backward 
direction. 
Inserting this result into Eq.~(\ref{e:glo:redshift_Omega}), we thus get, for an emission in the
forward direction
\[
  z_{\rm eq}^{\rm f} = \frac{\Gamma}{N} \left( 
  1 - \Omega \, \frac{Br}{N + Br\omega} \right) - 1 
  = \frac{\Gamma}{N+Br\omega} \bigg[ 1 - \underbrace{\frac{Br}{N}(\Omega-\omega)}_{U} \bigg]
    - 1 
  = \frac{1}{N+Br\omega} \frac{1-U}{\sqrt{1-U^2}} - 1 , 
\]
i.e. 
\be
  \encadre{ z_{\rm eq}^{\rm f} = \frac{1}{N+Br\omega} \sqrt{\frac{1-U}{1+U}} - 1 } .
\ee
In the right-hand side, all the functions are to be taken at the emission point. 
Similarly choosing the $+$ sign in (\ref{e:glo:chip_xip_NB}) leads to the redshift in the 
backward direction:
\be
  \encadre{ z_{\rm eq}^{\rm b} = \frac{1}{N-Br\omega} \sqrt{\frac{1+U}{1-U}} - 1 } .  
\ee
\begin{remark}
As a check of the above expressions, we may consider the flat spacetime limit:
$N=1$ and $\omega=0$. We then recover the standard Doppler formulas
\[
  z_{\rm eq}^{\rm f} = \sqrt{\frac{1-U}{1+U}} - 1 
  \qquad\mbox{and} \qquad
  z_{\rm eq}^{\rm b} = \sqrt{\frac{1+U}{1-U}} - 1 .
\]
They imply $z_{\rm eq}^{\rm f}\leq 0$, i.e. a blueshift (the emitter is moving towards the observer in the forward case)
and  $z_{\rm eq}^{\rm b}\geq 0$, i.e. a true redshift (the emitter is moving away from the observer in the backward case). 
\end{remark}

\section{Orbits around the star}

\subsection{General properties of orbits}

Let us consider a test particle $\Pp$ of mass $m\ll M$ orbiting the star. 
Let $\vv{v}$ be the 4-velocity of $\Pp$; its components with respect to QI
coordinates are 
\be \label{e:glo:v_comp}
	v^\alpha = \left( \frac{dt}{d\tau}, \frac{dr}{d\tau}, \frac{d\theta}{d\tau}, \frac{d\ph}{d\tau} \right) , 
\ee
where $\tau$ is the proper time of $\Pp$. The 4-momentum of the particle is
\be \label{e:glo:p_mv}
  \vv{p} = m \vv{v} . 
\ee
The 4-velocity normalization relation $\vv{v}\cdot\vv{v}=-1$ is equivalent to 
\be \label{e:glo:pp_m2}
	\vv{p}\cdot\vv{p} = - m^2 . 
\ee
Assuming that $\Pp$ is submitted only to the gravitation of the star, its worldline is a timelike geodesic of $(\M,\w{g})$. In particular $\wnab_{\vv{v}} \vv{v} = 0$ or 
equivalently 
\be \label{e:glo:p_part_geod}
  \wnab_{\vv{p}} \, \vv{p} = 0 .
\ee
Due to the symmetries of $(\M,\w{g})$, we have the same conserved quantities as those given
by Eqs.~(\ref{e:glo:cons_xi_p})-(\ref{e:glo:cons_chi_p}) in the photon case: 
\bea
   & & \encadre{ E := -\vxi \cdot \vv{p} = - p_t = \mbox{const} } \label{e:glo:part_E} \\
   & & \encadre{ L:= \vchi \cdot \vv{p} = p_\ph = \mbox{const} } . \label{e:glo:part_L}
\eea
Indeed the demonstration of (\ref{e:glo:cons_xi_p})-(\ref{e:glo:cons_chi_p}) 
[cf. Eq.~(\ref{e:glo:nab_xi_p_comp})] used only (i) the Killing character of 
$\vxi$ and $\vchi$ and (ii) the property (\ref{e:glo:p_part_geod}), which holds for a photon as well as for a massive particle. 
If the particle reaches spacelike infinity, where $\vxi \sim \vv{n}$ (the ZAMO 4-velocity), 
the constant $E$ is the particle energy as measured by the asymptotic ZAMO
Similarly, $L$ is the angular momentum of the particle with respect to the rotation axis as measured by the same observer: when $r\rightarrow +\infty$, 
$L \sim m r^2 \sin^2\theta \, d\ph/dt$ (assuming that asymptotically, the proper time of the
$\Pp$ coincides with $t$, i.e. that $\Pp$ is not relativistic). 

Everywhere is space, the energy of $\Pp$ as measured by the ZAMO is [cf. Eq.~(\ref{e:sym:xi_n_chi})]
\[
  E_{\rm ZAMO} = - \vv{n}\cdot\vv{p} = - \frac{1}{N} \left(\vxi + \omega\vchi\right)
  \cdot \vv{p} =  \frac{1}{N} \left( - \vxi \cdot \vv{p} - \omega \vchi \cdot \vv{p} \right) ,
\]
hence 
\be \label{e:glo:E_ZAMO}
  \encadre{ E_{\rm ZAMO} =  \frac{1}{N} (E - \omega L) } .  
\ee
\begin{remark}
Since $N\rightarrow 1$ and $\omega \rightarrow 0$ when $r\rightarrow+\infty$, it is clear on this formula that $E=E_{\rm ZAMO}$ if the particle reaches spatial infinity. 
\end{remark}

\subsection{Orbits in the equatorial plane}

We restrict ourselves to orbits in the equatorial plane: $\theta=\pi/2$. 
Then, according to Eq.~(\ref{e:glo:v_comp}) and (\ref{e:glo:p_mv}), 
$p^\theta = m d\theta/d\tau = 0$. Given the form (\ref{e:sym:metricQI}) of the metric
coefficients, this yields 
\[
	p_\theta = g_{\theta\mu} p^\mu = A^2 r^2 p^\theta = 0 . 
\]
Combining with (\ref{e:glo:part_E})-(\ref{e:glo:part_L}), we conclude that 
we have three constants among the four components $p_\alpha$: $p_t$, $p_\theta$ and $p_\ph$. 
The fourth component, $p_r$, is obtained from the normalization relation (\ref{e:glo:pp_m2}):
\[
  g^{\mu\nu} p_\mu p_\nu = -m^2 . 
\]
Substituting the values of $g^{\mu\nu}$ given by (\ref{e:sym:metric_up}), we get
\[
  - \frac{p_t^2}{N^2} - 2 \frac{\omega p_t p_\ph}{N^2}
  + \frac{p_r^2}{A^2} + \frac{p_\theta^2}{r^2 A^2} + 
  \left( \frac{1}{B^2r^2\sin^2\theta} - \frac{\omega^2}{N^2} \right) p_\ph^2 = -m^2 . 
\]
Since $\theta=\pi/2$, $p_t = -E$, $p_\theta=0$ and $p_\ph = L$, we get :
\be \label{e:glo:pr2}
  \frac{p_r^2}{A^2} = \frac{1}{N^2} (E-\omega L)^2 - m^2 - \frac{L^2}{B^2 r^2} . 
\ee
Now, from (\ref{e:sym:metric_down}), (\ref{e:glo:v_comp}) and (\ref{e:glo:p_mv}),
\[
  p_r = g_{r\mu} p^\mu = A^2 p^r = A^2 m v^r = A^2 m \frac{dr}{d\tau} . 
\]
Thus, we can rewrite (\ref{e:glo:pr2}) as
\be \label{e:glo:eff_pot}
  \encadre{ \frac{1}{2} \left( \frac{dr}{d\tau} \right) ^2 
	+ \mathcal{V}(r,\bar E, \bar L) = 0 }, 
\ee
where $\bar E$ and $\bar L$ are the constants of motion defined by 
\be
  \bar E := \frac{E}{m} \qquad \mbox{and} \qquad \bar L := \frac{L}{m} ,
\ee
and 
\be
  \encadre{ \mathcal{V}(r,\bar E, \bar L) := \frac{1}{2 A^2} 
  \left[ 1 -\frac{1}{N^2} \left( \bar E - \omega \bar L \right) ^2
	+ \frac{{\bar L}^2}{B^2 r^2} \right] } . 
\ee
In view of Eq.~(\ref{e:glo:eff_pot}), we conclude that the geodesic motion of a test particle in
the equatorial plane is equivalent to the Newtonian motion of a particle in a 
one-dimensional space (spanned by the coordinate $r$) under the effective potential ${\cal V}(r,\bar E, \bar L)$. 

\subsection{Circular orbits}

Circular orbits are defined by $r=\mathrm{const}$, i.e. by 
\bea
    & & \frac{dr}{d\tau} = 0 \label{e:glo:drdt_0} \\
    & & \frac{d^2 r}{d\tau^2} = 0 \label{e:glo:d2rdt2_0} .
\eea
\begin{remark}
Condition (\ref{e:glo:drdt_0}) is not sufficient to ensure a circular orbit, for it is satisfied at the periastron or apoastron of non-circular orbits. 
\end{remark}
From Eq.~(\ref{e:glo:eff_pot}), the two conditions (\ref{e:glo:drdt_0})-(\ref{e:glo:d2rdt2_0}) are equivalent to 
$\mathcal{V} = 0$ and $\dert{\mathcal{V}}{r}=0$, i.e. to 
\bea
  & & 1 -\frac{1}{N^2} \left( \bar E - \omega \bar L \right) ^2
	+ \frac{{\bar L}^2}{B^2 r^2} = 0 \label{e:glo:syst_circ1} \\
  & & \frac{\bar E-\omega \bar L}{N^2} 
    \left[ (\bar E-\omega\bar L)  \der{\nu}{r} + \bar L \der{\omega}{r} \right]
  - \frac{{\bar L}^2}{B^2 r^2} \left( \der{\beta}{r} + \frac{1}{r} \right) = 0 ,
   \label{e:glo:syst_circ2}
\eea 
where 
\be
    \nu := \ln N \qquad\mbox{and}\qquad \beta := \ln B . 
\ee

As we did for the fluid, let us introduce the particle angular velocity as seen by 
a distant observer,
\be
  \Omega_{\Pp} := \frac{d\ph}{dt} = \frac{v^\ph}{v^t}  ,
\ee 
as well as the particle velocity $\vv{V}$ measured by the ZAMO.
Then, similarly to Eqs.~(\ref{e:sym:u_circular}), (\ref{e:sym:u_GnU})
and	(\ref{e:sym:u_GnU}) we have
\be \label{e:glo:v_circ_comp}
  \vv{v} = v^t \left( \vxi + \Omega_{\Pp} \vchi \right),
\ee
\be
  \vv{v} = \Gamma_{\Pp} \left( \vv{n} + \vv{V} \right),
\ee
\be \label{e:glo:Gam_p_V}
  \Gamma_{\Pp} = N v^t \qquad \mbox{and}\qquad 
	\vv{V} =  \frac{1}{N} (\Omega_{\Pp} - \omega) \vchi  . 
\ee
As for (\ref{e:sym:U2}), the norm of $\vv{V}$ is $|V|$ with [cf. Eq.~(\ref{e:sym:U_Omeg_omeg}) with $\sin\theta=1$]
\be \label{e:glo:def_V}
	V := \frac{B}{N}(\Omega_{\Pp}-\omega) r .
\ee
The normalization relation $\vv{v}\cdot\vv{v}=-1$ is then equivalent to 
\be \label{e:glo:Gp_Lorentz}
  \Gamma_{\Pp} = \left( 1 - V^2 \right) ^{-1/2} . 
\ee

From Eq.~(\ref{e:glo:part_E}) and (\ref{e:glo:v_circ_comp}), we have
\[
  \bar E = - v_t = -g_{t\mu} v^\mu = 
   (N^2 - B^2 \omega^2 r^2) v^t + \omega B^2 r^2  v^t \Omega_{\Pp}
  = v^t \left[ N^2 + B^2 \omega r^2 (\Omega_{\Pp} - \omega) \right] . 
\]
Thanks to (\ref{e:glo:Gam_p_V}) and (\ref{e:glo:def_V}), we can write
\be \label{e:glo:bE_circ}
  \encadre{ \bar E = \Gamma_{\Pp} \left( N + B\omega r V \right) }. 
\ee
Next, from Eq.~(\ref{e:glo:part_L}) and (\ref{e:glo:v_circ_comp}), we have
\[
  \bar L = v_\ph = g_{\ph\mu} v^\mu = - \omega B^2 r^2 v^t 
  + B^2 r^2 v^t  \Omega_{\Pp}
  = v^t B^2 r^2 (\Omega_{\Pp} - \omega) , 
\]
i.e. thanks to (\ref{e:glo:Gam_p_V}) and (\ref{e:glo:def_V}),
\be \label{e:glo:bL_circ}
  \encadre{ \bar L = B r \Gamma_{\Pp} V }. 
\ee
Combining (\ref{e:glo:bE_circ}) and (\ref{e:glo:bL_circ}), we have
\be \label{e:glo:E_omegL}
  \encadre{ \bar E - \omega \bar L = \Gamma_{\Pp} N } . 
\ee
This result leads to the following expression for the energy of $\Pp$ as measured
by the ZAMO and given by Eq.~(\ref{e:glo:E_ZAMO}): 
\be
	\encadre{ E_{\rm ZAMO} = \Gamma_{\Pp} \, m } . 
\ee
We thus recover Einstein's mass-energy relation. This is not surprising since
$\Gamma_{\Pp}$ is the Lorentz factor of $\Pp$ with respect to the ZAMO. 

In view of the above relations, let us go back to the system 
(\ref{e:glo:syst_circ1})-(\ref{e:glo:syst_circ2}).
Thanks to (\ref{e:glo:E_omegL}) and (\ref{e:glo:bL_circ}), Eq.~(\ref{e:glo:syst_circ1})
is equivalent to $1-\Gamma_{\Pp}^2 + \Gamma_{\Pp}^2 V^2 = 0$. It is trivially satisfied since
$\Gamma_{\Pp}$ and $V$ are related by (\ref{e:glo:Gp_Lorentz}). 
Inserting (\ref{e:glo:E_omegL}) and (\ref{e:glo:bL_circ}) into Eq.~(\ref{e:glo:syst_circ2})
gives rise to a second order equation for $V$:
\be
  \left( \der{\beta}{r} + \frac{1}{r} \right) V^2
  - \frac{Br}{N} \der{\omega}{r} \, V - \der{\nu}{r} = 0 . 
\ee
This equation was first written by Bardeen (1970) \cite{Barde70a}.
It admits two solutions:
\be \label{e:glo:V_pm}
  \encadre{ V_\pm = \frac{ \frac{B r}{N} \der{\omega}{r}
  \pm \sqrt{ \frac{B^2 r^2}{N^2} \left(\der{\omega}{r}\right) ^2 
  + 4 \der{\nu}{r} \left( \der{\beta}{r} + \frac{1}{r} \right) } }{2
  \left( \der{\beta}{r} + \frac{1}{r} \right)} } . 
\ee
This formula gives the circular orbit velocity with respect to the ZAMO as a function of 
$r$. All the metric coefficients in the right-hand side are to be taken at $\theta=\pi/2$. 
The $+$ (resp. $-$) sign corresponds to a motion in the same sens (resp. opposite sens) than 
the stellar rotation: the orbit is then called \defin{direct} (resp. \defin{retrograde}).

At the Newtonian limit, $\omega \rightarrow 0$, $|\dert{\beta}{r}| \ll 1/r$ and
$\nu \sim \Phi/c^2$, so that (\ref{e:glo:V_pm}) reduces to 
\[
  V_\pm = c \frac{\pm \sqrt{ \frac{4}{c^2 r} \der{\Phi}{r} }}{\frac{2}{r}}
\]
i.e.
\be \label{e:glo:V_pm_Newt}
  \encadre{ V_\pm = \pm \sqrt{ r \der{\Phi}{r} } } _{\mbox{\footnotesize\  Newt.}} .
\ee
We thus recover the standard formula obtained by 
equating the gravitational acceleration, $-\der{\Phi}{r}$, and the centrifugal one,
$-V^2 / r$. In particular, in spherical symmetry, $\Phi = -GM/r$ and (\ref{e:glo:V_pm_Newt})
yields to the well known result : $V_{\pm} = \pm \sqrt{GM/r}$.

\begin{remark}
At the Newtonian limit, $V_- = - V_+$, whereas in the relativistic case, if the star is rotating
($\omega\not=0$), $V_- \not = - V_+$. 
\end{remark}

\subsection{Innermost stable circular orbit (ISCO)}

We have seen from Eq.~(\ref{e:glo:eff_pot}) that a circular orbit corresponds to an extremum
of the effective potential: $\dert{\mathcal{V}}{r} =0$. It is 
stable if, and only if, this extremum is actually a minimum, i.e. if, and only if,
\be
  \dderr{\mathcal{V}}{r} > 0 .
\ee 
Taking into account that $\mathcal{V}=0$ and $\dert{\mathcal{V}}{r}=0$ on a 
circular orbit, the above criteria is equivalent to 
\bea
  \lefteqn{ \frac{\bar E - \omega\bar L}{N^2} \left\{
  (\bar E-\omega \bar L) \left[ \dderr{\nu}{r} - 2 \left( \der{\nu}{r} \right) ^2 \right] + {\bar L} \left( \dderr{\omega}{r} - 4 \der{\nu}{r} \der{\omega}{r} \right)
  \right\} } \nonumber \\
  & & + \frac{{\bar L}^2}{B^2 r^2} \left[ - \dderr{\beta}{r}
  + \frac{4}{r} \der{\beta}{r} + 2 \left( \der{\beta}{r} \right) ^2 
  + \frac{3}{r^2} \right]
 - \frac{{\bar L}^2}{N^2} \left( \der{\omega}{r} \right) ^2 > 0 .
\eea
Substituting (\ref{e:glo:E_omegL}) for $\bar E - \omega\bar L$ and
(\ref{e:glo:bL_circ}) for $\bar L$, we get, after division by $\Gamma_{\Pp}^2$,
\bea
  \lefteqn{ \dderr{\nu}{r} - 2 \left( \der{\nu}{r} \right) ^2
  + \frac{ V B r}{N} \left( \dderr{\omega}{r} - 4 \der{\nu}{r} \der{\omega}{r} \right) }
  \nonumber \\
  & & + V^2 \left[ - \dderr{\beta}{r}
  + \frac{4}{r} \der{\beta}{r} + 2 \left( \der{\beta}{r} \right) ^2 
  + \frac{3}{r^2} \right]
  - \frac{V^2 B^2 r^2}{N^2}\left( \der{\omega}{r} \right) ^2 > 0 . \label{e:glo:stab_circ}
\eea
The marginally stable orbit, if any, corresponds to the vanishing of the left-hand side in the above inequality. This orbit is called the
\defin{innermost stable circular orbit (ISCO)}. To compute its location, one has to 
solve the equation formed by replacing the $>$ sign in (\ref{e:glo:stab_circ})
by an equal sign: 
\bea
  \lefteqn{ \dderr{\nu}{r} - 2 \left( \der{\nu}{r} \right) ^2
  + \frac{ V B r}{N} \left( \dderr{\omega}{r} - 4 \der{\nu}{r} \der{\omega}{r} \right) }
  \nonumber \\
  & & + V^2 \left[ - \dderr{\beta}{r}
  + \frac{4}{r} \der{\beta}{r} + 2 \left( \der{\beta}{r} \right) ^2 
  + \frac{3}{r^2} \right]
  - \frac{V^2 B^2 r^2}{N^2}\left( \der{\omega}{r} \right) ^2 = 0 . \label{e:glo:ISCO}
\eea
In this equation, $V$ is considered as the function of the metric potentials $N$, $B$ and $\omega$ given by (\ref{e:glo:V_pm}) and all the potentials are to be taken at
$\theta=\pi/2$. So, once the metric is known, 
Eq.~(\ref{e:glo:ISCO}) is an equation in $r$. If it admits a solution $r > r_{\rm eq}$
(i.e. outside the star), then the ISCO exists. Otherwise, all the circular orbits are 
stable down to the stellar surface. 

At the Newtonian limit ($\nu\simeq \Phi/c^2$, $|\beta| \ll 1$ and $\omega \rightarrow 0$), Eq.~(\ref{e:glo:ISCO}) reduces to
\be \label{e:glo:ISCO_Newt}
  \dderr{\Phi}{r} + \frac{3}{r} \der{\Phi}{r} = 0 ,
\ee
where we have used (\ref{e:glo:V_pm_Newt}) to express $V^2$. 

There are basically two causes for the existence of the ISCO:
\begin{itemize}
\item \emph{The star is highly relativistic:} for instance in spherical symmetry the metric outside the star is Schwarzschild metric (cf. \S~\ref{s:sym:outside}); if the star is compact enough, its surface is located under the famous $\rr = 6M$ which defines
the ISCO in Schwarzschild metric. Note that via the relation (\ref{e:eer:rr_r}) between the
areal coordinate $\rr$ and the isotropic coordinate $r$, $\rr = 6M$ corresponds to
$r = (5/2 + \sqrt{6})M$. 
\item \emph{The star deviates significantly from spherical symmetry:} in Newtonian gravity, 
all the orbits are stable around a spherically symmetric star. But if the gravitational potential deviates 
significantly from $\Phi = -GM/r$, an ISCO may exist, even in Newtonian gravity. 
This was shown explicitly for very rapidly rotating low-mass strange quark
stars in Ref.~\cite{ZduniG01}. For instance, assuming that the deviation from spherical symmetry
arises only from the mass quadrupole moment of the star $Q$ (this is the leading non-spherical term when $r\rightarrow +\infty$), we have\footnote{Equation~(\ref{e:glo:Phi_trunc}) is just Eq.~(\ref{e:eer:Phi_harm}) truncated at $\ell=2$
and with equatorial symmetry, so that the term $\ell=1$ vanishes.}
\be \label{e:glo:Phi_trunc}
  \Phi = -\frac{GM}{r} + \frac{GQ}{r^3} P_2(\cos\theta) , 
\ee
where $P_2(x) = (3x^2-1)/2$ is the Legendre polynomial of degree 2. Inserting the above expression of $\Phi$
in Eq.~(\ref{e:glo:ISCO_Newt}), setting $\theta=\pi/2$ and solving for $r$ leads to 
\be
  r_{\rm ISCO} = \sqrt{ \frac{3Q}{2M} } . 
\ee
\end{itemize}

\appendix

%
%

\chapter{Lie derivative} \label{s:lie}


\minitoc
\vspace{1cm}


\section{Lie derivative of a vector field}

\subsection{Introduction}

We have seen in \S~\ref{s:ein:cov_deriv} that the definition of the derivative of a vector 
field $\vv{v}$ on $\M$ requires some extra structure on the manifold $\M$. This can be
some \emph{connection} $\wnab$ (such as the Levi-Civita connection associated with
the metric tensor $\w{g}$), leading to the \emph{covariant derivative}
$\wnab\vv{v}$. Another extra structure can be some reference 
vector field $\vv{u}$, leading to the concept of derivative of $\vv{v}$ along $\vv{u}$; 
this is the \emph{Lie derivative} presented in this Appendix.  
These two types of derivative generalize straightforwardly to
any kind of tensor field\footnote{For the specific kind of tensor fields constituted by
antisymmetric multilinear forms, there exists a third type of derivative, which does not 
require any extra structure on $\M$: the \emph{exterior derivative}
(see the classical textbooks \cite{MisneTW73,Wald84,Strau04} or
Ref.~\cite{Gourg06} for an introduction).}. 

\subsection{Definition}

\begin{figure}
\centerline{\includegraphics[width=0.6\textwidth]{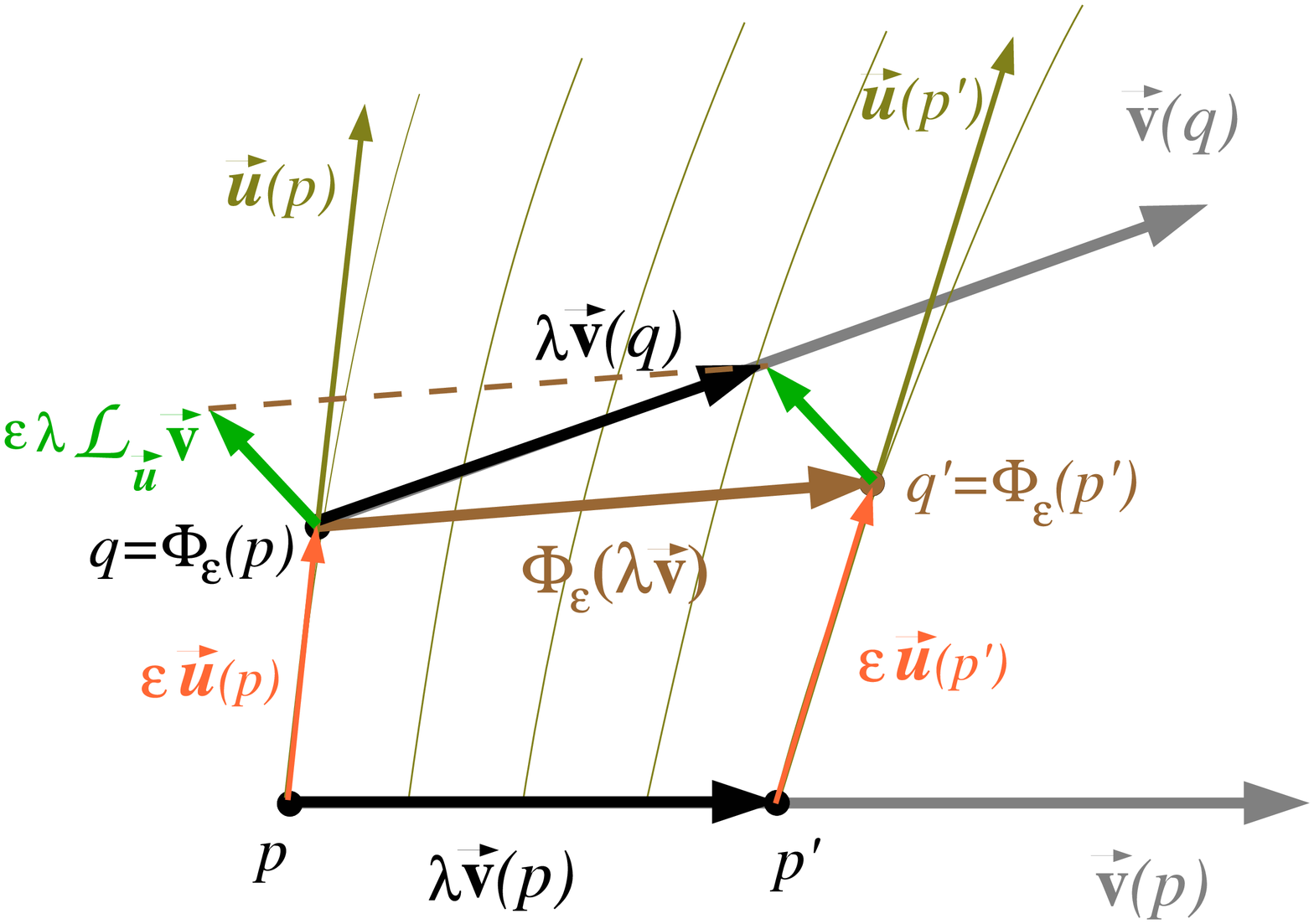}}
\caption[]{\label{f:lie:deriv} Geometrical construction of the Lie derivative of a
vector field: given a small parameter $\lambda$, each extremity of the arrow
$\lambda\vv{v}$ is dragged by some small parameter $\varepsilon$ 
along $\vv{u}$, to form
the vector denoted by $\Phi_\varepsilon(\lambda\vv{v})$. The latter is then compared with
the actual value of $\lambda\vv{v}$ at the point $q$, the difference (divided 
by $\lambda\varepsilon$) defining the Lie derivative $\Lie{\vv{u}}\vv{v}$.}
\end{figure}

Consider a vector field $\vv{u}$ on $\M$, called hereafter the \defin{flow}.
Let $\vv{v}$ be another vector field on $\M$, the variation of which is to be studied.
We can use the flow $\vv{u}$ to transport the vector $\vv{v}$ from one point $p$ to
a neighbouring one $q$ and then define rigorously the variation of $\vv{v}$
as the difference between the actual value of $\vv{v}$ at $q$ and the transported
value via $\vv{u}$. More precisely the definition of the Lie derivative of 
$\vv{v}$ with respect to $\vv{u}$ is as follows (see Fig.~\ref{f:lie:deriv}).
We first define the image $\Phi_\varepsilon(p)$ of the point $p$ by the transport by an infinitesimal ``distance'' $\varepsilon$ along the field lines of $\vv{u}$ as 
$\Phi_\varepsilon(p)=q$, where $q$ is the point close to $p$ such that
$\vp{p q}=\varepsilon\vv{u}(p)$.
Besides, if we multiply the vector $\vv{v}(p)$ by 
some infinitesimal parameter $\lambda$, it becomes an infinitesimal vector at $p$.
Then there exists a unique point $p'$ close to $p$ such that 
$\lambda\vv{v}(p)=\vp{p p}'$.
We may transport the point $p'$ to a point $q'$ along the field lines of
$\vv{u}$ by the same ``distance'' $\varepsilon$ as that used to transport
$p$ to $q$: $q'=\Phi_\varepsilon(p')$ (see Fig.~\ref{f:lie:deriv}). $\vp{q q}'$ is then an
infinitesimal vector at $q$ and we
define the transport by the distance $\varepsilon$ of the vector $\vv{v}(p)$ 
along the field lines of $\vv{u}$ according to
\be
	\Phi_\varepsilon(\vv{v}(p)) := \frac{1}{\lambda} \, \vp{q q}'.
\ee
$\Phi_\varepsilon(\vv{v}(p))$ is vector in $\T_q(\M)$. We may then subtract it from the
actual value of the field $\vv{v}$ at $q$ and define the \defin{Lie derivative}
of $\vv{v}$ along $\vv{u}$ by
\be
	\Lie{\vv{u}} \vv{v} := \lim_{\varepsilon\rightarrow 0} \frac{1}{\varepsilon}
	\left[ \vv{v}(q) - \Phi_\varepsilon(\vv{v}(p)) \right] .
\ee

If we consider a coordinate system $(x^\alpha)$ adapted to the
field $\vv{u}$ in the sense that $\vv{u}=\vpar_0$ where $\vpar_0$ is the first
vector of the natural basis associated with the coordinates $(x^\alpha)$ (cf. \S~\ref{s:ein:gr_space_time}), then
the Lie derivative is simply given by the partial derivative of the vector components
with respect to $x^0$:
\be \label{e:Lie_adapted}
	\left( \Lie{\vv{u}} \vv{v} \right)^\alpha = \der{v^\alpha}{x^0} .
\ee
In an arbitrary coordinate system, this formula is generalized to 
\be
	\encadre{ \Liec{u} v^\alpha = u^\mu \der{v^\alpha}{x^\mu}
	- v^\mu \der{u^\alpha}{x^\mu} }, 
\ee
where use has been made of the standard notation 
$\Liec{u} v^\alpha := \left( \Lie{\vv{u}} \vv{v} \right)^\alpha$.
The above relation shows that the Lie derivative of a vector with respect to another
one is nothing but the commutator of these two vectors:
\be
	\encadre{ \Lie{\vv{u}} \vv{v} = [\vv{u},\vv{v}] } . 
\ee

\section{Generalization to any tensor field}

The Lie derivative is extended to any tensor field by (i) demanding that for
a scalar field $f$, $\Lie{\vv{u}} f = \dd f(\vv{u})$ and (ii) using the Leibniz
rule. As a result, the \defin{Lie derivative} $\Lie{\vv{u}}\w{T}$ of a tensor field $\w{T}$ of type 
$\left({k \atop \ell}\right)$ is a tensor field of the same type, the components of which
with respect to a given coordinate system $(x^\alpha)$ are
\be
\Liec{u} T^{\alpha_1\ldots\alpha_k}_{\qquad\ \; \beta_1\ldots\beta_\ell}=
u^\mu \der{}{x^\mu} T^{\alpha_1\ldots\alpha_k}_{\qquad\ \; \beta_1\ldots\beta_\ell} 
- \sum_{i=1}^k T^{\alpha_1\ldots
\!{{{\scriptstyle i\atop\downarrow}\atop \scriptstyle\sigma}\atop\ }\!\!
\ldots\alpha_k}_{\qquad\ \ \ \  \  \  \beta_1\ldots\beta_\ell}
 \; \der{u^{\alpha_i}}{x^\sigma} 
+  \sum_{i=1}^\ell T^{\alpha_1\ldots\alpha_k}_{\qquad\ \; \beta_1\ldots
\!{\ \atop {\scriptstyle\sigma \atop {\uparrow\atop \scriptstyle i}} }\!\!
\ldots\beta_\ell} 
\; \der{u^{\sigma}}{x^{\beta_i}} . \label{e:Lie_der_comp}
\ee 
In particular, for a 1-form,
\be \label{e:Lie_der_1form}
	\Liec{u} \omega_\alpha = u^\mu \der{\omega_\alpha}{x^\mu}
	+ \omega_\mu \der{u^\mu}{x^\alpha} 
\ee
and, for a bilinear form,
\be \label{e:lie:der_bilin}
	\Liec{u} T_{\alpha\beta} = u^\mu \der{T_{\alpha\beta}}{x^\mu}
	+ T_{\mu\beta} \der{u^\mu}{x^\alpha} + T_{\alpha\mu} \der{u^\mu}{x^\beta} . 
\ee
Notice that, thanks to the symmetry properties of the Christoffel symbols defined by (\ref{e:ein:Christoffel}), the partial derivatives the above equations can be 
replaced by the covariant derivatives $\wnab$ associated with the metric $\w{g}$, yielding
\be
\Liec{u} T^{\alpha_1\ldots\alpha_k}_{\qquad\ \; \beta_1\ldots\beta_\ell}=
u^\mu \nabla_\mu T^{\alpha_1\ldots\alpha_k}_{\qquad\ \; \beta_1\ldots\beta_\ell} 
- \sum_{i=1}^k T^{\alpha_1\ldots
\!{{{\scriptstyle i\atop\downarrow}\atop \scriptstyle\sigma}\atop\ }\!\!
\ldots\alpha_k}_{\qquad\ \ \ \  \  \  \beta_1\ldots\beta_\ell}
 \; \nabla_\sigma u^{\alpha_i} 
+  \sum_{i=1}^\ell T^{\alpha_1\ldots\alpha_k}_{\qquad\ \; \beta_1\ldots
\!{\ \atop {\scriptstyle\sigma \atop {\uparrow\atop \scriptstyle i}} }\!\!
\ldots\beta_\ell} 
\; \nabla_{\beta_i} u^{\sigma} . \label{e:lie:der_comp_nab}
\ee 
\be \label{e:lie:der_1form}
	\Liec{u} \omega_\alpha = u^\mu \nabla_\mu \omega_\alpha
	+ \omega_\mu \nabla_\alpha u^\mu ,
\ee
\be \label{e:lie:der_bilin_nab}
	\Liec{u} T_{\alpha\beta} = u^\mu \nabla_\mu T_{\alpha\beta}
	+ T_{\mu\beta}\nabla_\alpha u^\mu + T_{\alpha\mu} \nabla_\beta u^\mu . 
\ee

%
%

\chapter{Lorene/nrotstar code} \label{s:lor}


\minitoc
\vspace{1cm}


The \texttt{nrotstar} code is a free software based on the C++ \textsc{Lorene} library
\cite{Lorene}. It comes along when downloading \textsc{Lorene}, in the directory
\texttt{Lorene/Codes/Nrotstar}. It is a descendant of \texttt{rotstar}, a previous 
\textsc{Lorene} code described in \cite{GourgHLPBM99} and used in many astrophysical studies (see e.g. \cite{HaensZBL09}, \cite{ZduniBHG08} and references therein). 

\begin{figure}
\centerline{\includegraphics[height=0.35\textheight]{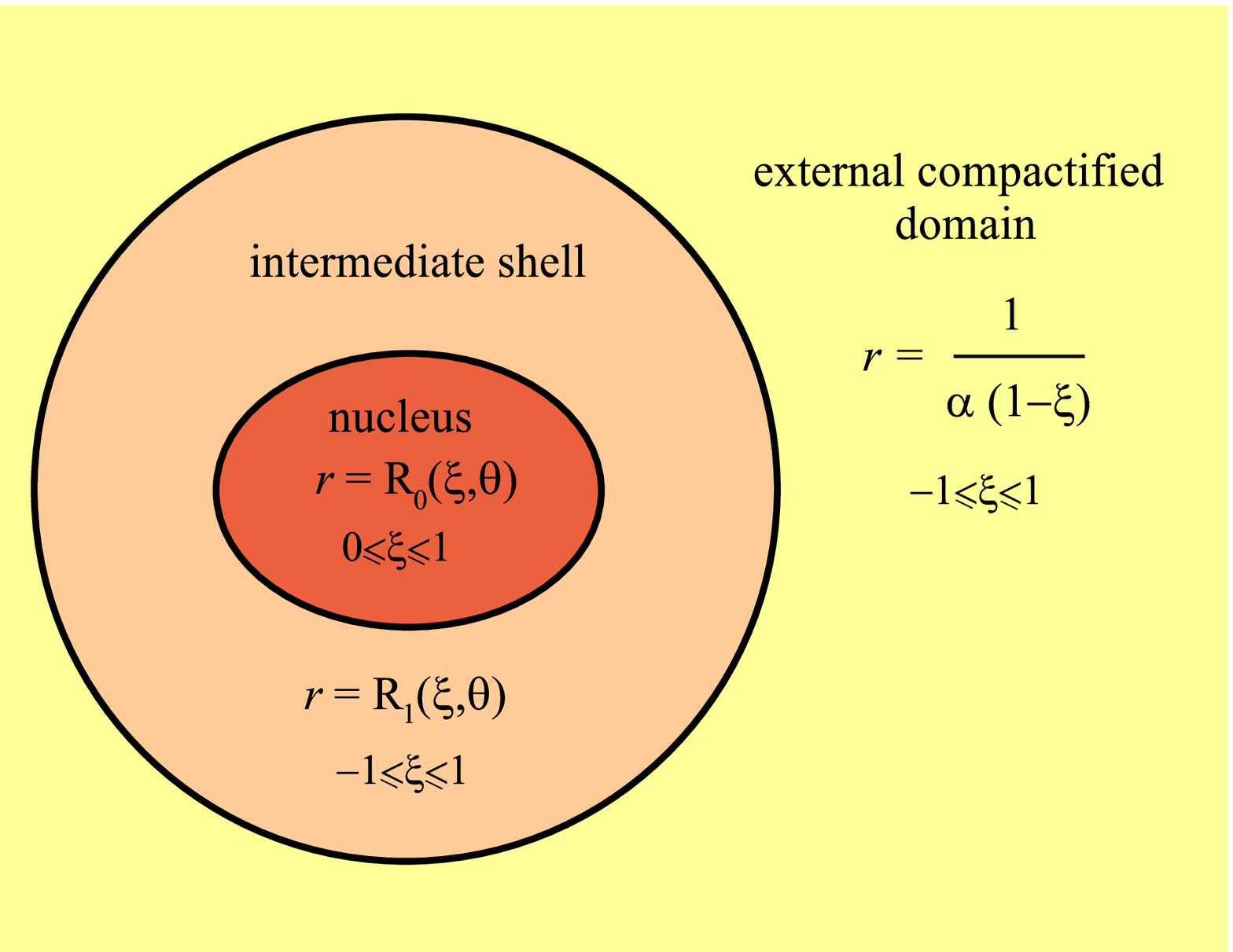}}
\caption{\label{f:lor:domaines} \small
Domains used in the \texttt{nrotstar} code.}
\end{figure}

\texttt{nrotstar} solves the Einstein equations (\ref{e:sym:eq_N})-(\ref{e:sym:eq_lnApnu}) via
the self-consistent-field method exposed in \S~\ref{s:eer:SCF}. For this purpose it uses a
multi-domain spectral method developed in \cite{BonazGM98}
and \cite{GrandBGM01}
(see \cite{GrandN09} for a review about spectral methods).
At least three domains are employed (cf. Fig.~\ref{f:lor:domaines}):
\begin{itemize}
\item a \emph{nucleus}, covering the interior of the star;
\item a \emph{intermediate shell}, covering the strong field region outside the star;
\item an \emph{external compactified domain}, extending to $r=+\infty$. 
\end{itemize}
But more intermediate shells can be used, either inside the star or outside it. For instance,
the nucleus can be dedicated to the fluid interior and a the first shell to the solid crust
\cite{ZduniHG01}, or two domains inside the star can be used to describe a phase transition 
\cite{ZduniHGB04}. 

\vspace{1.5cm}
\centerline{\emph{Not finished yet ! (to be continued...)}}


%
%


\end{document}